\documentclass[12pt]{article}
\usepackage{tabularx}
\usepackage[usenames,dvipsnames,table]{xcolor}
\usepackage[justification=centering]{caption}
\usepackage{graphicx}
\usepackage{float}

\usepackage{bm}
\usepackage{tikz}

\usepackage[legalpaper, margin=0.5in]{geometry}
\usepackage{array}
\usepackage{titling}
\usepackage{geometry}
\usepackage{MnSymbol}
\usepackage{multirow}
\usepackage{xcolor}
\usepackage{amsmath}
\usepackage{graphicx}
\usepackage{float}
\usepackage{hyperref}
\usepackage[]{minted}
\usepackage{enumerate}

\hypersetup{
    colorlinks=true,
    linkcolor=blue,
    filecolor=magenta,      
    urlcolor=blue
    }

\usepackage{pifont}

\usepackage{tabularx,ragged2e,booktabs,caption} 
\definecolor{miverde}{HTML}{00AA00}
\definecolor{miazul}{HTML}{0000AA}
\definecolor{migris}{HTML}{444444}
\definecolor{minaranja}{HTML}{E79000}
\definecolor{LightGray}{gray}{0.9}

\usepackage{amsthm}
\newtheorem{theorem}{Theorem}
\newtheorem{lemma}{Lemma}
\newtheorem{obs}{Observation}

\newtheorem{property}{Property}
\newtheorem{cor}{Corollary}

\long\def\comment#1\endcomment{}

\def\opt{{\rm OPT}}
\def\eop{\hfill$\fbox{}$\medskip}

\usepackage{algorithm}
\usepackage{algpseudocode}
\long\def\comment#1\endcomment{}
\usepackage{caption}

\makeatletter
\newenvironment{breakablealgorithm}
  {
   \begin{center}
     \refstepcounter{algorithm}
     \hrule height.8pt depth0pt \kern2pt
     \renewcommand{\caption}[2][\relax]{
       {\raggedright\textbf{\fname@algorithm~\thealgorithm} ##2\par}%
       \ifx\relax##1\relax 
         \addcontentsline{loa}{algorithm}{\protect\numberline{\thealgorithm}##2}%
       \else 
         \addcontentsline{loa}{algorithm}{\protect\numberline{\thealgorithm}##1}%
       \fi
       \kern2pt\hrule\kern2pt
     }
  }{
     \kern2pt\hrule\relax
   \end{center}
  }
\makeatother

\date{}

\begin{document}
\title{
Variable Parameter Analysis for Scheduling One Machine
}
\normalsize 


\author{Nodari Vakhania\\
Research Center in Sciences, State University of Morelos\\
e-mail: nodari@uaem.mx}
\maketitle


\begin{abstract} 
Many ${\cal NP}$-hard problems resist classical fixed-parameter 
tractability, motivating the need for a more flexible framework. 
We introduce Variable Parameter Analysis (VPA), a framework for analyzing 
${\cal NP}$-hard problems, in which the exponential dependence of the running time is confined to an instance-dependent parameter $\nu(I)$, rather than the input size $n=|I|$. A VP-algorithm identifies the structurally critical 
$\nu(I) < |I|$ objects of each instance $I$, yielding ``computationally tight'' bounds that reflect its true difficulty. 
We study the single-machine scheduling problem with non-simultaneously
released jobs to minimize the maximum job lateness, proposing an exact
VP-algorithm with a simplified upper bound for the runtime  
$\nu(I)!(n-\nu(I))O(n\log n)$. By empirical evaluation, 
$\nu(I)\ll |I|$, with the robustness factor $\nu(I)/|I|$ converging 
to zero as $|I|$ grows. This shows that VPA yields both sharper
theoretical bounds and practical efficiency across diverse instances. 
Based on VPA, we also design a polynomial time approximation scheme that
returns an $(1+1/k)$-approximation solution for an instance $I$
in a simplified runtime bounded by $\kappa(I)!\kappa(I)^2 O(n\log n)$, 
where $\kappa(I) < k$ is a variable parameter;  by empirical evidence
$\kappa(I) \ll k$  making 
instance-dependent factor $\kappa(I)!$ substantially smaller than 
$k^{O(k)}$, the factor in the earlier approximation schemes. 
As a side result of our VPA for the scheduling problem, we obtain a polynomial-time algorithm for the partial scheduling variant where at 
most $n/2$ jobs may be discarded, running in $(n-\nu(I))O(n\log n)$ time. 
Together, these results reveal that VPA, yielding computationally tight bounds, produces efficient algorithms, and explains empirical efficiency across diverse problem instances.
\end{abstract}







\section{Introduction}

A combinatorial optimization problem involves $n$ objects subject to constraints that define a feasible solution set, and seeks a solution that optimizes a given objective function. 
For an ${\cal NP}$-hard optimization problem, it is unlikely to exist an 
exact algorithm that runs in time polynomial in $n$. Traditionally, 
the runtime is expressed in terms of the size $n$ of the input, 
and so an exponential dependence is on $n$. The {\em Fixed Parameter 
Analysis} (FPA) restricts brute-force search to a subproblem of size 
$k$ (a {\em Fixed Parameter}), yielding algorithms with 
runtime $f(k) n^{O(1)}$  in two-dimensional setting $(I,k)$, where $I$ 
is the input of size $n$ (so the the size of the input $|I|$ 
becomes non-critical). The problems, for which this
approach works form the class of {\it FP-Tractable} (FPT) problems. 

Many weakly ${\cal NP}$-hard problems with numerical parameters admit 
pseudo-polynomial algorithms, but parameterizing by these numeric values is often impractical: binary encoding yields exponential dependence, and no reasonable fixed parameter exists for all problems. As a result, the 
hierarchy of non-FPT classes $W[i]$ within the class of non-FPT 
problems culminated with the 
class ${\cal XP}$, emerges. The limitation of FPA is that a fixed parameter 
typically captures one aspect of the problem regardless of the structural properties of individual instances. If there exist ``embarrassing'' instances, no reasonable fixed parameter can be found, even though most instances may be treatable.

To address these difficulties, we propose a {\em Problem Instance 
Dependent} (PID) analysis, based on a {\em Variable Parameter} (VP) 
determined per instance. {\em Variable Parameter Analysis} (VPA) 
studies each input $I$ individually, identifies the structurally 
critical $\nu(I)$ ``emerging objects'' that drive the intractability 
and confines the exponential dependence to them. A VP-algorithm arranges 
the remaining $|I|-\nu(I)$ objects in polynomial time and performs implicit enumeration only on the kernel sub-instance of size $\nu(I)$. 
This yields computationally tight bounds that reflect the true hardness of each instance.
Such a PID approach is highly advantageous for tackling the varying levels 
of difficulty across different instances of the same problem. Indeed, 
it is a well-known fact that the empirical performance of solvers for 
${\cal NP}$-hard problems varies significantly across different 
instance types. 
In Appendix A, we report particular values of the function $\nu(I)$
that we obtained for large number of randomly generated instances of the 
scheduling problem that we consider later. The parameter $\nu(I)$ was
typically far less than $n=|I|$,  and the {\em robustness factor}, 
the ratio $\nu(I)/|I|$, converged to 0 as $|I|$ was growing.  

Formally, a VP $\nu(I)$ is a function
that assigns to every instance $I_P\in {\cal I}(P)$ 
of problem $P$ a non-negative integer $\nu(I_P) < |I_P|$, 
where ${\cal I}(P)$ is the set of instances of the problem 
(for the simplicity, further we omit parameter $P$ in $I_P$). 

A VP-algorithm, for any instance 
$I\in {\cal I}(P)$, runs in time $f(\nu(I)) (|I|-\nu(I))^{O(1)}$, 
where $f$ is a computable function independent of $|I|$ and $\nu(I) < |I|$. 
Note that $\nu(I)$ is not part of the input $I$ but it is implicitly 
determined by $I$. Although the runtime expressions of FP and 
VP algorithms are syntactically similar, they are conceptually different: 
FP uses a fixed parameter $k$, a constant across all inputs, 
while VP uses an instance-dependent function $\nu(I)$   
(in this sense, VPA extends FPA).  

Ideally, $\nu(I)\ll |I|$ is the minimum size up to which the 
reduction of the problem instance is 
possible, i.e., a brute force search is unavoidable for 
any sub-instance of the same or smaller size, unless ${\cal P}={\cal NP}$. 
We do not include this minimality property of function $\nu(I)$ 
in the formal definition as proving it would require currently 
unavailable formal tools in fine-grained complexity theory -  
an open direction in the field. 
 
For illustration purposes, a very simplified 
VP-algorithm for a classical $W[2]$-hard domination problem 
in a graph can be as follows.  Given a graph $I$, 
we look for a smallest subset of vertices
in that graph such that any vertex, not from that set, is adjacent to
at least one vertex from the set. It is easy to see that 
a minimum dominating set of graph $I$ contains all  
support vertices of that  graph (ones adjacent to a 1-degree
vertex) and does not contain its 1-degree neighbors. Hence, we 
obtain the first kernel sub-instance by omitting all these vertices
from graph $I$. We recursively remove support vertices and their 
1-degree neighbors until none remain, leaving a kernel of $\nu(I)$ 
emerging objects. A brute-force search is then confined to this kernel.

In summary, Variable Parameter Analysis (VPA) provides a framework that adapts parameters to individual instances, confining exponential dependence to structurally critical ``emerging'' objects identified by a VP-algorithm. This approach yields computationally tight bounds that reflect the true hardness of each input, explains the empirical variability in solver performance, and extends the reach of parameterized complexity into domains where fixed parameters are impractical. In the next section, we  
situate VPA within the broader landscape of parameterized complexity by discussing earlier work relevant to VPA to highlight both the 
lineage and the novelty of our approach. Then we present
the single-machine scheduling problem and some relevant work. 
In the following subsections, we describe our algorithms, analyze 
their theoretical properties, and report empirical evidence supporting the efficiency of VPA.

\subsection{Earlier work relevant to VPA}

Parameterized complexity has developed several techniques to cope 
with ${\cal NP}$-hardness beyond classical FPT. {\em Kernelization} seeks polynomial‑time preprocessing rules that reduce an instance to a ``kernel'' sub-instance whose size is bounded by a function of the parameter. For many graph problems, kernelization has yielded kernels of size polynomial in the parameter, sometimes even linear. However, kernelization requires a fixed parameter chosen universally across all instances. If no such parameter exists, or if the kernel size bound is too large to be practical, kernelization provides limited insight. In contrast, VPA does not rely on a universal parameter: the kernel size $\nu(I)$ is determined per instance, reflecting its structural difficulty. Thus VPA can yield tighter bounds even when kernelization is ineffective.

{\em Graph-structural PID parameters} have also been used to create FPT algorithms; examples include vertex cover size, cutwidth, pathwidth, treewidth, tree depth, maximum leaf number, maximum matching, and others. Formally, given a parameter $p$, let $p(I)$ denote the number of vertices
corresponding to that parameter in graph $I$. Several problems that are 
non-FPT in general become FPT when restricted to graphs with bounded 
$p(I)$, with algorithms running in  $O(\alpha^k n^{O(1)})$ time. However, 
for most parameters $p$, deciding whether $p(I)\le k$ is  
${\cal NP}$-complete. Even when polynomial (e.g., edge cover number), 
the exponential base $\alpha$ is typically very large, making bounds 
of the form $\alpha^k$ of limited practical use.

{\em Above-guarantee parameterizations} define parameters relative to a natural lower bound or guarantee, measuring the difference between the solution size and the bound. For some problems, such bounds are derived relative to the input size $|I|$ (e.g., MaxSAT and MaxCUT), while for others only PID lower bounds are known. A classical example is Vertex Cover, where the maximum matching number or a linear programming relaxation provide lower bounds. This approach has been fruitful when the derived fixed parameters are ``small enough''. It works well whenever there exists a meaningful baseline that can be obtained relatively easily and the solution size above the trivial bound is relatively small. VPA differs fundamentally: it does not measure deviation from a guarantee, but instead identifies the structurally critical objects of each instance. The exponential dependence is confined to these emerging objects, which vary from instance to instance.

Thus VPA extends parameterized analysis by allowing the parameter itself to vary with the input. It expresses runtime in terms of an 
instance-sensitive VP $\nu(I)$, whereas FPA does so in terms of a fixed parameter $k$ that does not really reflect the intractable structure of instance $I$.  
Unlike FPA, VPA can be applied to any combinatorial optimization problem 
which structural properties differ widely across instances and 
the hardness is concentrated in a relatively small subset of 
structurally critical objects.  
By confining exponential dependence to these emerging objects, VPA provides computationally tight bounds that reflect the true difficulty of each instance, offering a unifying framework that extends beyond specific domains to the general theory of parameterized complexity.

\subsection{Scheduling problem} 

The single-machine scheduling problem with non-simultaneously 
released jobs that we study here can be
cast in two equivalent alternative forms. We have $n$ independent 
jobs from set $J$ that have to be scheduled non-preemptively on a 
single machine, which can process at most one job at a time. Each
job $j$ becomes available at its {\em release time} 
$r_j$ (it cannot be assigned to the machine before that time) and
needs to be processed uninterruptedly on the 
machine during $p_j$ time units, called the {\it processing time}. 
All job parameters are non-negative integer numbers.
In both formulations, a {\em feasible schedule} $S$ is a mapping that
assigns every job $j$ a starting time $t_j(S)$ on the machine so that 
the just stated resource, release and processing time restrictions 
are satisfied. 

In the first, more common setting of the problem, each job $j$ has
one additional parameter, {\it due date} $d_j$ (informally, it is 
the desirable time for the completion of job $j$). 
The penalty for late completion of job $j$ is measured by its 
{\em lateness} $$L_j = c_j(S) - d_j,$$ where $$c_j(S)=t_j(S)+p_j$$ is the {\em completion time} of job $j$ {\em on the machine}.
The objective is to find an 
{\it optimal schedule}, a feasible one with the minimum maximum 
job lateness $L_{\max}=\max_{j\in J} L_j$. In the standard scheduling 
notation \cite{Gr-et-al} it is abbreviated as $1|r_j|L_{\max}$: 
In the first field the single-machine environment is indicated, the second 
field specifies job parameters, and in the third field the objective criterion
is given (job processing times are not explicitly specified, and 
the $L_{\max}$ criterion implies jobs with due dates and so this
parameter is also omitted).

In a feasible schedule for $1|r_j|L_{\max}$ some jobs may 
have negative lateness, that causes an inconvenience in expressing 
an approximation ratio. This issue is naturally settled in the
alternative setting $1|r_j,q_j|C_{\max}$, in which the due date 
of each job $j$ is replaced by its {\em delivery time} $q_j$: 
once finished on the machine, job $j$ needs the additional time  $q_j$ to 
be {\em fully completed} (notice that $q_j$ contributes to the completion 
time of job $j$ but it does not require machine time, and hence
the machine can process another job during the delivery of job $j$);  
$$C_j(S)=c_j(S)+q_j$$ is the {\em full completion time} of 
job $j$ in schedule $S$ (note that unlike $c_j(S)$, $C_j(S)$ 
includes the delivery of job $j$). The objective is to find a 
feasible schedule $S$ minimizing the maximum job full completion 
time $C_{\max}(S)=\max_{j\in J} C_j(S)$, referred to  as the {\em makespan} 
(we note that in the settings without delivery times, 
``makespan'' $C_{\max}$ commonly refers to maximum completion time of a job 
{\em on the machine}). 

The problem is strongly $NP$-hard Garey and Johnson \cite{GJ}, important  
on its own, and serves as a foundation for multiprocessor and shop 
scheduling problems.  An instance of $1|r_j,q_j|C_{\max}$ can be 
transformed to a symmetric instance of $1|r_j|L_{\max}$ in linear 
time: Take a number 
$K \ge \max_j q_j$ and define due date of job $j$ as $d_j=K-q_j$. 
Vice-versa, given an instance of $1|r_j|L_{\max}$, take a magnitude 
$D \ge \max_j d_j$ and define job delivery times as $q_j=D-d_j$. 
An optimal solution to a problem instance of either of the two settings gives 
an optimal solution to the equivalent instance of the other setting 
Bratley et al. \cite{Br}. A detailed survey of related scheduling 
results is provided in the following section.

\subsection{Earlier results related to the 
scheduling problem $1|r_j,q_j|C_{\max}$} 

Jackson \cite{J} proposed an $O(n\log n)$  algorithm
for an optimal solution of problem $1||L_{\max}$ 
with simultaneously released jobs. 
Iteratively, the algorithm 
schedules the next unscheduled job with the 
largest delivery time (equivalently, with the smallest due date) 
at the earliest machine idle time. The extended version of this algorithm, 
described by Schrage \cite{Sch}  gives a 
2-approximation solution to scheduling problem $1|r_j,q_j|C_{\max}$: 
Iteratively, the current scheduling time $t$ is the maximum between
the minimum release of a yet unscheduled job and the completion time 
of the last scheduled job (if any). Among the jobs released by time 
$t$, the heuristic schedules a job with 
the largest delivery time (or the smallest due date). We use the extended 
algorithm as a schedule generator, referring to it as LDT-heuristic. 

Among the early  branch and bound algorithms for the problem, we mention 
those given by McMahon \& Florian \cite{Mc} and Carlier \cite{C2}. For the
later implicit enumerative algorithms, see Pan and Shi \cite{pan}, Della 
Croce \&  T'kindt \cite{fed}, and Gharbi \& Labidi \cite{semi}. 
Horn \cite{horn} observed that LDT-heuristic delivers an optimal solution 
if  all jobs have unit processing time (and job parameters are
integers). With real valued job release times and unit job processing times, the 
problem can be scaled to an equivalent one in which all job parameters are 
integers and job processing times are equal. While applying LDT-heuristic 
for the non-preemptive case, during the execution of a job another more 
{\it urgent} job (i.e., one with a larger delivery time, equivalently, a smaller due date) may be released, that 
makes the heuristic non-optimal. In the preemptive 
variant $1|r_j,q_j,pmtn|C_{\max}$ we allow to preempt the
former job and schedule the latter one at its release time. Then 
we can solve the preemptive problem optimally in $O(n\log n)$ time. 

The non-preemptive setting with equal length jobs can be 
solved in $O(n^2\log n)$ time. Garey, Johnson, Simons \& Tarjan  
\cite{GJS} developed a  
sophisticated data structures with a union and find tree with path compression 
and achieved an $O(n\log n)$ time complexity. 
For a more general 
setting, where every job processing time divides every larger job processing time, 
Vakhania  \cite{1div} gave an $O(n^2\log n \log p_{\max})$ time algorithm, where $p_{\max}$ is the maximum job processing time. 

Better than 2-approximation polynomial-time algorithms exist for problem
$1|r_j,q_j|C_{\max}$. Potts \cite{p} showed that by repeated application of 
LDT-heuristic $O(n)$ times, the performance ratio can be improved to 3/2, resulting in 
an $O(n^2\log n)$ time performance. Nowicki and Smutnicki \cite{nov} have proposed
another 3/2-approximation algorithm with time complexity $O(n \log n)$.  Hall and 
Shmoys \cite{HS} illustrated that the application of the LDT-heuristic to the original 
and a derived reversed problems leads to an improved approximation 
of 4/3 in runtime $O(n^2\log n)$. Besides, \cite{HS} describes two 
polynomial-time approximation schemes with time complexities $O(16^k (nk)^{3+4k})$ and $O(n\log n+n(4k)^{8k^2 + 8k+2})$. 
Mastrolilli \cite{mast} suggested another approximation scheme for 
a more general setting with identical processors, 
and showed that his algorithm  has an aggregated time complexity 
$O(nk^{O(1)}+k^{O(k)})= O(n+k^{O(k)}).$ These PTASs carry 
out a complete enumeration of all possible combinations, yielding a tight 
exponential-time dependence $O(k^k)$ (see Section 6 for a more  
detailed overview). 

As to the known parametrization results,  problem 
$1|r_j,q_j|C_{\max}$, it is FP-tractable with the numeric
parameters $p_{\max}$ and $d_{\max}$, where $p_{\max}$ and $d_{\max}$ 
are the maximum job processing time and due date, respectively
Vakhania and Werner \cite{tcs13}. Note however, that 
the fixed parameter tractability with these parameters is by 
their numeric values, but not by their binary encodings. 
Mnich and van Bevern \cite{PA over} 
overview some other FP-algorithms for scheduling problems. 
Nederlof and Swennenhuis \cite{partial scheduling}
conducted FPA for scheduling problems where the fixed parameter 
$k$ is the number of scheduled jobs and with  $C_{\max}$ 
objective criterion; a parameterized scheduling problem with 
such fixed parameter is referred to as {\em partial scheduling}.  

\subsection{Our contributions} 

We confine exponential dependence to instance-specific emerging jobs, yielding computationally tight bounds. Variable parameter analysis  is a general paradigm that adapts parameters to individual instances and confines exponential dependence to structurally critical $\nu(I)$ emerging objects. 
For the single-machine scheduling setting $1|r_j,q_j|C_{\max}$, 
we design an implicit enumeration algorithm whose complexity depends factorially on the number of  emerging jobs. Informally, an {\em emerging job} is one with wide time window that can delay urgent jobs with narrow time windows and thereby force the makespan. These jobs, identified via critical segments    (that we call kernels) in heuristic schedules, form the structurally hard 
core of instance $I$; the number of emerging jobs in $I$ is the 
variable parameter $\nu(I)$. 
We also obtain a polynomial-time algorithm for the variant where at most 
$n/2$ jobs may be discarded. Once a baseline guarantee (scheduling at least half the jobs) is imposed, the problem becomes polynomially solvable without further parameterization. Moreover, by construction, the algorithm omits only emerging jobs: in practice, discarding non-urgent jobs with wide time windows prevents delays of urgent ones, a relevant property in overloaded systems.
Finally, we develop a PTAS that improves earlier schemes by replacing the exponential factor $k^{O(k)}$ with a smaller, instance-dependent factorial bound $\kappa(I)!$:  in practice, $\kappa\ll k$ 
since the emerging jobs typically form a small subset of all
jobs (see Section 6 for empirical evidence). 
Hence VPA can yield a significant practical speedup.
In the next theorems we give simplified runtimes of our algorithms 
(see Theorems \ref{IEA} and \ref{PTAS} for precise refined runtime 
expressions). Throughout the paper, all multiplicative factors shown 
outside the $O(.)$ terms are upper bounds rather than exact 
equalities (tighter PID bounds may hold in practice). 

{\bf Theorem 1.1 (Exact Algorithm).}  
Problem $1|r_j,q_j|C_{\max}$ admits an exact algorithm with 
a simplified runtime $$\nu(I)!(n-\nu(I))O(n\log n),$$ where 
$\nu(I)$ is the number of emerging jobs in instance $I$.\eop 

{\bf Theorem 1.2 (Partial Scheduling). } 
The partial scheduling of the variant of $1|r_j,q_j|C_{\max}$ where at most 
$n/2$ jobs may be discarded, admits a polynomial-time algorithm with 
runtime $$(n-\nu(I)) O(n\log n).$$\eop 

{\bf Theorem 1.3 (Polynomial-Time Approximation Scheme).}  
Problem $1|r_j,q_j|C_{\max}$  admits a PTAS that returns a 
$(1+1/k)$-approximation solution  in a simplified runtime bounded 
by $$\kappa(I)!\kappa(I)(n-\kappa(I)) O(n\log n),$$ where 
$\kappa(I) < \nu(I)$ is a variable parameter with $\kappa(I) < k$.\eop

In summary, our work introduces Variable Parameter Analysis (VPA) as a general framework for instance-sensitive parameterized complexity that yields computationally tight bounds and practical efficiency across diverse inputs.
The distinction between fixed and variable parameters becomes particularly natural in the scheduling domain. Here, the structural hardness of an instance is not governed by a universal constant but by the number of emerging jobs 
$\nu(I)$, which varies across inputs. Urgent jobs with narrow time windows form kernels, while flexible jobs with wide time windows can delay them, creating bottlenecks that define $\nu(I)$ (for notational simplicity, we omit the argument $I$ in $\nu(I)$ from here on).
This perspective shows how VPA extends FPA: exponential dependence is confined to the few structurally critical emerging jobs, while the majority of jobs can be arranged in polynomial time. The remainder of the paper is organized as follows: Section 2 introduces preliminaries, Section 3 presents the implicit enumeration algorithm, Section 4 discusses partial scheduling, Section 5  develops the PTAS, and Section 6 concludes with remarks and future directions.
--------Appendix A presents a brief description of our experimental study, and
Appendix B gives alternative probabilistic runtime bounds.


\comment


Emerging jobs force a delay of other
jobs and make the problem instance intractable, 
in the sense that the derived instance without them becomes 
polynomially solvable. To identify the emerging jobs, we create 
a series of LDT-schedules and detect critical segments in 
these schedules that realize the makespan, referred to as 
kernels.\footnote{This concept of a kernel is different from
a kernel in FPA: the latter type of kernel consists of emerging 
objects in VPA.} Roughly, emerging jobs are those who 
force a delay of kernel jobs. 

Next, we specify basic steps of our IEA. 
At stage 0 it determines the kernels and the emerging 
jobs, and it decomposes each kernel into smaller atomic 
kernels and type (1.2) jobs. At stage 1, different ways of 
distribution of type (1) jobs in between atomic components are 
verified considering corresponding permutations of these jobs. 
In aggregated terms, IEA is as follows: 

\begin{algorithm}[H]
\renewcommand{\thealgorithm}{}
\caption{IEA}\label{alg:cap}
\begin{algorithmic}[1]
\State \textbf{input:} Set $J$ of $n$ jobs
\State $\sigma^*:=$ Stage 0($J$) \Comment{identify atomic kernels and emerging jobs 
and form partial schedule $\sigma^*$ consisting of all non-emerging jobs}
\State \textbf{for each} next permutation $\pi$, \textbf{call}
stage 1 \Comment{incorporate emerging 
jobs from permutation $\pi$ in schedule $\sigma^*$ according to their order in that permutation} 
\State \Return a best complete schedule created at stage 1
\end{algorithmic}
\end{algorithm}

From here on, we omit the argument $I$ in function $\nu(I)$
for notational simplicity. Both stages 0 and 1 in IEA run in time  
$(n-\nu) O(n \log n)$ and there are at most 
$\nu_{1.1}! \ \sigma_{K\in \cal K}\nu_{1.2}(K)!$ permutations of 
type (1) jobs to be considered, where $\nu_{1.1}=\nu_{1.1}(I)$ 
($\nu_{1.2}=\nu_{1.2}(I)$, respectively) is the number of type (1.1) 
(type (1.2), respectively) jobs in instance $I$). Hence, IEA runs in time 
$$\nu_{1.1}! \ \sigma_{K\in \cal K}\nu_{1.2}(K)! (n-\nu) O(n \log n).$$



If we consider different orders in which type (1.1) jobs can be distributed in 
between the kernels, and type (1.2) jobs, associated with a kernel, can be 
scheduled within the remaining jobs of that kernel, we obtain 
a worst-case running time $$O(n^2\log n + n\log n (\nu_{1.1}! +
p_{\max}^{1.1}(n-\nu)2^{\nu_{1.2}^{aver}(K)}))$$ or
$$O(n^2\log n + n\log n \ \nu_{1.1}!(n-\nu)2^{\nu_{1.2}^{aver}(K)})$$ 
of the IEA, where the variable parameter $\nu_{1.1}$ is the total 
number of type (1.1) jobs, $p_{\max}^{1.1}$ is the 
maximum processing time of a type (1.1) job, and $\nu_{1.2}^{aver}(K)$ is 
the average number of the type (1.2) jobs in a kernel. 
Using some ``metric'' properties of feasible schedules, we obtain an 
alternative worst-case time complexity expression 
$$O((n^2\log n + n\log n (((r_j+p_{\max}^{1.1})/p_{\min}^{1.1})! q_{\max}! + 
 \ (n-\nu) p_{\max}^{1.1}2^{(p_{\max}^{1.1}+P_{1.2})/{p_{\min}^{1.2}}})$$
without variable parameters, where $p_{\min}^{1.1}$ is the minimum processing time 
of a type (1.1) job, $q_{\max}$ is the maximum delivery time of a non-emerging
job,  $P_{1.2}$ is the average total processing time of the type (1.2) jobs
and $p_{\min}^{1.2}$ is the minimum processing time of a type (1.2) job, 
the last two magnitudes being taken over all kernels. 

This partial schedule is extended to less than $\nu!$ 
complete schedules containing all emerging jobs. Each 
complete schedule is generated by a polynomial time procedure that 
distributes the jobs from a dominant permutation of some emerging jobs
in between the jobs from the above partial schedule.

$p_{\min}(K)$ is the minimum processing time 
of a type (1.2) job from kernel $K$ , 
 $P_{1.2}(K)$ is the total processing time of jobs in set $J_{1.2}(K)$

The PTAS partitions the 
set of jobs into short and less than $k$ long jobs, and generates a
solution, at most $(1+1/k)$ times far from an optimal one. 
It can be implemented without using VPA, in which case  
it enumerates all possible $k!$ permutations
of the {\em long jobs}, yielding factor $k!$  in its runtime. 
So roughly, the factor $k^{O(k)}$ in the earlier schemes 
is replaced by $k!$. With VPA, $k!$, in turn, is replaced
by $$\kappa_{1.1}! \ \sigma_{K\in \cal K}\kappa_{1.2}(K)!,$$ where
$\kappa_{1.1}=\kappa_{1.1}(I)$ ($\kappa_{1.2}=\kappa_{1.2}(I)$, 
respectively) is the number of {\em long type (1.1) (long type (1.2) 
jobs}, respectively); $\kappa=\kappa(I)= \kappa_{1.1}+\kappa_{1.2}$ 
is the total number of long emerging jobs. This results in an 
overall runtime $$\kappa_{1.1}\kappa O(n\log n)  \kappa_{1.1}! \ \sigma_{K\in \cal K}\kappa_{1.2}(K)!.$$ Although $\nu$ is a variable parameter, 
for any fixed approximation factor in the PTAS, the parameters 
$\kappa_{1.1}$, $\kappa_{1.1}$ and $\kappa$ are constants. Notice 
that the use of VPA is crucial, as $\kappa<k$, and $\kappa$, 
being considerably smaller than $\nu$, will be substantially smaller 
than $k$ (see Appendix A for the experimental behavior of function $\nu$). 

\endcomment

\section{Preliminaries}

Several technical concepts and tools related to the scheduling problem have appeared in the literature, often under different notations. In this section we unify and refine these notions in the form needed for our analysis. While the definitions and proofs are specialized, they build on established ideas in the scheduling area, and the reader familiar with this domain will recognize their underlying structure.

During the construction of an LDT-schedule $S$, we iteratively update 
the current scheduling time $t$ as the largest magnitude between 
the completion time of the job scheduled the last so far and the 
release time of the earliest released yet unscheduled job.
Suppose $A\subset J$ and $I(i,S)$ be the execution interval of job $i$ 
in LDT-schedule $S$. Let $S(-i,+A)$ be an alternative complete LDT-schedule 
in which the interval $I(i,S)$  is reserved for the jobs from set $A$ 
being scheduled by LDT-heuristic; i.e., no job, different from one from
set $A$  can be included in schedule $S(-i,+A)$  
until LDT-heuristic includes the  last job from set $A$.
If set $A$ consists of a single job $j$ then we will just write $S(-l,+j)$. 

Suppose job $i$ precedes job $j$ in schedule $S$ and job $j$ is released
within the interval $I(i,S)$. Then we will say that job $i$ {\em pushes} 
job $j$ in schedule $S$. Note that if job $i$ pushes job $j$ in schedule 
$S$, then job $j$ is started at time $r_j$ in schedule $S(-i,+j)$. 

A largest component of an LDT-schedule $S$ is a {\it block}, a longest 
sequence of the successively scheduled jobs in that schedule,
such that, for each two neighboring jobs $i$ and $j$, job $j$ starts 
immediately at the completion time of job $i$ and $r_j(S) < c_i(S)$; 
if $j$ is the last job of the block, $r_l(S) \ge c_j(S)$, 
for any job $l$ succeeding job $j$ in schedule $S$. 

A block of an LDT-schedule $S$ may contain one or more kernels: 
A {\em kernel} $K$ is a longest LDT-sequence of successively scheduled 
jobs $(\{j_1,\dots,j_l\})$ (i.e., $q_{j_i}\ge  q_{j_{i+1}}$, for all 
$i=1,\dots,l-1$) such that $$C_{j_k}(S) = {\max}_{i\in J}\{C_i(S)\}.$$ 
We will refer to the job $j_k$, as the {\it overflow} job of kernel 
$K$ and  denote it by $o=o(K)$; $K$ may include more than one 
job realizing ${\max}_{i\in J}\{C_i(S)\}$, $o(K)$ is the last
scheduled one. Note that $K$ contains no idle-time interval 
(a {\em gap}) and that 
$q_{j_i}\le \sum_{\iota=i+1}^m p_{j_{\iota}} + q_{j_{m}}$, 
for all $i=1,\dots,k-1$ and $m=i+1,\dots,l$,  
as otherwise job $j_i$ would end the kernel.  
The overflow job $o(K)$ is either succeeded by a gap or  
by job $j$ with $C_j(S) < C_o(S)$ (and hence $q_j < q_o$). 

We denote by $K(S)$ the earliest kernel in schedule $S$, and we let
$r(K)=\min_{i\in K}\{r_i\}$ be the minimum release time of a job 
in $K$. Although a kernel and a block are defined as  
sequences of jobs in a schedule, abusing the terminology, we may 
also refer to these components as the corresponding job-sets.

Suppose in kernel $K=(\{j_1,\dots,j_l\})$, job $j_1$ starts after 
its release time. Then it must be pushed by a job $\lambda$ with 
$q_\lambda < q_o$ as otherwise $(\{\lambda,j_1,\dots,j_l\})$ would
form an LDT-sequence and hence be part of $K$. More 
generally, let $e$ be a job from the block to which 
kernel $K$ belongs, such that $q_e < q_o$ and $r_e < r(K)$. 
Then we will refer to it as an 
{\it emerging} job for kernel $K$, and to the above job $\lambda$  
as the {\it delaying} emerging job for that kernel. Job $\lambda$, 
being a most urgent job among the already released non-urgent emerging jobs,  
has a ``large enough'' processing time so that it forces a delay 
of the soon-to-be-released kernel jobs. A delaying emerging job
dominates the schedule as its long processing time creates 
a bottleneck that directly impact the urgent kernel jobs. 
If kernel $K$ possesses no delaying emerging job,  
LDT-heuristic will include the first job of that kernel at its 
release time $r(K)$ (kernel $K$ will start a new block).  
The following observation, on which we rely in our enumeration 
process, can be easily seen: 

\begin{obs}\label{inside}
A feasible schedule in which an emerging job for kernel $K$
is scheduled in between the jobs of that kernel can be transformed
into another feasible schedule with a no-larger makespan in which 
that job appears either before or after all jobs of kernel $K$.
\end{obs}

\smallskip

{\bf Example 0.}
To illustrate the above notions, consider an airline scheduling scenario. Each passenger corresponds to a job with predictable release and due times (arrival and departure). Most passengers are expected to have tight time windows and thus behave as urgent, non-emerging jobs. A smaller set of passengers with wide time windows can delay urgent ones; these correspond to emerging jobs. In this analogy, passengers with ``close enough'' arrival 
dates would normally form a block. Within a block, 
the urgent passengers naturally form kernels: the amount of time they can be right-shifted within the schedule is quite restricted. Identifying and isolating emerging jobs that may cause a delay to the urgent jobs captures the structural hardness of the instance. Once the kernels (urgent jobs) are separated from the non-urgent emerging jobs, the former can be scheduled efficiently. Note that it is natural to expect many more passengers with sufficiently narrow time windows than ones with wide time windows, i.e.,
 $\nu(I)\ll |I|-\nu(I)$. The number of emerging jobs in instance $I$ defines the variable parameter $\nu(I)$, which quantifies potential structurally critical parts of schedules and determines the exponential component of the runtime underlining our framework.\eop  

By modifying the original problem instance and then applying 
LDT-heuristic, different alternative LDT-schedules can be generated.
Given an LDT-schedule $S$ with kernel $K$ and an
emerging job $e$, in an alternative LDT-schedule denoted by $S_e$, 
job $e$ is said to be {\em activated}  for kernel $K$ if it 
is forced to be scheduled after all jobs of that kernel. We say
that job $e$ is {\em strongly activated} in schedule $S_e=S(-e,+K)$
if, in addition, all the emerging jobs that were scheduled 
after kernel $K$ in $S$ remain scheduled after that kernel. 

Observe that if job $e$ is strongly 
activated in schedule $S_e$, then job $e$ together with
all the emerging jobs for kernel $K$ scheduled after that kernel in
schedule $S$ are in the {\em state of activation} for kernel $K$ in
schedule $S_e$. If $l$ is the delaying emerging job, then 
the first included job $j$ of kernel $K$ starts at 
its release time $r_j=r(K)$ in  schedule $S_l$. If job $e$ is 
non-strongly activated,  LDT-heuristic may include some emerging job(s) 
before kernel $K$ in schedule $S_e$ and $t_j(S_e)$ can be larger than $r_j$ 
and $r(K)$. 

In terms of Example 0, activating an emerging job corresponds to the airline intentionally delaying service to the emerging passenger until all urgent passengers from the kernel are served. If there are emerging passengers that can be served quickly enough without critically delaying the urgent kernel passengers, they may still be served before the kernel passengers (note that the serving time of a passenger corresponds to the processing time of the job). Under strong activation, however, this flexibility is disallowed: all emerging passengers remain scheduled after the kernel. 
Activation models how emerging jobs are intentionally postponed relative to kernels, while the delaying job highlights when an emerging job's processing time is large enough to force such postponement. This
captures the structural interaction between emerging jobs and kernels that underlies the hardness parameter $\nu(I)$ providing the foundation for our exact and approximation algorithms.

In summary, Example 0 provides the guiding intuition: urgent passengers with narrow time windows form kernels, while flexible passengers with wide time windows correspond to emerging jobs. The notion of a delaying emerging job captures when such flexibility becomes structurally critical, and the activation/strong activation framework formalizes how emerging jobs interact with kernels in LDT‑schedules. Together, these concepts define the variable parameter $\nu(I)$, which isolates the structural hardness of the instance.

\subsection{Creating alternative LDT-schedules}

We start this subsection by formalizing some concepts that we 
illustrated on Example 0. 
Let $l$ be the delaying emerging job of kernel $K\in S$. In a
relatively easy scenario, the sequence of the jobs of kernel $K$ in both
schedules $S$ and $S(-l,+K)$ is the same. Otherwise, there occurs an 
{\em anticipated} job in schedule $S(-l,+K)$, a job rescheduled by
LDT-heuristic to an earlier position in schedule $S(-l,+K)$ compared
to its position in schedule $S$ (see the next subsection). 
If there is no anticipated job in schedule $S(-l,+K)$, 
the starting time of the last scheduled job of kernel $K$ in that schedule 
will be $$\delta(K,S) = c_l(S)-r(K)$$ time units less compared to $C_o(S)$.    
So $\delta(K,S)$ is the {\em delay} of kernel $K$ in schedule $S$, 
a forced right-shift imposed by the delaying emerging 
job $l$ on kernel $K$ in that schedule. 

Given an emerging job $e$, we 
construct  schedule $S_e$ ($S(-e,+K)$, respectively), applying 
LDT-heuristic to a modified problem instance, in which the release 
time of job $e$ (and that of any emerging job included 
after kernel $K$ in schedule $S$, respectively) 
is no less than that of any job of kernel $K$. Then by LDT-heuristic, 
job $e$ (and the above emerging jobs, respectively) 
will appear after all jobs of kernel $K$ in schedule $S_e$ ($S(-l,+K)$, 
respectively). 
Note that LDT-schedule $S$ can be restored by the {\em revision} of 
the activated job $e$, that is, by restoring the modified release time of 
the corresponding emerging job(s) and applying again  LDT-heuristic to the restored problem instance.

Our algorithms ``distribute'' emerging jobs in between kernels. 
An emerging job is to be 
included either before or after all jobs of the corresponding
kernel (Observation \ref{inside}). If a job $e$ is scheduled 
behind all jobs of kernel $K\in S$, 
it cannot be completed earlier than $p_e$ time later than the overflow 
job $o$ of that kernel completes. Since job $o$ has the minimum delivery 
time among all jobs of kernel $K$,  $q_e < q_o$ must hold (i.e., $e$
must be an emerging job) as otherwise the makespan in any resultant
feasible schedule will not be less than that in schedule
$S$. Based on the following lemma, we  
strengthen the definition of an emerging job reducing in this way
the set of emerging jobs due to the earlier definition:

\begin{lemma}\label{emerging job}
Among all emerging jobs for kernel $K$ in an LDT-schedule $S$,
it suffices to consider emerging jobs $e$  with 
$q_e < q_o - \max\{0,p_e -\delta(K,S)\}$.
\end{lemma}
Proof. In an LDT-schedule $S_e$ in which $e$ appears after $K$, we wish  
$C_e(S_e) < C_o(S)$ to hold. Expanding both sides of this inequality, 
we have $$t_e(S)+p_e+q_e = C_e(S_e) < C_o(S) = c_o(S) + q_o.$$
Since the last job of kernel $K$ can be shifted to the left by at
most $\delta(K,S)$ time units, $$t_e(S_e)\ge  c_o(S) - \delta(K,S).$$
Then 
$$c_o(S) -\delta(K,S) +p_e+q_e\leq t_e(S) + p_e + q_e < c_o(S) + q_o.$$
From 
$$c_o(S) -\delta(K,S) +p_e+q_e < c_o(S) + q_o$$ we have  
$$q_e < q_o - p_e + \delta(K,S) = q_o - (p_e-\delta(K,S))$$ given
that $p_e\ge \delta(K,S)$.\eop

\comment 

\begin{lemma}\label{no-emerg}
LDT-schedule $S$ is optimal if it contains a kernel $K$ with its
earliest scheduled job starting at time $\min_{i\in K}\{r_i\}$, 
i.e., there exists no delaying emerging job for that kernel. 
\end{lemma}

\endcomment 

The next property implicitly defines a lower bound $LB$ on the optimal schedule makespan $OPT$. 

\begin{property}\label{delay}
$\delta(K, S)< p_l$, hence $C_{\max}(S)- OPT < p_l$. 
 \end{property}
 
Thus $LB\ge C_{\max}(S)- p_l$. More generally, as it can be easily seen,  
$$LB \ge \max\{C_{\max}(\sigma)- p_l, P(J), \max_j\{r_j+p_j+q_j\}\},$$ 
where $P(J)$ is the total processing time of all jobs.

{\small {\bf Example.} 
We use the following problem instance with 13 jobs for the 
illustration purposes:\\ 
$r_1=0,\ p_1=12,\ q_1=11$, \ \ \ $r_2=2,\ p_2=2,\ q_2=50$, \ \ \ 
$r_3=5,\ p_3=3,\ q_3=48$,\\$r_4=8,\ p_4=5,\ q_4=44$,\ \ \ 
$r_5=13,\ p_5=4,\ q_5=43$,\ \ \ $r_6=1,\  p_6=7,\ q_6=41$, 
\\$r_7=32,\ p_7=10,\ q_7= 3$, \ \ \ $r_8=35,\ p_8=5,\ q_8=15$, \ \  
$r_9=37,\ p_9=4,\ q_9=12$,\\ $r_{10}=41,\ p_{10}=3,\ q_{10}=11$, \ \ \ 
$r_{11}=45,\ p_{11}=2,\ q_{11}=11$,  \\ $r_{12}=47,\ p_{12}=1,\ q_{12}=10$,
\ \ \ $r_{13}=58,\ p_{13}=2,\ q_{13}=2$.

The initial LDT-schedule $\sigma$ is depicted in Fig. 1. Primary kernel
$K^1=K(\sigma)$ consists of jobs $(2,3,4,5,6)$ and possesses the delaying 
emerging job 1, which is
the only emerging job for that kernel. Job 1 pushes job 2. 
The overflow job in kernel $K^1$ is job 6 with the full completion time 
$C_6(\sigma)= 33+41=74$, the makespan of schedule $\sigma$, and 
$\delta(K^1,\sigma)=1$. 

In LDT-schedule $\sigma_1$ of Fig. 2 (with an anticipated job 6) 
newly arisen primary kernel is $K^2=K(\sigma_1)=(8,9,10,11,12)$; 
it possesses the delaying emerging job 7; $\delta(K^2,\sigma_1)= 44-35=9$.   
The overflow job is 12 with $C_{12}(\sigma_1)=59+10=69$.\eop  }

This section's definitions establish the structural notions of 
kernels and emerging jobs that underlie the variable parameter $\nu(I)$. 
With this foundation, we now turn to Section 3, where Theorem \ref{IEA}
establishes the implicit enumeration algorithm and its computationally tight bound.

\begin{figure}
  \centering
    \includegraphics[width=\linewidth]{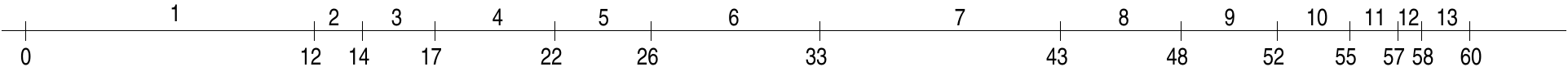} \\
  \tiny{\caption{Initial LDT-schedule $\sigma$}}
  \label{fig1}
\end{figure}

\begin{figure}
  \centering
    \includegraphics[width=\linewidth]{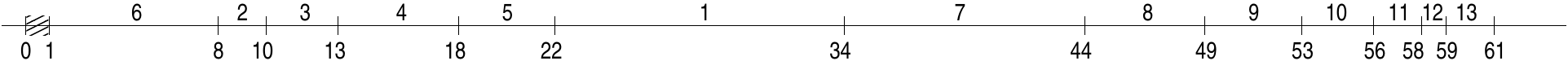} \\
  \tiny{\caption{LDT-schedule $\sigma_1$}}
  \label{fig2}
\end{figure}

\section{Implicit enumeration algorithm}

At stage  0, the implicit enumeration algorithm (IEA) extracts potential kernels and converts them into smaller atomic components. From these components, a partial schedule of all non-emerging jobs is constructed in polynomial time. Atomic components possess a crucial property: regardless of when the earliest released job starts, the LDT-heuristic produces the same optimal partial schedule for the jobs within that component. This property is particularly useful at stage 1, where the incorporation of an emerging job may push an atomic component.

\subsection{Kernel decomposition procedure}\label{dec}

First, we briefly describe how a kernel is decomposed into 
atomic components. Let $K$ be a kernel with the overflow job $o$ 
in an LDT-schedule $S$. Any job $e\in J$ with $r_e < r(K)$ and
$q_e < q_o - p_e$ is declared as a {\em type (1.1)} job for kernel 
$K$. 

Consider $K(+l)$, the fragment of schedule $S$ 
containing the delaying emerging job $l$ together with the jobs of
kernel $K$. We also use $K$ for the fragment of $K(+l)$ 
without job $l$. Define $K(-l)$ as a partial LDT-schedule 
obtained by applying LDT-heuristic to the jobs from $K\setminus \{l\}$.
Since in schedule $K(-l)$ kernel $K$ possesses no delaying emerging job, 
an anticipated job may occur in that schedule.  

If no anticipated job occurs, the procedure halts declaring 
$K$ as an {\em atomic kernel} (in this case, the 
sequence of jobs in the partial schedule $K$ coincides with that 
of the original kernel $K\in S$). Otherwise, the anticipated jobs in schedule 
$K(-l)$ are declared as {\em type (1.2)} jobs for kernel $K$ and 
are discarded (we will use just ``type (1)'' to
refer to both type (1.1) and type (1.2) jobs). 
The LDT-heuristic is then repeatedly applied to the 
reduced set of jobs with similar operations carried out as long as anticipated jobs appear. When no further anticipated jobs remain, the procedure returns the last created partial schedule $S^*[K]$. 

Schedule $S^*[K]$ consists of all jobs of kernel $K$ except the 
omitted anticipated jobs. These jobs are distributed across
one or more blocks - the {\em atomic kernels} of $K$. We  denote 
an atomic kernel of kernel $K$ by $K^*$ whenever the particular choice is immaterial. Note that the maximum job completion times in different atomic kernels are not necessarily equal, and that $S^*[K]$ respects the absolute scheduling horizon determined by the LDT‑heuristic.

Finally, the jobs of the atomic kernels from schedule $S^*[K]$ are declared 
as {\em type (2)} jobs. Intuitively, anticipated jobs are discarded at this stage because they would inevitably delay atomic kernels without changing 
their structural bottleneck. By removing them, the decomposition isolates  ``irreducible cores'' - atomic kernels, whose overflow jobs dictate the makespan. This ensures that the exponential dependence in the algorithm is confined to the structurally critical emerging jobs, while the rest can be handled in polynomial time. A formal description of this procedure, together with stage 0 that creates a partial schedule of non-emerging jobs, is given in the next subsection.


\begin{lemma}\label{lb}
(i) The decomposition procedure invoked for kernel $K$, runs in 
time $O(\mu^2\log\mu)$, where $\mu$ is the total number of jobs in kernel $K$.
(ii) The maximum job full completion time in schedule $S^*[K]$ is 
attained by the overflow job of some atomic kernel from that schedule
and it is a lower bound on the optimum schedule makespan.  
\end{lemma}
Proof. For part (i), note that the number of recursive calls is 
bounded by the number of anticipated jobs, which is less than $\mu$. 
At every iteration LDT-heuristic is applied with time complexity $O(\mu\log\mu)$. Hence, the overall complexity is $O(\mu^2\log\mu)$. 

For part (ii), by construction of schedule $S^*[K]$, its makespan
is attained by the overflow job of some atomic kernel $K^*$.  
This kernel possesses no delaying emerging job, so  
its first scheduled job starts at its release time,
and every job in $K^*$ has a delivery time at least 
that of its overflow job. By a simple interchange argument,
no reordering of the jobs of kernel $K^*$ can improve the makespan. 
Therefore, the makespan of $S^*[K]$ is a valid lower bound on 
the optimum schedule.\eop 

Let $C^*_{\max}(K)$, $K\in{\cal K}$, denote the maximum completion 
time attained by an overflow job of an atomic kernel in schedule 
$S^*[K]$. By Lemma \ref{lb} (ii),  
$C^*_{\max}(K)$ is a lower bound on $OPT$. For an atomic 
kernel $K^*$ of $K$, let $$\delta(K^*)=C^*_{\max}(K)-C_{\max}(K^*).$$ 
Then note that entire kernel $K^*$ can be right-shifted by up to $\delta(K)$ 
time units without surpassing a lower bound on the optimum makespan. 
At the same time, let $U$ be an upper bound on $OPT$. Then if kernel 
$K^*$ is right-shifted by more than $U-C_{\max}(K)$ time units, 
the resultant makespan will surpass $U$. We proved the next
corollary: 

\begin{cor}\label{shift}
In an optimal schedule, an atomic kernel $K^*$ is right-shifted by 
up to $\delta(K^*)$ and by no more than $U-C_{\max}(K)$ time units.\eop
\end{cor} 

{\small {\bf Example.}
The decomposition of kernel $K^1$ can be observed in figures 1, 2 and 3.
Schedule $K^1(+1)$ consisting of jobs 1 through 6, extends through the time
interval $[0,33)$ the LDT-schedule $\sigma$ of Fig. 1. 
This partial schedule corresponds to the initial iteration 0 in the decomposition 
procedure with the delaying emerging job $l=1$. At iteration 1, schedule 
$(K^1)_1$ is the fragment of the complete schedule $\sigma_1$ 
in Fig. 2 from time 1 to time 22, with an anticipated job 
$l_1=6$. Hence, the procedure continues with iteration 2. It 
generates partial schedule  $(K^1)_{1,6}$ of Fig. 3 that
occupies the interval $[2,17)$.  The overflow 
job in partial schedules $(K^1)_1$ and  $(K^1)_{1,6}$ is job 5.
There are two atomic kernels in schedule  $(K^1)_{1,6}$, 
$K_1=\{6\}$ and $K_2=\{3,4,5\}$; $K_2$ is the kernel in the schedule 
of Fig. 3. The decomposition procedure halts at iteration 2 with partial 
schedule $S^*[K^1]$ of Fig. 3 and with a single type (1.2) job 6 
associated with $K^1$.\eop }

\begin{figure}
  \centering
    \includegraphics[width=5.3 cm]{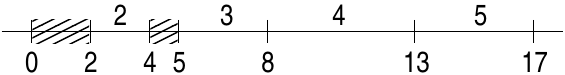} \\
  \tiny{\caption{The result of the decomposition of kernel $K^1$: Schedule $S^*[K^1]$}}
  \label{fig3}
\end{figure}

\subsection{Stage 0: Construction of a schedule with non-emerging jobs}

The main procedure of stage 0 is {\bf Algorithm 0}, which builds a partial 
LDT-schedule of non-emerging jobs by iteratively extracting and 
decomposing arisen kernels, and omitting the identified type (1) jobs. 
In more details, 
at an iteration $h$, the algorithm forms two schedules, $\sigma_h^T$ 
and $\sigma_h$; $\sigma_h^T$  is an auxiliary schedule used to build  
$\sigma_{h+1}$. Initially, set $\sigma_0^T:=\sigma$, $\sigma_0:=\emptyset$  we ${\cal K}_0:=\emptyset$. For $h\ge 1$, let ${\bf K}_h$ be the set of kernels in  $\sigma_{h-1}^T$. Update 
 ${\cal K}_h:= {\cal K}_{h-1} \cup {\bf K}_h$. For each 
$K\in {\bf K}_h$ we paste the partial schedule $S^*[K]$ into  
$\sigma_{h-1}$, producing $\sigma_h$. The
temporary schedule $\sigma_h^T$ is obtained by applying LDT-heuristic 
applied to the job set 
$J\setminus (\bigcup_{K\in {\cal K}_h}{J_{1.1}(K)})\setminus
(\bigcup_{K\in {\cal K}_h}{J_{1.2}(K)})$, where  
 $J_{1.1}(K)$ and $J_{1.2}(K)$ denote the set of type (1.1) 
 and type (1.2) jobs, respectively,  for kernel $K$. 

At each iteration, we define the corresponding {\em configuration} as the collection of kernels detected so far together with their associated type (1.1), type (2) and type (1.2) jobs (the latter identified in the decomposition procedure). The final configuration of stage 0 consists 
of the set of kernels ${\cal K}_{h^*}$, referred to as {\em primary 
kernels}, and the initial job partition. Here, $h^*$ denotes the last 
iteration in Algorithm 0 - the earliest iteration in which no kernel 
possesses a delaying emerging job in schedule $\sigma_{h^*}$.  
In this initial partition, all jobs not previously classified (i.e.,
non-type (1)-(2) jobs), are declared as type (4) jobs.  
At the last iteration, schedule $\sigma_{h^*}$ is extended by 
all type (4) jobs using the LDT-rule (see line 12 in the formal description
of {\bf Algorithm 0}). 

A sequence of type (4) jobs from schedule $\sigma_{h^*}$ may be 
converted into a primary kernel at stage 1. 
Although a primary non-atomic kernel does not exist as 
a single unit once it has been decomposed into atomic kernels, we still distinguish it conceptually at stage 1.
At any point during stage 1, $\cal K$ denotes 
the set of all the primary kernels detected so far.
Note that a primary kernel is disjoint (as a job set) from any 
kernel in $\cal K$ formed earlier.

Thus partial schedule $\sigma_{h^*}$, hereafter denoted $\sigma^{*}$, 
consists of a sequence of atomic kernels interleaved with zero or more sequences of type (4) jobs. Observe that at any iteration $h$, 
no overlap can occur between jobs already in $\sigma_{h-1}$ and newly included jobs, since the time intervals of different atomic kernels are disjoint.

\smallskip

Below we give a formal description of stage 0, including that of the
decomposition procedure from the previous subsection and an auxiliary 
routine that augments the current set of primary kernels.
We represent each job $j$ by a {\bf Job} structure with three fields, 
and we maintain two global variables during stage 0: the partial schedule for the current iteration, $\sigma^A$, and the current set of primary
kernels, $\cal K$.

{
\begin{algorithm}[H]
{\small
\floatname{algorithm}{{\small Structure {\bf Job}:}}
\renewcommand{\thealgorithm}{}
\caption{}\label{alg:cap}
\begin{algorithmic}
\State $\quad$ type: integer; \Comment{one of the types (1.1), (1.2), (2), (4)}
\State $\quad$ kernel: integer; \Comment{the primary kernel to which job $j$ belongs}
\State $\quad$ E: list of integers; \Comment{a connected list of atomic/primary 
kernels for which $j$ is an emerging job}
\end{algorithmic}
}
\end{algorithm}

\begin{algorithm}[H]
{\small
\floatname{algorithm}{{\small Global variables}}
\renewcommand{\thealgorithm}{}
\caption{}\label{alg:cap}
\begin{algorithmic}
\State $\mathcal{K} := \{\} $ \Comment{current set of primary kernels}

\end{algorithmic}
}
\end{algorithm}

\begin{algorithm}[H]
{\small
\floatname{algorithm}{{\small Decompose}}
\renewcommand{\thealgorithm}{}
\caption{{\small$(K,Pos_{K} = \{Pos_K(j_1),\dots,Pos_K(j_{|K|})\})$\\
\textcolor{gray}{$K$ is the set of jobs of the primary kernel which is to be decomposed}\\
\textcolor{gray}{$Pos_K$ is the set of current positions of jobs in $K$}}}\label{alg:cap} 
\begin{algorithmic}[1]
\State $S^*[K] := LDT(K)=(j_1,j_2,\dots,j_{|K|})$ \Comment{LDT-schedule 
for the jobs in $K$}
\State $pos = 1$
\While{$pos \not = |K|$} \Comment{there can be anticipated jobs}
\State $pos = 1$
\For{$i=1$ to $|S^*[K]|$} \Comment{determine and omit anticipated jobs}
{
\If{$Pos_K(j_i) > i$} \Comment{$j_i$ is anticipated}
\State $j_i.type := 1.2$
\State $K := K \setminus \{j_i\}$
\Else \Comment{$j_i$ is not anticipated}
\State $Pos_K(j_i) := pos$
\State $pos := pos+1$
\EndIf
}
\EndFor
\State $S^*[K] := LDT(K)=(j_1,j_2,\dots,j_{|K|})$ \Comment{LDT-schedule 
for the jobs in $K$}
\EndWhile
\State \Return $S^*[K]$ \Comment{this is a sequence of atomic components}
\end{algorithmic}
}
\end{algorithm}


\begin{algorithm}[H]
{\small
\floatname{algorithm}{{\small Add\_Primary\_Kernel}}
\renewcommand{\thealgorithm}{}
\caption{{\small$(K,\sigma_h)$\\\textcolor{gray}{$K$ is a new primary kernel to be decomposed}\\
\textcolor{gray}{$\sigma_h$ is the current schedule where $K$ arrived}}}\label{alg:cap}
\begin{algorithmic}[1]
\State $\mathcal{K} := \mathcal{K} \cup \{K\}$
\State $J_{1.1}(K) := \{e \in \sigma_h | r_e < r(K) $ and $q_e< q_o - \max\{0,p_e-\delta(K,\sigma_h)\}\}$  \Comment{find the 1.1 type jobs for $K$}
\State \textbf{for each} job $e$ in $J_{1.1}(K)$ \textbf{do} $e.type:=1.1$
\State \textbf{for each} job $e$ with $e.type=1.1$ {\textbf do} \Comment{search for the primary kernels for which $e$ is an emerging job}
\State $\quad$\textbf{for each} primary kernel $K\in\mathcal{K}$ {\textbf do}
\State $\quad\quad$ \textbf{if} $e$ is an emerging job for $K$ \textbf{then} add $K$ to the list $e.E$
\State $\quad$\textbf{end for}
\State \textbf{end for}
\State $Pos_K = \{Pos_K(j_i) := i \;|\; i = 1,\dots, |K|$ for all $j_i\in K\}$ \Comment{set the positions of jobs of kernel $K$}
\State $S^*[K] := {\bf Decompose}(K,Pos_K)$
\State \textbf{for each} job $j\in K$ \textbf{do}
\State $\quad$ $j.kernel := K$
\State $\quad$ \textbf{if} $j.type=1.2$ \textbf{then} \Comment{for each type 1.2 job search for the atomic kernels for which it is an emerging job}
\State $\quad$$\quad$ \textbf{for each} atomic kernel $K^*\in S^*[K]$ {\bf do}
\State $\quad$$\quad\quad$ \textbf{if} $j$ is an emerging job for $K^*$ \textbf{then} add $K^*$ to $j.E$
\State $\quad$$\quad$ \textbf{end for}
\State $\quad$ \textbf{end if}
\State \textbf{end for}
\State \textbf{for each} $j\in S^*[K]$ \textbf{do} $j.type:= 2;$
\State \textbf{return} $S^*[K]$
\end{algorithmic}
}
\end{algorithm}


\begin{breakablealgorithm}
{\small
\floatname{algorithm}{{\small Algorithm\_0}}
\renewcommand{\thealgorithm}{}
\caption{{\small $(J,\sigma_{h-1}^T,\sigma_{h-1})$ \\
\textcolor{gray}{$J$ is the current set of all jobs except type (1) jobs} 
}}\label{alg:cap}
\begin{algorithmic}[1]
\State $\mathbf{K_h} :=\{K \in \sigma_{h-1}^T | K $ has a delaying emerging job $\} $
\If{$\mathbf{K_h} = \emptyset $ } \Comment{All kernels are atomic}
\State $\sigma_{h^*} := \sigma_{h-1}^T$
\State \Return $\sigma_{h^*}$
\EndIf
\State $\sigma_h := \sigma_{h-1}$ \Comment{initial assignation of $\sigma_h$}
\State \textbf{for each} $K \in \mathbf{K_h}$ \textbf{do}
\Comment{decompose each kernel $K \in \mathbf{K_h}$}
\State $\quad$ $S^*[K] := $ {\bf Add\_Primary\_Kernel}$(K,\sigma_{h-1}^T)$
\State $\quad$ $\sigma_h := \sigma_h \cup S^*[K]$
\State $\quad$ $J_{1.2}(K) := K \cap S^*[K]$
\State \textbf{end for}
\State $\sigma_h^T := $ the LDT schedule constructed for the jobs in $J\setminus\bigcup_{K\in{\cal K}_h}J_{1.1}(K)\setminus \bigcup_{K\in{\cal
K}_h}J_{1.2}(K)$ in which no type (4) job is included within $S^*[K]$
\State $\sigma^* := {\bf Algorithm}\_{\bf 0}( J\setminus\bigcup_{K\in{\cal K}_h}J_{1.1}(K)\setminus\bigcup_{K\in{\cal K}_h}J_{1.2}(K), \sigma_h^T, \sigma_h)$
\State \Return $\sigma^*$
\end{algorithmic}
}
\end{breakablealgorithm}

\smallskip  
{\bf Example.} Fig. 4 represents  schedule $\sigma^*$ delivered   
at stage 0. It has three atomic kernels $K_1=\{2\}$, $K_2=\{3,4,5\}$
and $K_3=\{8,9,10,11,12\}$.  $J_{1.1}(K_1)=J_{1.1}(K_2)=\{1\}$, and 
$J_{1.1}(K_3)=\{7\}$; $J_{1.2}(K^1)=\{6\}$ and $J_{1.2}(K^2)=\emptyset$; 
there is only one type (4) job 13.\eop 
 
\begin{figure}
  \centering
    \includegraphics[width=\linewidth]{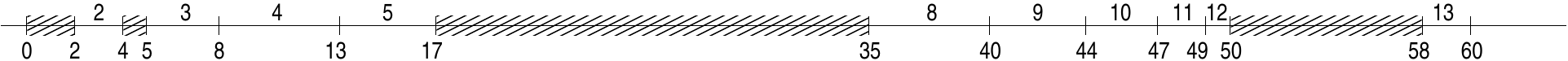} \\
  \caption{Schedule $\sigma^*$}
  \label{fig5}
\end{figure}

\begin{lemma}\label{stage 0 complex}
Stage 0 returns partial LDT-schedule $\sigma^{*}$ with 
$C_{\max}(\sigma^{*})\le OPT$ in time $(n-\nu) O(n \log n)$.   
\end{lemma}
Proof. For a worst-case running-time estimate assume each iteration detects exactly one kernel and every detected kernel contains $\mu$ jobs. Let $x$
be the number of iterations. Since there are \(n-\nu\) non-emerging jobs in total, we have \(\mu x \le n-\nu\), hence \(x \le (n-\nu)/\mu\). By Lemma \ref{lb} (i) the decomposition procedure invoked at a single iteration runs in time $O(\mu^2\log\mu)$. Therefore, the total time over all $x$ iterations is $x O(\mu^2\log\mu)$.
Substituting the bound on $x$ gives $(n-\nu) O(\mu\log\mu)$. 
Using $\mu \le n$  yields the stated worst-case bound.
As to the makespan bound, by construction of $\sigma^*$, no  type (4) job attains the maximum completion time and the inequality in the lemma follows from Lemma \ref{lb} part (ii).\eop

\comment 


----move later--
We will refer to a complete LDT-schedule with a gap before 
the first scheduled job of kernel $K$ as  a {\em loose} schedule for 
that kernel; otherwise, we will say that it is {\em tight} 
for kernel $K$. LDT-schedule $\sigma$ is tight for kernel $K$ if this
kernel possesses the delaying emerging job; schedule $S(-e,+K)$ (in which
job $e$ is strongly activated for kernel $K$) is loose 
for kernel $K$; LDT-schedule $S_l$ can be loose or tight. ---

define $S(-A,+B)$ --- move later where needed--
Let $A$ be ----a subset of the set of type (1.1) jobs for kernel $K$. 
Recall that in LDT-schedule $S(-e,+A)$ job $e$ is included after all 
jobs from set $A$. Note that if $r(A)+p(A) \ge r(K)$, schedule 
$S(-e,+A)$ is tight for kernel $K$ and job $e$ is rescheduled after
all jobs of that kernel in that schedule. In particular, if $p(A)<p_e$, 
the first job of kernel $K$ will start earlier in schedule $S(-e,+A)$
than in schedule $S$. 

---if no IA(b1) with a type (1.1) job of $K$ occurs, we don't need to take care about type (1.1) jobs 
scheduled after $K$ ---

\endcomment 

\subsection{Stage 1: Incorporating type (1) jobs} 

Let $\pi$ be a permutation of type (1.1) jobs and  $\psi(K)$ a 
permutation of the jobs from set $J_{1.2}(K)$, for each  $K\in {\cal K}$.
An $O(n)$ time greedy procedure \textbf{Incorporate\_Permutation}, 
a subroutine of stage 1, produces a complete schedule that respects 
these permutations by extending the partial schedule $\sigma^*$. 

The routine proceeds as follows. It schedules type (1.1) jobs according 
to $\pi$  and, for each $K$, schedules the jobs in $J_{1.2}(K)$
according to $\psi(K)$. Each job is placed at the earliest available 
idle time so that no avoidable gap is left. When the first job of 
$\psi(K)$ is released, it is inserted immediately after the completion 
of the currently processed job from $\pi$, if any; if the
two jobs are simultaneously released, the job from  $\psi(K)$ is
scheduled. No type (1.1) job is started until all jobs $\psi(K)$ are
finished. If the next  type (1) job to be included overlaps a job in 
$\sigma^*$, the overlapping jobs in $\sigma^*$ are right-shifted 
accordingly, and the  type (1) job is placed at the next available idle time.
We implement the schedule as a linked list of blocks (atomic kernels and
inserted jobs) with an index to the next idle slot; each job is inserted 
once and each right-shift updates contiguous blocks in amortized constant
time, yielding overall $O(n)$ time. 

We denote by $\sigma(\pi,\psi)$ the so constructed complete schedule
and by $(\pi,\psi)$ the corresponding {\em combined permutation}, in which
the jobs from each permutation $\psi(K)$ are inserted within the
permutation $\pi$ according to the above described greedy algorithm,
where $\psi$ is a vector of permutations $\psi(K)$, 
$K\in {\cal K}$. Thus the initial LDT-schedule 
$\sigma$ is $\sigma(\pi^*,\psi^*)$, where $\pi^*$ and
$\psi^*(K)$, $K\in {\cal K}$, are the {\em steady} permutations 
naturally created in that schedule. We give a formal description 
of the procedure below, where 

We denote the schedule produced by this routine as $\sigma(\pi,\psi)$. 
The {\em combined permutation} $(\pi,\psi)$ is the sequence obtained 
by inserting, for each $K$, the jobs of  $\psi(K)$ into $\pi$ 
according to the greedy insertion rule above; here $\psi$ denotes 
the vector of permutations $\psi(K)$, $K\in {\cal K}$. Note that
the initial LDT-schedule $\sigma$ can be written as 
$\sigma(\pi^*,\psi^*)$, where $\pi^*$ and
$\psi^*(K)$, $K\in {\cal K}$, are the {\em steady} permutations 
induced by $\sigma$. 

\smallskip 

The following subroutine of stage 1 incorporates a combined
permutation of emerging (type (1)) jobs into the partial schedule 
$\sigma^*$ of non-emerging jobs produced at stage 0. 
In the pseudocode below $\pi(i)$ and $\psi(K,i)$ 
denote the $i$-th job of the corresponding permutation.

\begin{breakablealgorithm}
{\small
\floatname{algorithm}{{\small Incorporate\_Permutation}}
\renewcommand{\thealgorithm}{}
\caption{{\small$(\sigma^*,(\pi,\psi))$}}\label{alg:cap}
\begin{algorithmic}[1]
\State $t:=0$
\For{$i = 1$ to $|\pi|$}
\State $t:=$ next gap in $\sigma^*$ 
\State $t_{\pi(i)}(\sigma^*) := t$ \Comment{schedule job $\pi(i)$ in $\sigma^*$ at time $t$}
\State right-shift jobs scheduled after job $\pi(i)$
\If{job $\pi(i)$ pushes an atomic kernel $K^*$}
\State $K :=$ the primary kernel of $K^*$
\For{$j = 1$ to $|\psi(K)|$}
\State $t:=$ next gap in $\sigma^*$ 
\State $t_{\psi(K,j)}(\sigma^*) := t$ \Comment{schedule job $\psi(K,j)$ in $\sigma^*$ at time $t$}
\State right-shift jobs scheduled after job $\psi(K,j)$
\EndFor
\EndIf
\EndFor
\While{exists an unconsidered $\psi(K)$} \Comment{no type (1.1) job pushed the first atomic kernel of kernel $K$}
\For{$j=1$ to $|\psi(K)|$}
\State $t:=$ next gap in $\sigma^*$ 
\State $t_{\psi(K,j)}(\sigma^*) := t$ \Comment{schedule job $\psi(K,j)$ in $\sigma^*$ at time $t$}
\State right-shift jobs scheduled after job $\psi(K,j)$
\EndFor
\EndWhile
\State \Return $\sigma^*$ \Comment{complete schedule respecting permutations $\pi$ and $\psi(K)$}
\end{algorithmic}
}
\end{breakablealgorithm}

\medskip

It suffices to consider only complete schedules that respect combined
permutations. Using combined permutations reduces the theoretical 
worst-case term $(\nu_{1.1}+\sum_{K\in \cal K} \nu_{1.2}(K))!$, which 
arises from enumerating all permutations of the  type (1) jobs, to the
product $\nu_{1.1}! \ \sigma_{K\in \cal K}\nu_{1.2}(K)!$. Note that 
the latter term is an exact reduction from the above full term. 
(Recall $\nu_{1.2}(K)=|J_{1.2}(K)|$ and $\nu_{1.1}=|J_{1.1}|$.)  

\smallskip

Below we give a formal description of the main procedure 
and that of stage 1. For each combined permutation of type (1) jobs,
the main procedure invokes stage 1 and returns a solution with 
minimum makespan produced by stage 1. 

\begin{algorithm}[H]
{\small
\floatname{algorithm}{{\small Main}}

\renewcommand{\thealgorithm}{}

\caption{}\label{alg:cap}

\begin{algorithmic}[1]

\State $J := $ the set of the given $n$ jobs

\State $\sigma^* :=$ \textbf{Algorithm\_0}$(J,\sigma,\emptyset)$


\State $\sigma_{best} := \textbf{LDT}(J)$ \Comment{complete schedule
with minimum makespan so far}

\While{$(\pi,\psi) \not= nil$} \Comment{while exists an unverified
combined permutation of type (1) jobs}

\State $\sigma_{best} :=
min\{\sigma_{best},$\textbf{Stage\_1}$(\sigma^*,(\pi,\psi))\}$

\State $(\pi,\psi) :=$ next combined permutation of type (1) jobs or
$nil$ if all permutations were considered

\EndWhile

\State \textbf{output} $\sigma_{best}$

\end{algorithmic}
}
\end{algorithm}

\comment 

\subsection{Kernel interactions}

In an LDT-schedule, the set of type (1.1) jobs of any kernel 
is naturally partitioned into two subsets: ones 
which are not and ones which are in the state of activation for that kernel
(the ones preceding and succeeding, respectively, that kernel). If we
strongly activate a type (1.1) job, the former set is reduced, while
the latter one is extended. 
Let $E'$ ($E$, respectively) be a set of {\em some} type (1.1) jobs in 
the state of activation (not in the state of activation, respectively)
for kernel $K$ in LDT-schedule $S=\sigma(\pi,\psi)$. 
We can form an alternative to $\pi$ permutation of type (1.1) jobs
by interchanging the roles of the jobs from sets $E$ and $E'$ 
{\em swapping} these sets in schedule $S$; that is, the jobs
of set $E$ become in the state of activation and the ones from set
$E'$ not in the state of activation for kernel $K$ in the resultant 
LDT-schedule, denoted by $S_{E,E'}$  (we omit argument $K$ for the
simplicity). When we write $S_E$ we just 
express the fact that the jobs in set $E$, not being in the state of
activation for kernel $K$ in schedule $S$, become in the state of
activation  for kernel $K$ in schedule $S_E$.  For example, if
the delaying emerging job $e$ is strongly activated for kernel $K$
in schedule $S$ then the resultant LDT-schedule is $S_e=S_{\{e\},\emptyset}$. 
Note that schedule $S_{E,E'}$ can be created by the corresponding 
modification of the release time of the jobs in sets $E$ and $E'$ and 
the application of LDT-heuristic to the modified instance.  

Set swapping is similarly defined for type (1.2) jobs and applied 
locally for the corresponding atomic kernel(s) of a 
corresponding primary kernel $K$. 
In this case, the jobs from sets $E$ and $E'$ are subsets of 
set $J_{1.2}(K)$ and are defined individually for each atomic kernel 
$K^*$ of kernel $K$. Note that initially, in LDT-schedule $\sigma$, 
for any primary kernel $K$ with the delaying emerging job, 
all jobs from set $J_{1.2}(K)$ are in the state of activation 
for the corresponding atomic kernel(s) (so for any atomic kernel 
$K^*$, $E=\emptyset$ and $E'$ consists of all type (1.2) emerging 
jobs for kernel $K^*$). Note that, in general, set 
$E$ consists of either type (1.1) or (1.2) jobs as no type (1.1) 
job needs to be included in between the jobs of any primary kernel.  

Set swapping is a convenient way for altering combined permutations
of type (1) jobs. It may change a current kernel layout. 

\endcomment

{\bf Extended kennels.} 
At a given state of the computation, any kernel that arises and is 
distinct from every primary kernel in $\cal K$ and from every atomic 
kernel (viewed as a job sequence) will be called a {\em newly arisen}
kernel. A newly arisen kernel that contains all jobs of an atomic or 
primary kernel $K$ is called an {\em extended} kernel of kernel $K$  
and denoted by $\bar K$. Besides a type (1) job for $K$, $\bar K$
may include jobs of some succeeding atomic kernels and  type (4) jobs. 
Note that no type (1) job for kernel $K$ that belongs to  $\bar K$ 
(a former emerging job becomes a kernel job) can be an emerging job 
for $\bar K(E)$ or for any following kernel. 

\begin{lemma}\label{IAb}   
Let $S$ be an LDT-schedule that respects a combined permutation 
$(\pi,\psi)$. If an extended kernel $\bar K$ arises in $S$, then  
no other schedule that respects the same combined permutation can achieve 
a strictly smaller makespan. 
\end{lemma}
Proof. If  $\bar K$ contains a type (4) 
job, the algorithm will declare the corresponding new primary 
kernel(s) and update the job partition, producing a new combined 
permutation (see stage 1 description, line 15). Hence we may assume 
$\bar K$ contains no type (4) job. Let $e$ be a type (1) job for 
kernel $K$ that belongs to $\bar K$, and $j$ be any job scheduled 
after kernel $K$ and before $e$ in $S$. Since $e$ is released 
no later than the jobs of $K$, we have $q_j \ge q_e$, while the 
jobs of $\bar K$ are scheduled in LDT-order. Therefore,  the maximum 
full completion time of a job in $\bar K$ (which equals the makespan 
of $S$) cannot be reduced unless an emerging job $e'\not\in \bar K$ 
is activated for $\bar K$. Any such $e$ must 
appear before $K$ (and hence before job $e$) in $S$, but 
if activated it would be inserted after $e$. So the resultant 
schedule would not respect permutation $\pi$ and the lemma follows.\eop

Thus once an extended kernel $\bar K$ forms under a fixed combined
permutation, any improvement would require activating an out-of-order
emerging job, which violates the permutation and therefore cannot 
reduce the makespan.

\comment
A rise of a secondary kernel $\hat K$ may yield ``local'' (within the
scope of kernel $K$) rearrangement of the jobs in set $J_{1.2}(K)$,
whereas that of an extended kernel a ``global'' job rearrangement. 

Either (1) job $e$ was activated in one of the (earlier) generated LDT-schedules respecting permutation $\pi$ or (2) not. 
For case (1), let $\sigma'(\pi)$ be the first generated LDT-schedule
respecting permutation $\pi$ in which job $e$ was activated, say, for kernel $K''$. 
We show that there can exist no delaying emerging job for kernel $K'$. Indeed, 
suppose $\epsilon$ is such a job. Since there is a gap immediately before kernel 
$K''$ in schedule $\sigma'(\pi)$ and in any further generated LDT-schedule respecting permutation $\pi$, the delaying emerging job for kernel $K'$ must be included 
between kernel $K''$ and job $e$ in schedule $\sigma'(\pi)$. But no job included 
between kernel $K''$ and job $e$ can  be a type (1.1) job from $\pi$ since 
otherwise schedule $\sigma'(\pi)$ would not respect permutation $\pi$ 
(since job $e$ was included before such job in an earlier created
schedule respecting permutation $\pi$). Hence,  $\epsilon$ does not exist.  
\endcomment


\comment

For a better intuition and understanding, we first
give a greedy, referred to as Algorithm 0.0, that forms a skeleton 
of IEA. Unlike {\bf Algorithm 0}, Algorithm 0.0 does not omit type (1) jobs: 
Starting from the schedule $\sigma$, repeatedly, it strongly activates the 
delaying emerging job for the earliest arisen (primary) kernel in the current 
LDT-schedule as long as there exist that job.

\begin{lemma}\label{basic}
If during the execution of Algorithm 0.0 no extended kernel arises, it 
creates an optimal schedule in $O(n^2\log n)$ time.
\end{lemma}
Proof. Since during the execution of the algorithm no extended kernel 
arises, neither a type (1) nor a type (4) job 
becomes an overflow job. Hence, the overflow job in the last 
generated LDT-schedule belongs to some primary kernel $K\in\cal K$.
This kernel  possesses no delaying emerging job (as the delaying
emerging job of all arose primary kernels were strongly activated) 
and hence the earliest scheduled job of that kernel starts at time $r(K)$. 
Furthermore, since no extended kernel of an atomic kernel of kernel $K$ 
arose, the jobs of kernel $K$ are processed in (the same as in 
schedule $\sigma$) LDT-order. 
Hence, the full completion time of the last scheduled job of kernel
$K$ is the minimum possible (see Lemma \ref{lb}) and the last created
LDT-schedule is optimal. The algorithm runs in $O(n^2\log n)$ time 
since $|{\cal K}|\le n-\nu < n$ and each delaying 
emerging job is activated by LDT-heuristic applied to the 
corresponding modified problem instance (see Section 2).\eop

Due to Lemma \ref{basic}, we may complete Algorithm 0.0 specifying how 
we deal with extended kernels. An extended kernel  $\bar K(\{e\})$ may 
occur in Algorithm 0.0 when it strongly activates the delaying emerging  
job $e$ for primary kernel $K$ in a current LDT-schedule $S$
(see Lemma \ref{basic}). If this happens, Algorithm 0.0 cannot 
proceed, as kernel $\bar K(\{e\})$ possesses no delaying emerging 
job (due to the gap arisen before kernel $K$ in schedule $S_e$). 
IEA works in a bit different way. It first detects potential kernels 
and decomposes the non-atomic ones creating schedule $\sigma^*$ 
and an initial configuration  at stage 0. Then it distributes
type (1) jobs in between the atomic kernels. 

Having a partial schedule with only atomic kernels is convenient,
as a complete LDT-schedule including an atomic kernel without
delaying emerging job is optimal (see Lemma \ref{lb}). If there
is no such kernel, a swap of a  pair of sets $E$ and $E'$ can be 
carried out to verify an alternative combined permutation. 
In a typical such swap,  
as in Algorithm 0.0, $E=\{e\}$ and $E'=\emptyset$, where
$e$ is the delaying emerging job of the first (atomic) kernel
in a current updated schedule $S$ (note that this swap does not
change the current permutation). If in LDT-schedule $S_e$ no 
extended kernel arises, it is optimal (this can be seen
similarly as in Lemma \ref{basic}). Otherwise (after updating
the current configuration if applicable), alternative swaps can
repeatedly be carried out to create alternative permutations
of type (1) jobs (note that the involved atomic kernels will 
entirely be shifted to the right or to the left). 

Instead of dealing with swaps, it is easier to 
enumerate combined permutations. In practice, a swap-based implicit
enumeration is more efficient, though it yields the
same worst-case bound $\nu_{1.1}! \ \sigma_{K\in \cal K}\nu_{1.2}(K)!$
as a straightforward method that just enumerates all possible 
combined permutations of type (1) jobs. 
*dominance*This bound is too rough as there are a number of conditions
that allow to discard possible swaps/permutations.  
For example, due to job release times, 
some permutations might be dominated by others as they would yield
non-active complete schedules (in such a schedule, a job can be restarted 
earlier without delaying some other jobs). Corollary \ref{shift} also
restricts whole set of beneficial swaps.  There are optimality
conditions and other dominance relations that allow to discard 
potential swaps/permutations. An intelligent way of selection
of subsets of type (1) jobs for each next swap would determine
a particular order in which permutations of type (1) jobs
can be enumerated. Since these results take a considerable space 
and here we are mostly interested in theoretical worst-case bounds, 
they are omitted. 

\endcomment

Now we describe how stage 1 proceeds when  invoked for a combined 
permutation $(\pi,\psi)$. First it constructs the complete 
schedule $\sigma(\pi,\psi)$ and then verifies the previously 
established structural conditions.  A new primary kernel $K$ 
consisting of type (4) jobs may arise either as an independent 
kernel or as part of an extended kernel. In either case $K$ 
is declared primary, decomposed into atomic kernels, and the corresponding segment of the current LDT-schedule is replaced by the partial schedule 
$S^*[K]$; the job partition and the current configuration are updated as 
in procedure {\bf Add\_Primary\_Kernel}.

\comment 

\begin{itemize}  
\item 
Step 1 

IF an atomic kernel without delaying emerging job
is a kernel in schedule $\sigma(\pi,\psi)$  

THEN return that schedule
\{it is optimal by Lemma \ref{lb}\};  HALT IEA

\item  Step 2

IF there is an extended kernel in schedule $\sigma(\pi,\psi)$ without 
delaying emerging job 

THEN \{this kernel contains an activated type (1) job\} 
return the best created schedule \{Lemma \ref{IAb}\}

\item  Step 3 

determine the first kernel $K$ in schedule $\sigma(\pi,\psi)$ 
\{kernel $K$ possesses the delaying emerging job\} 

IF kernel $K$ contains a type (4) job 

THEN  \{either $K$ is anew kernel consisting of only type (4) jobs 
or it is an extended kernel\}
call the decomposition procedure for each consecutive sequence 
of the type (4) jobs belonging to kernel $K$; update correspondingly 
the current set of primary kernels, schedule $S'(\pi,\psi)$ and current 
job partition; repeat from step 1 

ELSE 
$\sigma(\pi,\psi):=\sigma_l(\pi,\psi)$, where  the
delaying emerging job $l$ of kernel $K$ is strongly activated 
for the last atomic kernel of kernel $K$; repeat from step 1.  
\end{itemize}

\endcomment 

\begin{breakablealgorithm}
{\small
\floatname{algorithm}{{\small Stage\_1}}
\renewcommand{\thealgorithm}{}
\caption{{\small$(\sigma^*,(\pi,\psi))$}}\label{alg:cap}
\begin{algorithmic}[1]
\State $\sigma^{stage1} := $ {\bf Incorporate\_Permutation}$(\sigma^*,(\pi,\psi))$ 
\State $\sigma_{min} := \sigma^{stage1}$ \Comment{the best generated complete schedule so far}
\While{$true$}
\State $\sigma_{min} := \min\{\sigma_{min},\sigma^{stage1}\}$ 
\State $lemma$ \ref{IAb} $:= false$
\State \textbf{foreach} kernel $K$ in $\sigma^{stage1}$
\State $\quad$ \textbf{if} $K$ is atomic with no delaying emerging job \textbf{then}
\State $\quad$$\quad$ \textbf{output} $\sigma^{stage1}$ and \textbf{halt VP-Algorithm}  \Comment{$\sigma^{stage1}$ is optimal by Lemma \ref{lb}}
\State $\quad$ \textbf{endif}
\State $\quad$ \textbf{if} $K$ is an extended kernel without a type $(4)$ job \textbf{then} $lemma$ \ref{IAb} $:= true$
\State \textbf{end foreach}
\State \textbf{if} $lemma$ \ref{IAb} $= true$ \textbf{then} \textbf{return} $\sigma_{min}$ \Comment{Lemma \ref{IAb}}
\State $K := $ the first kernel in $\sigma^*$
\If{$K$ is an extended kernel}\Comment{decompose each segment of type (4) jobs}
\State \textbf{for each} segment $K(4)$ of type (4) jobs \textbf{do} \textbf{call Add\_Primary\_Kernel}$(K(4),\sigma^{stage1})$
\ElsIf{$K$ consists of only type (4) jobs}
\State \textbf{call Add\_Primary\_Kernel}$(K,\sigma^{stage1})$
\Else 
\State $l :=$ delaying emerging job of $K$
\State $r_l := r(K^*)$ \Comment{strongly activate job $l$ for the last atomic kernel $K^*$ of $K$}
\EndIf
\State $\sigma^{stage1} := $ \textbf{Incorporate\_Permutation}$(\sigma^*,(\pi,\psi))$
\EndWhile
\State \Return $\sigma_{min}$
\end{algorithmic}
}
\end{breakablealgorithm}

\begin{lemma}\label{stage 1}
Stage 1 invoked for a combined permutation $(\pi,\psi)$
produces a schedule with minimum makespan among all schedules that respect 
$(\pi,\psi)$ and runs in time $(n-\nu)O(n\log n)$. 
\end{lemma}
Proof. Building $\sigma(\pi,\psi)$  by the greedy routine 
{\bf Incorporate\_Permutation} requires placing the $\nu$ type (1) jobs
and possibly right-shifting at most $n-\nu$ jobs from  $\sigma^*$
that needs time $O(n)$.  The number of iterations
in the main procedure  cannot be more than the number
of kernels, that cannot be more than $n-\nu$. At
each iteration, the strong activation of the corresponding delaying 
emerging job takes time $O(n\log n)$ (that of the LDT-heuristic). 
The total cost of the 
current configuration updates for all the arisen primary
kernels is  $O(n-\nu)(n\log n)$ (see the proof of
Lemma \ref{stage 0 complex}), and this term
is summed up with $(n-\nu) O(n\log n)$. Correctness
follows from Lemma \ref{lb} (ii) and Lemma \ref{IAb}.\eop

\begin{theorem}\label{IEA}
IEA generates an optimal solution to problem $1|r_j,q_j|C_{\max}$ 
in time $$(n-\nu)O(n\log n) \nu_{1.1}! \ \prod_{K\in \cal K}\nu_{1.2}(K)!.$$
\end{theorem}
Proof. For any primary kernels $K, K'$, the jobs 
of $J_{1.2}(K)$ must appear before any job from 
set $J_{1.2}(K')$ whenever $K$ precedes $K'$. Besides, no 
type (1.1) job is to be inserted in between the jobs of 
any primary kernel $K$, including the jobs of $J_{1.2}(K)$,
and no type (1) job is to be inserted in between the 
jobs of any atomic kernel. Hence, it suffices to enumerate the 
combined permutations, which exact number is 
$\nu_{1.1}! \ \prod_{K\in \cal K}\nu_{1.2}(K)!$.  
Since all combined permutations are enumerated, 
one of the complete schedules produced 
by stage 1 must be optimal by Lemma \ref{stage 1}. The 
time complexity follows as  stage 1 runs
in time $(n-\nu)O(n\log n)$ (see again Lemma \ref{stage 1}).\eop


\comment

======== move to Appendix B===========

----We already argued why the algorithm is optimal. As to the time 
complexity, the number of iterations, where a new kernel $K$ is
detected such that it has no intersection with some earlier kernel, 
cannot clearly be more than $n-\nu_{1.1}$ (an upper bound on the total 
number of such kernels). At each of these iterations, for the activation 
of the corresponding delaying emerging job, LDT-heuristic with cost  
$O(n\log n)$ is invoked, while the decomposition procedure has the 
cost $O(i^2\log i)$, where $i=|K|$ (Lemma \ref{lb}). 
For the purpose of this estimation, we may assume that each of 
these $n-\nu_{1.1}$ kernels have the same number $x\le n-\nu_{1.1}$ 
of jobs (this will maximize the overall cost). Then the 
total number of iterations is bounded by $(n-\nu_{1.1})/x$ and the total 
cost in all iterations is $O(n\log n(n-\nu_{1.1})^2\log(n-\nu_{1.1}))$ 
under the assumption that none of the instances IA(b1), IA(b2) or IA(a)
occurs, 
i.e., none of the above kernels has an intersection with an 
earlier detected kernel. Otherwise one of the behavior
alternatives IA(b1), IA(b2) or IA(a) occurs. These alternatives, 
in the worst case, will yield the enumeration of all possible combined 
interchanges, which number is bounded by $2^\nu$. If in none
of the created LDT-schedules there is no suitable due to IA(b1) sequence 
of type (1.1) jobs, an alternative bound $q_{\max}+\Delta p^{(1.1)}_{\max}$
holds (Corollary \ref{dom b1 bound}).\eop

The total  length of type (1.1) jobs that need to be activated for an 
occurrence of an IA(a) can be bounded as follows. 
By strongly activating a type (1.1) job $e$ for kernel $K$, we
guarantee that the earliest released job from set $K$ will be included at
its release time in schedule $S_e$. If $K$ is an atomic kernel,  
the full completion time of the last scheduled job of kernel $K$ is the
best possible (see Lemma \ref{lb}) and schedule $S_e$ is optimal if 
an IA(a) with that kernel arises. However, the gap that may arise 
before kernel $K$ by the activation of job $e$ may force the rise of
another kernel behind kernel $K$. So, intuitively, we may prefer
to activate the jobs from a subset $E$ of type (1.1) 
jobs with ``small enough'' total length. To this end, the next lemma, 
stating that there is a natural limit on the maximum
possible left-shift for the jobs in kernel $K$, is useful
as the total length of the jobs in a dominant set $E$ can 
correspondingly be bounded. Let $k$ be the latest scheduled candidate 
job to be included in set $E$, and let $a$ be the earliest scheduled 
non-emerging job, succeeding job $k$ in schedule $S$.

\begin{lemma}\label{left-shift}  
Kernel $K$ can be left-shifted by at most $c_k(S)-r_a < p_{\max}$ time units 
in schedule $S(-k,+a)$ compared to schedule $S$. 
\end{lemma}  
Proof. Since $k$ is the latest scheduled candidate  job, among all jobs of 
set $E$, its strong activation provides the maximum possible left-shift for 
the successively scheduled jobs. Note that job $a$ (with $r_a > t_k(S)$) 
is to be kept scheduled before kernel $K$, and it  
may start $c_k(S)-r_a$ time units earlier in schedule  $S(-k,+a)$ 
than in schedule $S$. Hence, the earliest job of kernel $K$ will 
start at most $c_k(S)-r_a$ time units earlier in schedule $S_k$ than 
the earliest scheduled job of that kernel started in schedule $S$. 
Finally, note that $c_k(S)-r_a < p_{k}$ since $r_a > t_k(S)$.\eop  

As before, $E$ and $E'$ stand for the sets of some jobs in the state of 
activation and not in the state of activation, respectively, for a 
primary/atomic kernel $K$ in an LDT-schedule $S$. 

\begin{lemma}\label{halt 1.1}
Let $e$ be a type (1.1) job for a primary/atomic 
kernel $K$ in LDT-schedule $S_E$, in which an IA(b1) with that
job occurs, let $E$ and $E'$ be a swapping pair and let $e'$ be 
a least urgent job from set $E'$. Further, suppose:

if $P(E') \ge P(E)$ then $$q_{e'} \ge q_e - (P(E') - P(E)),\ \ and $$

if $P(E') < P(E)$ then 
$$q_{e'} \ge q_e + (P(E) - P(E')).$$ 

Then the current best makespan cannot be improved and IEA can halt. 
\end{lemma}
Proof. The jobs from $J_{1.1}$ which are in the state of activation
for kernel $K$, in particular, those from set $E$, are included in 
LDT-order in schedule $S=S_E$ since all of them are released before 
the start time of kernel $K$. Hence $e$ is the last least urgent 
job from set $E$ included in schedule $S_E$. Likewise, a
least urgent job $e'$ from set $E'$ will be included the last in
in schedule $S(E,E')$. If $P(E') \ge P(E)$ then 
$c_{e'}(S(E,E'))\ge c_{e}(S(E))$, and by the second inequality, 
the full completion time of job $e'$ in schedule $S(E,E')$
will be no-less than that of job $e$ in schedule $S_E$. If 
$P(E') < P(E)$ then job $e'$ will complete 
$P(E) - P(E')$ time earlier and by the fourth inequality, 
the resultant makespan cannot again be less. So there 
exists no improving swapping and the lemma follows.\eop

...now, $P(E')$ cannot be much less than $P(E)$ as otherwise IA(a) 
with kernel $K$ will occur. 

The above lemma motivates the following definition where we keep
the notation used in that lemma. Again, we assume
that an IA(b1) in LDT-schedule $S_E$ occurs. 
We call {\em set $E'$ suitable due to IA(b1)} if for any $e'_i\in E'$, 
\\ if $c_{e'_i}(S_{E'}) \le c_{e}(S_{E})$ then 
$q_{e'_i} < q_{e} + (c_{e}(S_{E})-c_{e'_i}(S_{E'}))$, and 
\\ if $c_{e'_i}(S_{E'}) > c_{e}(S_{E})$ then 
$q_{e'_i} < q_{e} - (c_{e'_i}(S_{E'}) - c_{e}(S_{E})$.\\  
If $E'$ is suitable due to IA(b1), we will refer to schedule $S_{E'}$ and 
to set $E'$ as {\em dominant by IA(b1)} for kernel $K$. 
The next corollary follows from Lemma \ref{halt 1.1}. 
\begin{cor}\label{due IAb1}
It suffices to carry out -----combined interchanges for dominant due to IA(b1) 
sets for repeated occurrences of IA(b1) for a kernel $\bar K$, $K\in \cal K$. 
\end{cor}

Let $\Delta p^{(1.1)}_{\max}$ be the maximum difference 
between the processing times of type (1.1) jobs and let $q_{\max}$ 
be the maximum delivery time of a type (1.1) job. 
 
\begin{cor}\label{dom b1 bound}
For any kernel $K\in\cal K$, there are no more than
$q_{\max}+\Delta p^{(1.1)}_{\max}$ LDT-schedules dominant by IA(b1). 
\end{cor}
Proof. Follows from the fact that in each next created LDT-schedule
dominant by IA(b1) for kernel $K$, the sum of the corresponding parameters 
of the last scheduled job of the current dominant set $E$ is at least one less 
than that of the previously verified set.\eop 

\medskip

{\bf Kernel updates at stage 1.}
Recall that if no IA(b1) with an extended kernel $\bar K(E')$ schedule 
$S_{E,E'}$ occurs, then an EIA (b1) in that kernel arises. If 
the overflow job $o\in \bar K(E')$ is a type (4) job, then the corresponding
sequence of type (4) jobs defines a new primary kernel; the current set  
$\cal K$ is updated, the decomposition procedure for that kernel is
carried out, the set of type (1.2) jobs is updated respectively, and the 
fragment with that kernel in schedule $S_{E,E'}$ is replaced by the 
corresponding sequence of atomic kernels, similarly as at stage 0.

\medskip

In the following, with a gap before kernel 
$K$ in schedule $S_E$ we mean a gap in or after the execution 
interval of a job from set $E$ in schedule $S$. 

\begin{lemma}\label{EIA(b1) opt}
If in LDT-schedule $S_E$ with an EIA (b1) there is no gap 
before kernel $K$ and there exists no type (1.1) job for kernel 
$K$, then an optimal schedule is among the already created ones. 
\end{lemma}
Proof. The overflow job $o'\in \bar K$ in schedule $S_E$ 
is either a job from a primary kernel succeeding kernel $K$ or it
is a type (4) job succeeding the overflow job $o\in K$ of schedule 
$S$. Since there is no gap before kernel $K$ in schedule
$S_E$, the type (1.1) jobs of schedule $S$, the jobs of kernel $K$ and
all the succeeding jobs including job $o'$ scheduled in LDT-order
belong to the same block. Then the maximum job completion time 
in that kernel cannot be reduced since there exists no emerging job 
for kernel $\bar K$ and the lemma follows.\eop

\begin{cor}\label{EIA(b1) emerg}
If in LDT-schedule $S_E$ with an EIA (b1) there is no gap before 
kernel $K$, then for any LDT-schedule $S_{E'}$ with a makespan
better than that of schedule $S_E$, set $E'$ must contain a 
type (1.1) job for kernel $\bar K$.
\end{cor}
Proof. Note that by LDT-heuristic,  any $j\in E'$ which is a type (1.1) job for 
kernel $\bar K$ will be included behind all jobs of kernel $\bar K$ in schedule 
$S_{E'}$, i.e., it is in the state of activation for that kernel in that schedule. 
At the same 
time, from Lemma \ref{EIA(b1) opt}, the current makespan can only be 
potentially improved by activating a type (1.1) job for kernel $\bar K$.\eop 

Next, consider the case where there arises a gap before the jobs of kernel 
$K$ in schedule $S_E$. 

\begin{lemma}\label{EIA(b1) dom 1}
Suppose there is a gap before kernel $K$ in LDT-schedule $S_E$ with an 
EIA(b1) and no job from set $E'$ gets activated for kernel $\bar K$ in 
schedule $S_{E'}$. Then LDT-schedule $S_{E'}$ with $P(E')\ge P(E)$ is 
dominated by schedule $S_{E}$ and can be discarded.  
\end{lemma}
Proof.  Any job scheduled between a claimed gap in schedule $S_E$ and 
the overflow job $o'\in \cal K$ 
is no-less urgent than job $o'$ and hence is to remain before $o'$.  
By the hypothesis, set $E'$ includes no type (1.1) job for kernel $\bar K$
as otherwise it would have been activated for kernel $\bar K$ in schedule 
$S_E$ (see the proof of Corollary \ref{EIA(b1) emerg}); i.e., all jobs
scheduled before job $o'$ in schedule $S_{E}$ will remain before that job
in schedule $S_{E'}$. Then clearly, $c_{o'}(S_{E'})$ cannot be less than  
$c_{o'}(S_{E})$ if the total length of the gaps ......before kernel $K$ 
in schedule $S_{E'}$ is no less than that that in schedule $S_E$, and
$S_{E'}$ will be dominated by $S_{E}$. Now observe that the 
total length of the gaps in schedules $S_E$ and $S_{E'}$ before kernel $K$ 
is proportional to the total length $P(E)$ and $P(E')$ of the jobs in sets 
$E$ and $E'$. Correspondingly, for any non-dominated 
schedule $S_{E'}$,  $P(E')<P(E)$ must hold.\eop

\begin{cor}\label{EIA(b1) dom num}
The total number of dominant LDT-schedules $S_E$  from Lemma 
\ref{EIA(b1) dom 1}  is less than $p_{\max}$. 
\end{cor}
Proof. Let $e$ be the delaying emerging job in schedule $S$. Recall that
the gap that arises in schedule $S_e$ before kernel $K$ is smaller than $p_e$. 
By Lemma \ref{EIA(b1) dom 1}, for the next generated dominant LDT-schedule 
$S_E$, $p(E) < p_e$. 
We obtain the corollary by repeating similar argument for any two 
successive dominant LDT-schedules that potentially may be created.\eop

--Recall that any type (1.1) job 
for kernel $\bar K$ from a set $E$ becomes activated in schedule $S_E$ for 
that kernel (see the proof of Corollary \ref{EIA(b1) emerg}), which reduces
to one of the already studied cases.

\medskip


A detailed 
description and additional dominance properties are omitted due to
the space limitations.

======== END move to Appendix B=========== 
\endcomment

\comment

{\bf Bounds on the schedules with repeated IA(a) and IA(b2).} 
Now we derive the already mentioned upper bound on the total number of  
complete schedules that IEA can crate for repeated IA(a)/IA(b2). 
Let us consider a time interval where the jobs from a dominant 
subset $J' \in J_{1.2}(K)$ (to be included before the atomic kernel $K^*$) 
are to be scheduled. There are less than 
$p_{\max}^{1.1}(K)$ possible starting times for such an interval. 
Indeed, recall that, initially, the delaying emerging job pushes the jobs 
of kernel $K$, and that the forced delay for the first scheduled job of 
kernel $K$ varies from 1 to $p_{\max}^{1.1}-1$ (see Property \ref{delay}).   
If the delaying emerging job is activated for kernel $K$, an earliest 
released job from set $J_{1.2}(K)$ can start at its release time, i.e., it
is the earliest possible starting time for a time interval where the jobs 
from a subset $J'$ can be included. As we just saw, there are less 
than $p_{\max}^{1.1}(K)$ time moments at which such a time 
interval can start, further referred to as a 
{\em configuration for kernel} $K$.

As it is easy to see, to complete schedules respecting 
different permutations of type (1.1) jobs the same configuration may 
correspond. ccccc Whenever in a call of stage 1 for some permutation of type (1.1) 
jobs an IA(a) or an IA(b2) with a kernel $K$ occurs, the next subset of set 
$J_{1.2}(K)$ for the corresponding configuration is created (unless the
optimality condition from Lemma \ref{halt IEA} is 
satisfied). If all the dominant subsets for that configuration are 
already verified (in this case we let LDT$_K=NIL$ for that 
configuration), IEA halts and 
returns the generated schedule with the minimum makespan. 
It follows that the total number of  complete schedules with 
repeated instances of alternatives (a) and (b2) for kernel $K$ can
be bounded by $p_{\max}^{1.1}2^{\nu_{1.2}(K)}$, and the total 
number of complete schedules for repeated IA(a)/IA(b2) is
respectively $$\sum_{K\in \cal K} p_{\max}^{1.1}2^{\nu_{1.2}(K)}.$$  
Since there are at most $(n-\nu)$ kernels, we have an alternative
bound $$(n-\nu)p_{\max}^{1.1} 2^{\nu_{1.2}^{aver}(K)},$$ 
where a variable parameter $\nu_{1.2}^{aver}(K)$ is the average 
number of type (1.2) jobs in a kernel.  

Let  $p_{\min}(K)$ ($p_{\min}^{1.2}$, respectively) be
the minimum processing time of a job in set $J_{1.2}(K)$ (the 
minimum processing time of a type (1.2) job taken over all kernels, 
respectively),  $P_{1.2}(K)$ and $P_{1.2}$,
respectively, be the total processing time of jobs in set $J_{1.2}(K)$
and the average total processing time of the type (1.2) jobs 
in a kernel, respectively. In the 
following lemma, besides the above derived two bounds, we give 
alternative bounds containing no exponent in any variable parameter.

\begin{lemma}\label{compact}
The total number of complete schedules in which repeated instances
of alternatives (a) and (b2) with a kernel $K\in\cal K$ occur is bounded by
$$p_{\max}^{1.1} 2^{\nu_{1.2}(K)}$$ or alternatively by
$$p_{\max}^{1.1} 2^{(p_{\max}^{1.1}+P_{1.2}(K))/{p_{\min}^{1.2}(K)}}.$$
Hence, the total number of complete schedules in which repeated 
instances of alternatives (a) and (b2) occur is bounded by 
$$(n-\nu)p_{\max}^{1.1} 2^{\nu_{1.2}^{aver}(K)}$$ or alternatively, by
$$(n-\nu)p_{\max}^{1.1}2^{(p_{\max}^{1.1}+P_{1.2})/{p_{\min}^{1.2}}}.$$
\end{lemma}
Proof. We already showed the bounds 1 and 3. To see why bound 2 holds, 
consider again the interval where the jobs from set $J_{1.2}(K)$, to be 
included before the atomic kernel $K^*$, can be scheduled. The 
length of this interval is less than $p_{\max}^{1.1}+P_{1.2}(K)$ since
the length of the gap that arises after the activation of the delaying 
emerging job of kernel $K$ is less than $p_{\max}^{1.1}$ whereas the
total length of the jobs that can be included in that interval is no
more than $P_{1.2}(K)$. Then the total number of jobs that can be 
included within the interval is clearly bounded by
$(p_{\max}^{1.1}+P_{1.2}(K))/{p_{min}^{1.2}(K)}$ and the bound 2 follows. 
The bound 4 follows since the total number of kernels in set ${\cal K}$ 
is upper bounded by $n-\nu$ and there are no more than $p_{\max}^{1.1}$ 
possible configurations for any kernel.\eop

\subsection{Formal description of stage 1} 

In the formal description of stage 1, LDT$_{1.2}(K)$ stands for the 
complete schedule generated by the procedure for the next enumerated subset 
in the current configuration (recall from Section 3.2.1 that this subset is 
of the type $J''\setminus \{k\}$ ($J''$, respectively) for the loose (tight, 
respectively) cases, see Section 3.2.1); we let LDT$_{1.2}(K):=$NIL if all dominating 
subsets for the corresponding configuration are already enumerated. It 
straightforwardly follows from this section's earlier results and the construction
of stage 1, that for a permutation of type (1.1) jobs, it creates a complete 
schedule with the minimum makespan respecting that permutation.

+++++++****************************************

Suppose we
succeeded to construct a  {\em fully loose} LDT-schedule, one obtained
by strong activation of the delaying emerging job of each arisen kernel, 
such that  
The next corollary easily follows: 
\begin{cor}\label{fully loose}
A fully loose schedule is optimal. 
\end{cor}

\section{IEA: Stage 0}

Based on the initial job partition, at
stage 0 of IEA an auxiliary partial schedule without emerging 
jobs is created. At the initial iteration 0, for each kernel $K$ 
from LDT-schedule $\sigma$, the segment of kernel $K$ is replaced by 
the partial schedule $S^*[K]$ and LDT-heuristic is newly applied 
to the set of jobs 
$J\setminus\cup_K \{J_{1.1}(K)\}\setminus\cup_K K\}$, where 
$J_{1.1}(K)$ is the set of type (1.1) jobs in schedule $\sigma$ 
associated with kernel $K$. This results in a partial LDT-schedule 
$\sigma^0$ of iteration 0 (note that $\sigma^0$ does not include 
the type (1.2) jobs while it includes type (2) 
and type (3) jobs of each detected kernel). Iteratively, 
at iteration $h\ge 1$, for each kernel $K$ from schedule $\sigma^{h-1}$ 
similar operations are carried out and LDT-schedule $\sigma^h$ of 
iteration $h$ is similarly obtained. 
The procedure halts at iteration $h^*$ if the overflow job in 
schedule $\sigma^{h^*-1}$ is from schedule $S^*[K]$, for some 
kernel $K$ detected at iterations $0,\dots,h^*-1$. The number of 
iterations is clearly bounded by $n-\nu$, where 
$\nu=\nu_{1.1}+\nu_{1.2}$ is the total number 
of type (1.1) and type (1.2) jobs. Hence, the
whole procedure takes $O((n-\nu)n\log n)$ time. 

In IEA, a current state of computations is determined by the set 
${\cal K}$ of kernels detected so far and the corresponding job 
partition. Note that a kernel in schedule $\sigma^{h^*-1}$ is an 
atomic kernel of a kernel from set ${\cal K}$, and that this
schedule contains all jobs from set $J$ except the omitted jobs from 
sets $J_{1.1}$ (the set of all type (1.1) jobs) and $J_{1.2}(K)$ 
(the set of type (1.2) jobs associated with 
kernel $K$), $K\in {\cal K}$. The non-type (1-3) jobs from schedule 
$\sigma^{h^*-1}$ are declared as {\em type (4)} jobs. This completely
defines the initial job partition (note that a job from set $J_{1.1}$ 
can be associated with one or more kernels from set ${\cal K}$).   
We will further refer to schedule $\sigma^{h^*-1}$ as $\sigma(2,3,4)$.  

=========red=============
note that  a type (4) job from the initial 
job partition is scheduled either before  the earliest 
kernel from set ${\cal K}$ or between two partial schedules $S^*[K']$ and 
$S^*[K'']$, for two neighboring kernels $K'$ and $K''$ from set  
${\cal K}:= {\cal K}^{\xi + |{\bf K}^\xi|}$ or after the last of these 
partial schedules in schedule $\sigma^{\xi+|{\bf K}^\xi|}$.  
Note that every kernel in this schedule is an atomic 
kernel of the kernel $K^h$ of some iteration $h$, and that it
contains all jobs from set $J$ except the type (1) jobs 
from the initial job partition. 

$J_{1.1}$ and  $J_{1.2}(K)$ will stand for the  set of type (1.1) jobs
and  the set of type (1.2) jobs associated with kernel $K$, respectively.  
Observe that with every kernel $K\in {\cal K}$ its own type (1.1), 
type (1.2), type (2) and type (3) jobs are associated, whereas  
a type (1.1) job can be associated with one or more (successive) kernels
(with the ones for which it is an emerging job). Note that schedule 
$\sigma(2,3,4)$ contains no type (1.1) or (1.2) job. 
A type (4) job can be associated with a single kernel or with a pair 
of two neighboring kernels, depending on its  position. If a type (4) job $j$ 
is scheduled between two adjacent partial schedules  $S^*[K']$ and 
$S^*[K'']$ (before the first or after the last such partial schedule),   
then it cannot be an emerging job for none of the corresponding kernels; 
hence, it will be scheduled in between these two partial schedules (before 
the first and after the last such partial schedule, respectively). 
The jobs in set $J_{1.2}(K)$ are to be included immediately before, 
within or immediately after the jobs of schedule $S^*[K]$ and before 
the partial schedule $S^*[K']$ corresponding to the kernel $K'$ 
immediately succeeding kernel $K$ (see also Lemma \ref{distr}). 

In the lemma below $\nu=\nu_{1.1}+\nu_{1.2}$ is the total number 
of (type (1.1) and type (1.2)) emerging jobs. 

\begin{obs}\label{no delay}
The partial schedule $\sigma(2,3,4)$ is a well-defined feasible 
schedule obtained in less than $n-\nu$ iterations in time .
\end{obs}
Proof. Since schedule $\sigma(2,3,4)$ contains no type (1.1) and type (1.2) 
job, there occurs a gap before any partial schedule $S^*[K]$ for any 
detected kernel $K$. Hence, no overlapping of any job scheduled by 
LDT-heuristic with any of these partial schedules 
at any iteration may occur. Clearly,  
$h^* < n-\nu$ since less than $n-\nu$ kernels may occur.\eop
\endcomment

\comment  
\begin{lemma}\label{partition}
The sets of type (1)-(4) jobs form a partition of set $J$.
\end{lemma}
Proof. It is straightforward to see that any job is one of the types (1)-(4).
At the same time, these sets may have no (non-empty) intersection. We show 
that the set of type (1.1) jobs may have no intersection with the sets of the
other types of jobs. Suppose $e$ is a type (1.1) job for a kernel $K$ and
it is also part of a succeeding kernel $K'$. So, job $e$ gets included behind 
the jobs of kernel $K$. Since job $e$ is released before any job of kernel $K$, it
is no more urgent than any  job included after kernel $K$ and preceding job $e$. 
Since job $e$ belongs to kernel $K'$, there may exist no emerging job for kernel $K'$
scheduled after kernel $K$. But this implies that kernel $K$ cannot exist, i.e., the
set of jobs in $K$ and $K'$ must form part of a single kernel (to which job $e$
belongs). Thus either $e$ is not an emerging job or it does not belong to 
any kernel, i.e.,  there may exist no intersection of the set of type (1.1) 
jobs with the sets of type (1.2) and type (2)-(3) jobs.  The remaining 
implications can straightforwardly be seen.\eop
\endcomment 

\comment
\begin{lemma}\label{distr} 
There is an optimal schedule $S_{\opt}$ in which:

(a) Any type (4) job is included  between the intervals of adjacent partial 
schedules $S^*[K']$ and $S^*[K'']$, before the interval of the first of
these partial schedules or after that of the last of them. 
In particular, there is no intersection of the execution interval of a type 
(4) job in schedule $\sigma(2,3,4)$ with the time interval 
of any  type (2) and type (3) job from that schedule. 

(b2) If a type (1.1) job is an emerging job for two or more kernels 
$K^\iota,\dots, K^{\iota+l}$, then it is scheduled either before or 
after partial schedules $S^*[K^\iota],\dots,S^*[K^{\iota+l}]$ in schedule $S_{\opt}$; 
a type (1.1) job  which  is not an emerging job for kernel $K^\lambda$ 
does not appear after the jobs of partial schedule $S^*[K^\lambda]$ in that schedule. 

(c) A type (1.2) job associated with a kernel $K$ is scheduled within the 
interval of partial schedule $S^*[K]$ or after that interval  before 
the interval of the following partial schedule, whereas a job not from
kernel $K$ is not to  be included in between the jobs of that kernel.  

(d) A type (2) or type (3) job associated with a kernel $K$ can be scheduled after 
the initial time interval of partial schedule $S^*[K]$ only if it gets pushed by 
a type (1.1) job or/and by a type (1.2) job associated with kernel $K$.  
\end{lemma}
Proof. Part (a) holds as no type (4) job can be an emerging job for the 
corresponding kernel: a type (4) job scheduled between partial 
schedules $S^*[K']$ and $S^*[K'']$ is less urgent than any job from schedule
$S^*[K']$ and it is more urgent than any job from schedule $S^*[K'']$ (otherwise 
it would be a type (1.1) job for kernel  $K''$). Hence, it cannot be 
included after partial schedule $S^*[K'']$ in schedule $S_{\opt}$. Likewise,
there can be no benefit in including such jobs in between 
the jobs of these two partial schedules. The type (4) jobs included after the last partial 
schedule $S^*[\hat K]$ in schedule $\sigma(2,3,4)$ can be included after all jobs of that 
sequence (since they are less urgent than all jobs from schedule $S^*[\hat K]$ and hence 
there will be no benefit in rescheduling them earlier). This proves part (a). Part (b2), 
stating that the type (1.1) jobs can be dispelled in between the corresponding kernel 
sequences, easily follows. As to part (c), note that no type (1.2) job associated with
kernel $K^\iota$ is released before the interval of partial schedule $S^*[K']$ and 
it cannot be scheduled after the (initial) execution interval of that kernel without
causing the increase in the makespan. At the same time, only a type (1.1) job may 
potentially be scheduled  in between the jobs of kernel $K$. But this would yield
a forced left-shift of a more urgent job from the same kernel and cannot yield a
smaller makespan. Part (d) follows  since only a type (1.1) job can potentially be 
included before the initial time interval of partial schedule $S^*[K]$ and only 
a type (1.2) job can potentially be included within this time interval.\eop


\begin{lemma}\label{lb2}
The makespan of partial schedule $\sigma(2,3,4)$ is a lower bound on the 
optimal schedule makespan. 
\end{lemma}
Proof. By Lemma \ref{lb} and the fact that two kernel segments do not overlap 
in schedule $\sigma(2,3)$, $\sigma(2,3)$ is a feasible partial schedule such that
the maximum full job completion time in it is a lower bound on the optimal schedule 
makespan.  We show that this magnitude cannot be surpassed by any other job from 
schedule $\sigma(2,3,4)$. Indeed, any job $j\in \sigma(2,3,4)$ is either from 
partial schedule $S^*[K]$, for some $K\in {\cal K}$,  or it is a type (4) job. 
In the latter case, our claim follows since job $j$ does not belong to any kernel. 
If now $j\in S^*[K]$, then our claim follows from the fact that no type (4) job may 
push job $j$ in schedule $\sigma(2,3,4)$. Indeed, job $j$ may potentially be pushed 
only by a type (4) job in schedule $\sigma(2,3,4)$. But any type (4) job that was
originally scheduled before the delaying emerging job of kernel $K$, completes 
before the starting time of that job. But the latter job is omitted in schedule 
$\sigma(2,3,4)$ and hence no job can push job $j$. It follows that the full 
completion time of job $j$ is a lower bound on the optimum schedule makespan.\eop

\section{IEA: Stage 1}

At stage 1, type (1.1)  jobs from set $J_{1.1}$ 
are incorporated into the partial schedule $\sigma(2,3,4)$, and  
jobs from set $J_{1.2}(K)$ are incorporated locally into partial schedule 
$S^*[K]$, for each kernel $K\in {\cal K}$.

\subsection{Generating complete LDT-schedules}

We create an initial permutation of type (1.1) jobs and initial 
permutations of the jobs from $J_{1.2}(K)$, $K\in {\cal K}$, 
in a natural way while we generate our first complete LDT-schedule 
at stage 1. We basically use LDT-heuristic while extending the 
partial schedule $\sigma(2,3,4)$ with type (1) jobs. Iteratively, 
let the scheduling time $t$, at which the next type (1) job is included, 
be the earliest idle-time moment in the current augmented schedule such 
that there is yet unscheduled type (1) job released by that time. Among 
all such jobs,  ties are broken by selecting one with the maximum delivery time  
and further ties are broken by selecting any shortest job; if the so selected 
job overlaps  with a job from schedule $\sigma(2,3,4)$, the latter and 
the following jobs are right-shifted correspondingly. In this
way, the jobs from set $J_{1.2}(K)$ are included within the interval 
of schedule $S^*[K]$, for every $K\in {\cal K}$, since they
become released within that time interval and are more urgent 
than any released yet unscheduled job. The so formed partial schedule of 
the jobs of each kernel $K$ will contain no gap. Hence it has the minimum 
length among all partial schedules consisting of the jobs of kernel $K$
(imposing the smallest possible delay for the succeeding jobs). 

We denote the complete feasible schedule created as above
by $\sigma(\pi^*,\psi^*)$ (note that it is tight for all kernels 
$K\in \cal K$). The order of type (1.1) jobs (type (1.2) 
jobs from each set $J_{1.2}(K)$, respectively) in that schedule defines 
the {\em steady} permutation $\pi^*$ (permutations $\psi^*(K)$, for 
every $K\in {\cal K}$, respectively). Similarly, we create a complete 
LDT-schedule $\sigma(\pi,\psi)$ {\em respecting} given permutation $\pi$
of type (1.1) jobs and permutations $\psi(K)$, $K\in {\cal K}$, of type 
(1.2) jobs according to the order of these jobs in these permutations
while keeping the processing order of the jobs from schedule 
$\sigma(2,3,4)$ unchanged. 

===============RED===========================
The following dominance relation naturally arises from the restrictions 
imposed by job release times. 
Suppose while constructing a complete schedule respecting  permutation
$\pi=\{i_1,\dots,i_{\nu_{1.1}}\}$ there occurs a gap at iteration $\iota$, 
i.e., job $i_\iota$ is not released by that time, but there exists
$\iota'>\iota$ such that job $i_{\iota'}$ is released within that gap. 
Consider now an alternative permutation 
$\pi'=\{i_1,\dots,i_{\iota'},\dots, i_{\iota},\dots,i_\nu\}$ in which the
positions of the two jobs are interchanged and suppose that the 
makespan of the partial schedule respecting permutation 
$\{i_1,\dots,i_{\iota'},\dots, i_{\iota}$ is not more than that of
the partial schedule respecting permutation 
$\{i_1,\dots,i_{\iota},\dots, i_{\iota'}$. Then we will say that 
permutation $\pi$ is {\em dominated by release time} by permutation 
$\pi'$.  The next observation is easily seen.

\begin{obs}\label{dom by release}
A complete schedule respecting a permutation dominated by release time 
by another permutation can be discarded.
\end{obs}

\comment 
Initially at iteration 0, $\sigma(\pi,\psi,0):=\sigma(2,3,4)$; at iteration 
$\iota\ge 1$, schedule $\sigma(\pi,\psi,\iota)$ is obtained from schedule 
$\sigma(\pi,\psi,\iota-1)$ by including job $i_\iota$ at the earliest idle-time 
interval at (or) after time $r_{i_\iota}$ that occurs in schedule 
$\sigma(\pi,\psi,\iota-1)$ behind the jobs $i_1,\dots,i_{\iota-1}$; if the 
overlapping with an earlier included job from schedule $\sigma(\pi,\psi,\iota-1)$ 
occurs, the latter job and following ones are right-shifted by the (minimal) 
required amount of time units so that no overlapping of the execution 
intervals of these jobs occurs (the processing order of these jobs in schedule 
$\sigma(\pi,\psi,\iota-1)$ being kept unchanged). Whenever for the next kernel 
$K\in {\cal K}$, the time interval of the partial schedule $S^*[K]$
intersects with that of the last type (1.1) job  included so far or with the last
right-shifted job from schedule $\sigma(2,3,4)$, the jobs from the permutation 
$\psi(K)$ are similarly incorporated into the schedule $S^*[K]$ (no other job
being considered until the last job of kernel $K$ is scheduled). 

As we describe
a bit later, instead of dealing with permutations $\psi(K)$ of the jobs
from set $J_{1.2}(K)$, we will form 
subsets of these jobs and will schedule them by LDT-heuristic, forming in 
this way a corresponding partial schedule containing all jobs of kernel $K$.
This partial schedule will be incorporated into the formed complete schedule
on the fly, for each kernel $K\in {\cal K}$.

\subsection{Different types of kernels}  

In a created complete LDT-schedule $S$ at stage 1, a new kernel 
(one not in the current set of kernels ${\cal K}$) 
consisting of only type (4) jobs will be referred to 
as a {\em non-secondary} kernel. 
If the earliest occurred new kernel 
in schedule  $\sigma(\pi,\psi)$ is non-secondary, the current configuration 
is updated respectively and stage 1 is repeatedly invoked for the updated 
configuration. 
Note that, if a new kernel is not non-secondary, it must contain the job 
of some kernel(s) from set ${\cal K}$ (note that if it contains a single job 
of a kernel, it will contain all the remaining jobs from that 
kernel). Suppose, such a kernel, besides the jobs of a single kernel from set 
${\cal K}$, contains at least one more job, and let $K\in {\cal K}$ be the last 
kernel contained in that new kernel. Then this new kernel will be 
referred to as a {\em secondary} kernel of kernel $K$ and will commonly 
be denoted by $\bar K$. If a non-secondary kernel $K'$ is not new, it 
consists of the jobs of a single kernel $K\in {\cal K}$ (as job sets, 
the two kernels are the same, but they may be different as job sequences).

=================RED=================
\begin{obs}\label{delaying}
In an LDT-schedule $S$:\\
(i) A non-secondary kernel $K$  possesses the delaying emerging job, 
which is either a type (1.1) or a type (4) job from the current job partition.\\
(ii) The delaying emerging job of a secondary kernel can be only a type (1.1) 
job.\\
(iii) The overflow job of a non-secondary kernel is either a type (1.1) 
or a type (4) job, whereas that of a secondary kernel $\bar K$ is either the 
overflow job of an atomic kernel $K^*$ or a type (1.1) or a type (1.2) job. 
\end{obs}
Proof. The overflow job $o$ from a non-secondary kernel $K$ cannot  
be an emerging job, neither it can be a job from any kernel from 
${\cal K}\setminus K$. Hence, $o\in \sigma(2,3,4)$ and the completion time 
of job $o$ should have been increased in schedule $S$ compared to that
in schedule $\sigma(2,3,4)$. This increase could have only be caused by 
the type (1.1) emerging job(s) included before the jobs of kernel $K$. 
Either the last scheduled of these jobs is the delaying emerging job
for kernel $K$ or it pushes a (former) type (4) job that converts to a
type (1.1) job in the updated job partition. Part (i) follows.
To see part (ii), note that any type (4) job from the same block as 
kernel $K$ is no-less urgent than a job of that kernel. Hence, it cannot 
be an emerging job for kernel $K$, whereas neither a type (2) nor a type 
(3) job can be an emerging job for any kernel. 

The first claim of part (iii) easily follows since no job of a kernel 
from set ${\cal K}$ can belong to a non-secondary kernel. As to  
the second claim of part (iii), first note that no type (4) job scheduled 
after kernel $\bar K$ will be right-shifted in schedule $S$ by more time 
units than the last scheduled job of kernel $K$, and hence it cannot become 
the overflow job in the former kernel. xxxxx We claim that neither a type (3) 
job $i$ associated with kernel $K$ may become  the overflow job in 
kernel $\bar K$. It will suffice to show that job $i$, 
pushed by a type (1.2) job $j$ associated with kernel $K$, cannot be 
converted to an overflow job in schedule $S$. 
Suppose 
first that there is a gap between job $i$ and an atomic kernel $K^*$ in 
schedule   $S$. Then clearly, the claim holds as
the overflow job in schedule $S^*[K]$ is from  kernel $K^*$. Suppose
there is no such gap. Since $i$ is a type (3) job, the last job of 
kernel $K^*$ should have been the overflow job in an earlier generated
complete schedule, whereas there was no gap 
between job $i$ and kernel $K^*$ in that schedule. 
Then if job $i$ is pushed by job $j$, 
the jobs between  job $i$ and kernel $K^*$ will be right-shifted by the
same amount. Then the last job of kernel $K^*$ will remain to 
be the overflow job in that schedule. This completes the proof of part (iii). \eop

\subsubsection{Arrangement of the jobs of a secondary kernel}\label{ab}

Arrangement of the jobs of a secondary kernel is a most time consuming
part of the algorithm. 

========================RED=======================
Observe that a permutation of type (1.1) jobs imposes a specific 
time moment when the first job of kernel $K\in {\cal K}$ can start. 
For the LDT-schedule $S$ of the previous section, it was the time 
moment $c_e(S)$. Due to Property \ref{delay}, a forced delay imposed by a
permutation of type (1.1) jobs in an LDT-schedule respecting a given 
permutation of type (1.1) jobs for the first scheduled job of kernel $K$, 
from here on referred to as a {\em configuration of that kernel},  
can be any magnitude between 1 and $p_e-1$, where $e$ is the delaying
emerging job of kernel $K$. 
In (a tight) LDT-schedule $S$, the configuration for kernel $K$ is 
$\delta(K,S)$, whereas in (a loose) schedule $S_e$, it is 0 (as an 
earliest released job of kernel $K$  can start at its release time
without aby delay). Note that both above schedules respect the same
permutation of type (1.1) jobs. 

For each configuration of kernel $K$, some optimal arrangement of the
jobs of that kernel corresponds. Note that, since kernel $K$ is an 
LDT-sequence, this sequence is optimal for the 
$\delta(K,S)$-configuration of that kernel. However, if kernel $K$ is
not an atomic kernel, this sequence may not be optimal for the 
$\delta$-configuration of that kernel, for some $\delta< \delta(K,S)$.  
In this subsection, we describe how IEA manages possible job rearrangements 
of the jobs of kernel $K$ for a given configuration of that kernel.  

Suppose in schedule $S$ the first job of an atomic kernel $K^*$ does
not start at time $r(K^*)$ and that there occurs an IA(a) with kernel 
$K$ in that schedule (i.e., $K^*$ is a kernel in schedule $S$). Then 
any LDT-schedule $S'$ with $\delta(K^*,S') \ge \delta(K^*,S)$ will 
have a makespan at least as large as that of schedule $S$ since in
that schedule the overflow job from kernel $K^*$ will be right-shifted 
by  $\delta(K^*,S') - \delta(K^*,S)$ time units, i.e., will start
these time units later compared to its
starting time in schedule $S$. We will refer to such a schedule as 
one {\it dominated due to IA(a)} (by schedule $S$):  

\begin{obs}\label{dom atomic}
A schedule dominated due to IA(a) can be discarded.\eop
\end{obs}

xxxx needed???? At stage 1, we let $\beta(K):= \delta(K^*,S)$ whenever a complete 
LDT-schedule $S$ is created, in which an IA(a) for kernel $K$ with  
$\delta(K^*,S) < \beta(K)$ occurs (the initial value for the
global variable $\beta(K)$ that extends through all calls of stage 1 
is set to $\infty$).  

Suppose now an IA(b2) in schedule $S$ with kernel $K$ occurs, where
$i$ is the corresponding overflow job  from set $J_{1.2}(K)$. A makespan 
of an LDT-schedule $S'$, in which the last scheduled type (1.2) job 
$k\in J_{1.2}(K)$ with $q_k\ge q_i$ completes not earlier than job $i$ 
completes in schedule $S$ cannot be smaller than that of schedule $S$. 
In this case, we will say that schedule $S$ is {\em dominated due to 
an IA(b2)} (by schedule $S$): 

\begin{obs}\label{steady IAb} 
A schedule dominated due to IA(b2) can be discarded.\eop 
\end{obs}

\begin{obs}\label{non-new non-secondary kernel}
If a non-secondary kernel $K'$ in an LDT-schedule $S$ is not new, then 
an IA(a) or an IA(b2) with that kernel in schedule $S$ occurs.
\end{obs} 
Proof. From the hypothesis, kernel $K'$ consists of only the jobs of a 
single kernel  $K\in {\cal K}$. 
As $K'$ is a kernel in schedule $S$, it contains an overflow job of
that schedule. No type (3) job from kernel $K$ may be that overflow job 
(see the proof of part (iii) of Observation \ref{delaying}). Hence, it 
is either a type (2) job or a type (1.2) 
job from set $J_{1.2}(K)$. The observation follows.\eop
========================================================

---
{\bf Bounds for dominant permutations of type (1.1) jobs.}
Suppose a call of stage 1 completes for a permutation 
of type (1.1) jobs. By the construction of IEA, stage 1 
will again be invoked for some other permutation of type 
(1.1) jobs only if either the former call halted due to the condition 
in Lemma \ref{IAb} or it halted with an IA(a) or an IA(b2) once all  
dominant subsets of type (1.2) jobs were enumerated, i.e., 
with LDT$_K=NIL$ (steps (3.1) and (3.2.1), respectively, in the 
formal description of stage 1, see the next subsection) ---or 
----a new kernel . In the
following, we derive upper bounds on the total number of permutations
of type (1.1) jobs that can be created due to either of these two 
possibilities. 
----
\comment
We keep track of the current $q_e$ from Lemma \ref{dom 3.1} and 
Corollary \ref{dom 3.1 bound}  
 for kernel $K$ using a global variable $\gamma(K)$.
Initially, we let $\gamma (K):=\infty$, for every $K\in {\cal K}$.---

It keeps track of our  
dominance relations on the set of permutations of type (1.1) jobs using the  
global variables, which are set initially to $\infty$, for all 
$K\in {\cal K}$ and are updated each time a complete 
schedule respecting the next dominant  permutation is generated. 
Recall that in a complete schedule respecting a dominant by 
Observation \ref{dom atomic} permutation $\rho$ for kernel $K$, the forced delay for 
atomic kernel $K^*$ should be less than $\beta(K^*)= \delta(K^*,\sigma(\rho))$. 
At stage 1, while generating tight schedule $\sigma(\rho)$, for any atomic kernel 
$K^*\in {\cal K}$, if a forced right-shift for the first 
job of that kernel turns out to be more than $\beta(K^*)$, the current construction 
is altered by creating a loose for kernel $K$  complete schedule 
$\sigma_{d_{1.1}}(\rho,\psi(K'))$ in which the type (1.1) delaying emerging job $d_{1.1}$ 
is rescheduled after that kernel (see Observation \ref{dom atomic}). Similarly, in a 
tight complete schedule for kernel $K$ respecting permutation $\rho$, dominant by 
IA(a) for kernel $K$, 
the forced delay for the first 
scheduled job $j$ of xxxx kernel $\bar K$ must be less than  
$\beta(K)=\delta(\bar K,\sigma(\rho))$. While generating a tight
schedule $\sigma(\rho)$, for every kernel $K\in {\cal K}$ with
$\beta(K)<\infty$, if a forced right-shift for the first job of that kernel turns out
to be more than $\beta(K)$, a loose for kernel $K$ complete schedule respecting 
permutation $\rho$ in which the delaying emerging job is rescheduled after kernel 
$\bar K$  is created (see case (b2) just before  Observation \ref{dominant K}).

(0) if in fully loose no IA(b1)/ IA(b2) / IA(a) occurs return it (it is optimal)

If IA(b1) with dej job $l$ occurs, interchange and go to (1) 
If IA(a) occurs, create the next subset

================
(0) Generate schedule $\sigma(\pi^*,\psi^*)$ \{respecting the steady permutation 
$\pi^*$ and steady permutations $\psi^*(K)$,  $K\in {\cal K}$\}; 
$\sigma(\pi):=\sigma(\pi^*,\psi^*)$ 

(1) if there is an atomic kernel in schedule $\sigma(\pi)$  with no delaying 
emerging job, output $\sigma(\pi)$ and halt VP algorithm 

(2) let $K'$ be the first kernel in schedule $\sigma(\pi)$ 
\{either $K'=\bar K$, i.e., $K'$ is a secondary kernel of kernel $K\in{\cal K}$
or $K'=K^*$, i.e., $K'$ is the atomic kernel of kernel $K\in{\cal K}$\}

(2.1) if $\sigma(\pi)$ is loose for kernel $K$ schedule or such
schedule was already generated (at this or a previous call of stage 1),
proceed with step (3.1)

(2.2) if no loose for kernel $K$ schedule was earlier generated, 
let $e$ be the type (1.1) delaying emerging job for kernel $K'$ in 
schedule $\sigma(\pi)$; proceed with two alternative complete schedules:
 (i) the tight for kernel $K$ schedule  $\sigma(\pi)$
\{in which  job $e$  pushes the first job of kernel $K'$\},
and (ii) the loose for kernel $K$ schedule $\sigma(\pi):=\sigma_{e}(\pi,\psi(K'))$
\{in which job $e$ is rescheduled immediately after kernel $K'$\};
proceed with step (3.1) for case (i) and step (1.1) for case (ii)


(3.1) if $K'$ contains a type (1.1) job, output a created complete schedule with 
the minimum makespan and ----halt for permutation $\pi$   

(3.2) if an IA(a) or an IA(b2) with kernel $K'$ occurs  

(3.2.1) if  LDT$_{1.2}(K)=$NIL, output a created complete schedule with the 
minimum makespan and halt for permutation $\pi$

(3.2.2) if  LDT$_{1.2}(K)\ne $NIL, $\sigma(\pi):=$LDT$_{1.2}(K)$ and proceed with step (1.1) 
\endcomment

\comment

(3.2) if an IA(b2) with a job from set $J_{1.2}(K)$ occurs  

(3.2.1) if there is yet unconsidered dominated subset of jobs in $J_{1.2}(K)$: 
form a tight complete schedule respecting the next such subset; update $\sigma(\pi)$ 
to that schedule and proceed with step (1.1) 

(3.2.2) if all dominated subsets of jobs in $J_{1.2}(K)$ were already considered:
set $\alpha(K)$  to the minimum starting time of the first job from set  
$J_{1.2}(K)$ among all created tight schedules for kernel $K$; output a created 
schedule with the minimum makespan and halt for permutation $\pi$

(3.3) if an IA(a) with kernel $K'$  occurs 

 (3.3.1)  if the first scheduled job of the atomic kernel $K^*$ starts at its release time, 
output schedule $\sigma(\pi)$ and halt VP-algorithm \{Lemma \ref{lb}\}
 
 (3.3.2)  if the first scheduled job of the atomic kernel $K^*$ starts after its 
release time, carry out steps (3.3.2.1) and (3.3.2.2) in two alternative branches:

(3.3.2.1) create a loose for kernel $K'$ complete schedule $\sigma_{d_{1.1}}(\pi,\psi(K'))$ 
in which the type (1.1) delaying emerging job $d_{1.1}$ is rescheduled after that kernel;
$\beta(K'):=\delta(K',\sigma(\pi))$; 
$\sigma(\pi):=\sigma_{d_{1.2}}(\pi,\psi(K'))$ and proceed with step (1.1) 

(3.3.2.2) create a loose for kernel $K^*$ complete schedule $\sigma_{d_{1.2}}(\pi,\psi(K'))$ 
in which the type (1.2) job $d_{1.2}$ scheduled immediately before the atomic kernel $K^*$ 
is rescheduled immediately after that kernel; if a repeated IA(a) with kernel $K$ in that 
schedule occurs, output schedule $\sigma_{d_{1.2}}(\rho,\psi(K'))$  and halt VP-algorithm
\{Lemma \ref{lb}\}; if an IA(b2) in schedule $\sigma_{d_{1.2}}(\rho,\psi(K'))$ 
occurs, proceed with step (3.2.1), else: $\beta(K'):=\delta(K',\sigma(\pi))$; 
$\sigma(\pi):=\sigma_{d_{1.2}}(\pi,\psi(K'))$ and proceed with step (1.1) 

(4) if kernel $K'$ is non-secondary: \{update the current configuration\}
 ${\cal K}:={\cal K}\cup  K'$; xxxx update job partition -------- , 
 ---complete $\pi$--- ;
 \{invoke the decomposition procedure for kernel $K'$\}
update schedule $\sigma(\pi)$ by replacing in it the segment containing  kernel 
$K'$ with partial schedule ---$S^+[\psi^*(K'),\pi]$   and 
repeat from step (1.1) with the updated $\sigma(\pi)$. 
 
\comment 

----to the proof -----
\begin{obs}\label{lb3}
In schedule $\bar\sigma^{i}$, the maximum full completion time of a job from  
partial schedule $S^*[K^{i-1}]$ is a lower bound on the optimal schedule makespan.
\end{obs} 
Proof. By the construction, the first job of the secondary kernel $\bar K^{i-1}$ starts 
at its release time in schedule  $\bar\sigma^{i}$. Then in that schedule, the jobs of 
kernel $\bar K^{i-1}$ will be rescheduled as in the partial schedule $S^*[\bar K^{i-1}]$  
and the observation follows from  Lemma \ref{lb}.\eop 

There may exist 
at most $n-\nu-1$ kernels in the LDT-schedule of each iteration, whereas the same 
delaying emerging job may be activated at most once for the same kernel. Since 
the number of the delaying emerging jobs is bounded  
by the total number of the type (1) jobs, the total number of the activations 
cannot exceed $(n-\nu)\nu$. Then after less than $(n-\nu)\nu$ iterations, 
a complete schedule with no delaying emerging job will be created.\eop
\endcomment

\comment
\begin{cor}\label{stage 2}
For any permutation $\pi$, stage 1 creates a complete schedule 
with the minimum makespan respecting that permutation in time
$O(n\log n \ p_{\max}(n-\nu)2^{(p_{\max}+P_{1.2})/{p_{\min}}}$. 
\end{cor}
Proof. The makespan minimality follows from this 
subsection's earlier results. As to the time complexity, at step (4), the cost 
of declaring each non-secondary kernel and replacing the corresponding schedule part 
is absorbed in the cost of stage 0. While generating each complete 
schedule respecting permutation $\pi$, the overall cost is $O(n\log n)$ (that
of LDT-heuristic). The total number of the created schedules for repeated instances 
of alternatives (a)/(b2), for  all kernels and possible 
configurations is upper bounded by $p_{\max} 2^{(p_{\max}+P_{1.2})/{p_{\min}}}$
Lemma \ref{compact}. In particular, the 
total number of generated complete schedules for permutation $\pi$ is upper 
bounded by $O(n\log n \ p_{\max}(n-\nu)2^{(p_{\max}+P_{1.2})/{p_{\min}}}$.\eop

\subsection{Overall VP-algorithm}

We first summarize the iterative step of stage 0. 
Recall that, initially,  stage 0  forms the steady 
permutation $\pi^*$ and  calls stage 1 for that permutation that finds
a complete schedule with the minimum makespan respecting this permutation.  
Iteratively, each time stage 1 completes for a current permutation $\pi$, 
stage 0 resumes if stage 1 did not halt the whole algorithm. This happens
if either (1) step (3.1) (Lemma \ref{IAb}) or (2) step (3.2.1) is entered. Stage 0 then 
generates the next dominant permutation of type (1.1) jobs, which is created at 
stage 0 depending on whether the last call of stage 1 has halted at step (3.1) 
or at step (3.2.1) and calls stage 1 for that new permutation, which is 
determined as follows:

(1) Stage 1 halts for permutation $\pi$ at step (3.1): if there exists a yet 
untried candidate job $e'$ with $q_{e'}< \gamma(K)$, call stage 1 for a dominant
permutation $\rho$ (obtained from permutation $\pi$ by interchanging  jobs $e$ and 
$e'$, see Corollary \ref{dom 3.1 bound})

(2) Stage 1 halts for permutation $\pi$ at step (3.2.1): create the next dominant
(by IA(a)/IA(b2))   permutation $\rho=(\rho_1,\dots,\rho_l,\dots,\rho_{\nu_{1.1}})$ 
(such that in schedule $\sigma(\rho)$ no more than $(r_j+p_{\max})/p_{\min}$ jobs 
from permutation $\pi$ are included before the jobs of kernel $K$ and 
$q_{\rho_i} < q_o$, for all $i=l+1,\dots,\nu_{1.1}$, see  
the proof of Observation \ref{dom 10}). 


\begin{theorem}\label{Phase 1}
IEA generates an optimal solution to problem 
$1|r_j,q_j|C_{\max}$ in time 
$$O(n\log n(n-\nu_{1.1})^2\log(n-\nu_{1.1}) 2^{\nu})$$ or in time
$$O(n\log n(n-\nu_{1.1})^2\log(n-\nu_{1.1})(q_{\max}+\Delta p^{(1.1)}_{\max}))$$
if there occurs no suitable due to IA(b1) sequence of type (1.1) jobs. 
\end{theorem}
Proof. We already argued why the algorithm is optimal. As to the time 
complexity, the number of iterations, where a new kernel $K$ is
detected such that it has no intersection with some earlier kernel, 
cannot clearly be more than $n-\nu_{1.1}$ (an upper bound on the total 
number of such kernels). At each of these iterations, for the activation 
of the corresponding delaying emerging job, LDT-heuristic with cost  
$O(n\log n)$ is invoked, while the decomposition procedure has the 
cost $O(i^2\log i)$, where $i=|K|$ (Lemma \ref{lb}). 
For the purpose of this estimation, we may assume that each of 
these $n-\nu_{1.1}$ kernels have the same number $x\le n-\nu_{1.1}$ 
of jobs (this will maximize the overall cost). Then the 
total number of iterations is bounded by $(n-\nu_{1.1})/x$ and the total 
cost in all iterations is $O(n\log n(n-\nu_{1.1})^2\log(n-\nu_{1.1}))$ 
under the assumption that none of the instances IA(b1), IA(b2) or IA(a)
occurs, 
i.e., none of the above kernels has an intersection with an 
earlier detected kernel. Otherwise one of the behavior
alternatives IA(b1), IA(b2) or IA(a) occurs. These alternatives, 
in the worst case, will yield the enumeration of all possible combined 
interchanges, which number is bounded by $2^\nu$. If in none
of the created LDT-schedules there is no suitable due to IA(b1) sequence 
of type (1.1) jobs, an alternative bound $q_{\max}+\Delta p^{(1.1)}_{\max}$
holds (Corollary \ref{dom b1 bound}).\eop

{\bf A probabilistic estimation.} Now we give an intuitive  probabilistic
estimation of the running time of our algorithm. Assume that job delivery 
times are uniformly distributed within the interval $[0,q_{max}]$.  
Suppose, once stage 1 completes with the steady 
permutation $\pi^*$, stage 0 creates another (dominant) permutation $\rho$ of 
type (1.1) jobs and calls stage 1 for that permutation. Such an event may occurs when 
either step (3.1) or step (3.2.2) was entered during the previous call of stage 1 so
that stage 0 generated a dominant permutation $\rho$. Suppose the previous call ended
by the condition at step (3.2.2) with secondary kernel $\bar K$ and the type (1.2)
overflow job $o$. By Observation \ref{dom 10}, for any type (1.1) job $e'$ scheduled 
after the jobs of kernel $K$ in schedule $\sigma(\rho)$, $q_{e'}< q_o$. Assuming that 
there are just 2 such type (1.1) jobs (in average, one may expect to have about 
$\nu_{1.1}/2$ of them), the probability that  $\rho$ is a dominant permutation is 
$$\frac{q_{\max}}{2(q_{\max}+1)}\frac{q_{\max}}{2(q_{\max}+1)}< 1/4.$$ So the probability 
that $\mu$ such dominant permutations will be considered is less than 
${1}/{4^{\mu}}$ (which is close 0 for $\mu=\nu_{1.1}!$). If the previous call of 
stage 1 ended by the condition at step (3.1) with secondary kernel $\bar K$ and 
the type (1.1) overflow job $e$, for any type (1.1) job $e'$ included in 
schedule $\sigma({\rho})$ after the jobs of kernel $K$, $q_{e'} < q_e$ 
(the proof of Corollary xxxxxx \ref{dom 3.1}). Similarly to
the above case, the probability that two consecutive  dominant permutations will be
generated is less than $1/4$. 

\begin{theorem}\label{permutation}
The probability that stage 0 will create $2\mu$ consecutive dominant permutations 
of type (1.1) jobs is less than $\frac{1}{4^{\mu}}$.\eop
\end{theorem}

\endcomment

{\small
\paragraph {\bf Example.} We illustrate the implicit enumeration algorithm 
(IEA) on the running instance from Figures~1--6.
There are two type (1.1) jobs, 1 and 7, and one type (1.2) job 6, 
so relevant combined permutations are $((1,7),(6))$ and $((7,1),(6))$. 

Stage 1, invoked for (the steady) permutation $((1,7),(6))$, 
first creates schedule $\sigma$ that coincides with that of Fig. 1;
$\bar{K_1}=\{2,3,4,5,6\}$ is an extended kernel of the atomic 
kernel $K_1=\{2\}$ in $\sigma$ with the delaying emerging job 1. 
Hence this job is activated and an
alternative schedule $\sigma^1$, that coincides with that of Fig. 2,
is created. An atomic kernel $K_3=\{8,9,10,11,12\}$ has the 
delaying emerging job 7 in $\sigma^1$, which is also activated. The resultant
schedule $\sigma^{1,7}$ is depicted in Fig. 5. In $\sigma^{1,7}$,
6 is the delaying emerging job for kernel $\bar{K_1}$. Job $6$ is
activated and schedule $\sigma^{1,7,6}$ of Fig. 6 is created. 
There arises a new extended kernel $\bar{K_2}$ of the atomic kernel 
$K_2=\{3,4,5\}$ in schedule $\sigma^{1,7,6}$;  $\bar{K_2}$ has no delaying
emerging job and stage 1 halts for permutation $((1,7),(6))$
by Lemma \ref{IAb}. 

The second combined permutation \(((7,1),(6))\) is dominated by \(((1,7),(6))\): job~7 is not an emerging job for the first two atomic kernels in the latter order, and scheduling 7 before 1 cannot improve the makespan. In general, many permutations can be discarded by dominance relations (e.g., release‑time constraints or Corollary~\ref{shift}), so complete enumeration of all combined
permutations of type-(1) jobs is unnecessary. One can also impose a priority order on permutations to avoid exploring dominated orders.

Finally, observe that schedules \(\sigma^{1,7}\) and \(\sigma^{1,7,6}\) are optimal for this instance. Thus, out of the \(13! = 6{,}227{,}020{,}800\) feasible schedules for the 13 jobs, IEA verifies only four complete schedules to reach the optimum. This concrete run demonstrates how activation and dominance pruning confine the exponential search to a tiny set of structurally critical jobs,
making the algorithm effective in practice. \(\square\)  }

\comment 
As it is easy to see, the second combined permutation $((7,1),(6))$
is dominated by $((1,7),(6))$: 7 is not an emerging job
for the first two atomic kernels and scheduling it before 
job 1 cannot be beneficial. In general, one can avoid a complete
enumeration of all combined permutations of type (1) jobs. 
There are a number of conditions
that allow to discard dominated permutations.  
For example, as in this example, due to job release times, 
some permutations might be dominated by others as they would yield
non-active complete schedules (in such a schedule, a job can be restarted 
earlier without delaying some other jobs). Corollary \ref{shift} also
restricts potentially non-dominated permutations. There are optimality
conditions and other dominance relations that allow to discard 
potential permutations. One can also determine
a particular priority order in which permutations of type (1) jobs
are enumerated. Since these results take a considerable space 
and they do not improve our theoretical worst-case bounds, 
they are omitted. 

Finally, we observe that Schedules $\sigma^{1,7}$ and $\sigma^{1,7,6}$
are optimal. In summary, from $13!=6227020800$ 
possible feasible schedules of the 13 jobs, IEA verified only 4
complete schedules.\eop

Consider a small scheduling instance with 13 jobs. Naively, one would expect to explore all possible schedules. Our implicit enumeration algorithm (IEA), however, prunes almost the entire search space. For example, jobs 1 and 7 act as delaying emerging jobs for certain kernels, and by activating them we generate alternative schedules that capture the structural bottlenecks. Job 6 then emerges as another delaying job, leading to a refined schedule. At each stage, dominated permutations are discarded: some orders are provably non‑optimal because release times or structural dominance relations make them redundant.

The striking outcome is that instead of verifying billions of permutations, IEA needs to check only four complete schedules to reach the optimum. This illustrates how activation and dominance pruning confine exponential dependence to a handful of structurally critical jobs, while the rest can be arranged efficiently.

\begin{figure}
  \centering
    \includegraphics[width=\linewidth]{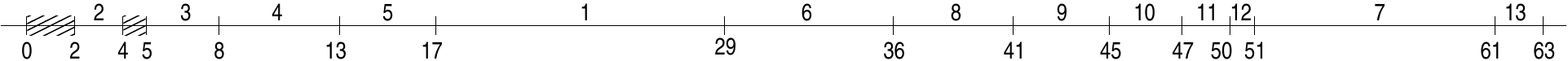} \\
  \caption{LDT-schedule $\sigma_1((1,6,7),3)$}
  \label{fig6}
\end{figure}

\endcomment

\begin{figure}
  \centering
    \includegraphics[width=\linewidth]{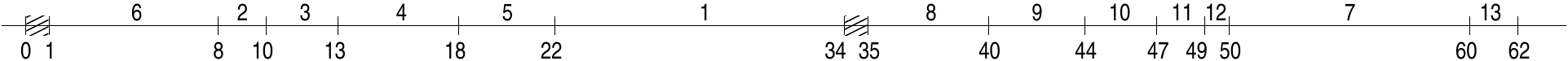} \\
  \caption{LDT-schedule $\sigma^{1,7}$}
  \label{fig5}
\end{figure}

\begin{figure}
  \centering
    \includegraphics[width=\linewidth]{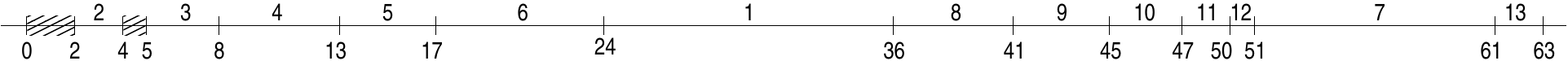} \\
  \caption{LDT-schedule $\sigma^{1,7,6}$}
  \label{fig7}
\end{figure}

\section{Partial scheduling}

As mentioned in the Introduction,  parameterized partial
scheduling with the fixed parameter equal to the number of
scheduled jobs was earlier studied. In this section, 
we show that there is no  need in 
parameterization of our scheduling problem if we allow
to omit at most $\omega\le n/2$ jobs. In particular, 
using previous section's results, we obtain a polynomial 
algorithm for that setting.

The $\omega$ jobs that we may omit are type (1) jobs, determined
as follows. Recall that partial schedule $\sigma^*$ contains
all non-emerging jobs, which number is $n-\nu$. Moreover, 
it is optimal for these $n-\nu$ jobs by Lemma \ref{lb}. We
can obtain a parameter that does not depend on $\nu$ 
omitting just delaying emerging jobs. 

\begin{obs}\label{dej}
An instance of problem $1|r_j,q_j|C_{\max}$ with $n$ jobs may
possess at most $n/2$ delaying emerging jobs.
\end{obs}
Proof. Let $I(n)$ be an instance of problem $1|r_j,q_j|C_{\max}$
with $n$ jobs and $n/2$ kernels, each kernels consisting of a single
job and possessing a delaying emerging job. As it is easy 
to see, among all instances of the problem with $n$ jobs, 
instance $I(n)$ maximizes the number of delaying emerging 
jobs. The observation 
follows since this instance has $n/2$ delaying emerging jobs.\eop

\begin{theorem}\label{fpt}
There exists an $(n-\nu) O(n \log n)$ algorithm for problem 
$1|r_j,q_j|C_{\max}$ that schedules $n-\omega$ jobs,
where $\omega\le n/2$.
\end{theorem}
Proof. We prove the theorem presenting the following  
algorithm that obtains an optimal solution consisting 
of all jobs except delaying emerging jobs. First, we carry out stage 0 
of IEA and obtain set of primary kernels $\cal K$, job partition and 
partial schedule $\sigma^*$ consisting of all non-emerging jobs. 
Let $x$ be the number of atomic kernels in schedule $\sigma^*$,  
i.e., there are at most $n-x$ type (1) jobs. 
We extend schedule $\sigma^*$ with these $n-x$ jobs by LDT-heuristic, 
similarly as in Algorithm 0, so that each time the next scheduled 
type (1) job pushes the first job of an atomic kernel, it is omitted. 
There are $n-\omega$ jobs in the so obtained schedule, where 
$\omega\le n/2$ by Observation \ref{dej}. This schedule is optimal 
by Lemma \ref{lb}. The construction cost is dominated
by that of stage 0, which is $(n-\nu) O(n \log n)$ (see Lemma 
\ref{stage 0 complex}). The theorem is proved.\eop

\smallskip 

{Example 00. (Partial Scheduling in Airline Systems).  } 
Consider again the airline scheduling scenario where an overloaded flight can accommodate only some part of the passengers. The scheduling task then becomes one of partial scheduling: selecting a subset of passengers to board while discarding the rest. In this setting, the guarantee is that at least 
$n/2$ passengers are scheduled: once this baseline guarantee is imposed, the problem becomes polynomially solvable without further parameterization.

From the structural perspective, the algorithm discards only emerging 
jobs - passengers with wide time windows who are flexible but can delay urgent ones. By omitting such non-urgent passengers, the schedule avoids delays to urgent passengers (kernels) with narrow time windows. Thus, the hardness is confined to the emerging jobs, while the urgent passengers can 
be scheduled efficiently. The non-urgent passengers have enough large 
departure times so that they can be served after all urgent passengers in
the next scheduling horizon. 

This example shows that partial scheduling becomes tractable once the structural hardness is isolated in emerging jobs, aligning practical intuition with the theoretical framework of above-guarantee parameterizations. It underlies our formal result: once the structurally 
hard emerging jobs are isolated, the partial scheduling variant of 
$1|r_j,q_j|C_{\max}$ with the guarantee that at least half the jobs are scheduled admits a polynomial-time algorithm, as established in Theorem
\ref{fpt}.

In summary, the airline analogy illustrates partial scheduling as the case of overloaded flights, where at least half the passengers can be accommodated. By discarding emerging jobs (passengers with wide time windows who could delay urgent ones) the schedule avoids bottlenecks and ensures that all urgent kernel jobs are served efficiently. This perspective shows that the structural hardness lies entirely in the emerging jobs, while the urgent passengers remain tractable. Consequently, partial scheduling admits a polynomial-time solution once the above-guarantee baseline is imposed, as
formalized in Theorem \ref{fpt}.\eop

\section{Polynomial time approximation scheme}

\subsection{Relevant properties}

Let $LB$ be a lower bound on the optimal schedule makespan $OPT$, $P$
be the total processing time of all jobs, 
and let $l$ be the delaying emerging job for a kernel in the initial 
LDT-schedule $\sigma$. The next property shows that for certain values 
of $k$ an $(1+1/k)$-approximation is guaranteed a priori.

\begin{property}\label{approx} 
If $k \le LB/p_l$, then $|S|/OPT < (1+1/k)$.
\end{property}
Proof. By Property \ref{delay}, 
$|S|/OPT < (OPT+p_l)/OPT=1+p_l/OPT\le 1+p_l/LB\le 1+1/k.$\eop

Call a job $j$ {\em short} if $p_j\le LB/k$; otherwise, call it  
{\em long}. If the delaying emerging job is short, its contribution 
to the makespan is negligible relative to $LB$, so 
the schedule is within a factor $(1+1/k)$ of optimal:

\begin{property}\label{appr-s}
If an LDT-schedule contains a kernel whose delaying emerging job 
$l$ is short, then the schedule is an $(1+1/k)$-approximation.
\end{property}
Proof. Identical to the proof of Property \ref{approx}, using 
$p_l\le LB/k$ for the last inequality.\eop

\begin{property}\label{kappa-long} 
There are fewer than $k$ long jobs.\eop 
\end{property}
Proof. If there were at least $k$ long jobs, their total processing 
time would exceed $k(LB/k)=LB$ contradicting $LB\ge P$.\eop


During the construction of an LDT-schedule $S$, let $S^t$ be the partial 
LDT-schedule at  scheduling time $t$, and let $j_t$ be the job 
scheduled at time $t$. We call $t$ a {\em conflict} scheduling time in  
$S^t$ if some job $j$ with $q_j > q_{j_t}$ is released
during the execution interval of job $j_t$ 
(so $j_t$ is pushing a more urgent job $j$). 

\begin{lemma}
If no conflict scheduling time occurs during the construction of an 
LDT-schedule $S$, then at every scheduling time $t$, no job released 
within the execution interval of $j_t$ can start a kernel in $S$.
Hence $S$ is optimal.
\end{lemma}
Proof. If no conflict occurs, no kernel in $S$ has a delaying emerging job; schedules with this property are known to be optimal.\eop



\subsection{The PTAS in aggregated terms}

Our PTAS can be implemented with or without the VPA enhancement. 
VPA groups short-job placements so that only permutations of the 
long jobs must be enumerated; intuitively, short jobs are handled 
implicitly and do not contribute to the exponential enumeration factor. 
Without VPA the algorithm must enumerate all $k!$ permutations of the 
$k$ long jobs, producing an exponential factor $k!$ in the running
time. With VPA this factor is replaced by at most 
$$\kappa_{1.1}! \ \prod_{K\in\mathcal{K}}\kappa_{1.2}(K)!,$$ where
$\kappa_{1.1}$ ($\kappa_{1.2}$, respectively) is the number of 
{\em long} type (1.1) ({\em long} type (1.2) jobs of
primary kernel $K$); 
$$\kappa= \kappa_{1.1}+\sum_K \kappa_{1.2}(K)$$  is  the total 
number of {\em long emerging} jobs. Note that $\kappa < k$ since
long emerging jobs form a subset of all long jobs 
(see also Lemma \ref{kappa-long}). In practice, $\kappa\ll k$ 
since the emerging jobs typically form a small subset of all
jobs (see Section 6 for empirical evidence).
Hence VPA can yield a significant practical speedup. In what follows, 
we present the VPA implementation; the simpler variant without VPA 
follows by, roughly, replacing $\kappa_{1.1}$ and $\kappa_{1.2}(K)$ 
with $k$ and by enumerating all $k!$ permutations.

We retain the notation of Section 3, but restrict it to {\em long} 
emerging jobs. Thus  permutation $\pi$ denotes the permutations 
of {\em long} type (1.1) jobs, and for each $K\in\cal K$, $\psi(K)$
denotes  a permutation of the {\em long} type (1.2) jobs from 
$J_{1.2}(K)$.  

The PTAS has two phases. Phase 0 runs stages 0 and 1 of IEA to 
construct the partial schedule of all the long jobs respecting 
a combined permutation of type (1) jobs. First, {\bf Algorithm\_0} 
is applied to only the {\em long non-emerging} jobs to obtain a 
partial LDT schedule $\sigma^*$. As in IEA, $\sigma^*$
is augmented with {\em long emerging} jobs according to 
a combined permutation $(\pi,\psi)$ ($\psi$ is a vector of 
permutations $\psi(K)$, $K\in\cal K$,  of all long type (1.2) jobs).  
The resultant partial schedule $\sigma(\pi,\psi,\kappa)$ contains
all $\kappa$ long jobs. Phase 1 extends and refines
$\sigma(\pi,\psi,\kappa)$ into a complete schedule using 
the procedure {\bf BASIC} described in the next section: short jobs are 
inserted (or handled implicitly) so that the $(1+1/k)$-approximation 
guarantee is preserved. The factorial product above is an upper bound 
on the number of combined permutations to be examined; the overall running time is this upper bound times the per-permutation cost of
procedure {\bf BASIC}.

Below we describe procedure {\bf PTAS}, which is the main procedure. 
First it creates the 
initial job partition and schedule $\sigma^*$ consisting of only long 
non-emerging jobs. Then it incorporates permutations of long emerging
jobs declaring new kernels if they occur and updating current configuration. 
The procedure {\bf BASIC} is repeatedly called to create a schedule 
with the smallest makespan respecting a current combined permutation 
of long type (1) jobs. All subroutines in procedure {\bf PTAS} are
described in the following sections. 

\smallskip

Below we describe procedure {\bf PTAS}, the algorithm's top-level routine.
At Phase 0 it creates schedule $\sigma^*$ that consists of only long 
non-emerging jobs using previous section's {\bf Algorithm\_0}. 
Next, for each combined permutation of the 
long emerging jobs, {\bf PTAS} incorporates these emerging jobs into the partial schedule $\sigma^*$, declares new kernels when they arise, and updates the current configuration accordingly, as it was described in Section 3.
The subroutine {\bf BASIC}, described in the next subsection, is invoked repeatedly: for a fixed combined permutation of the long emerging jobs, {\bf BASIC} builds a schedule that respects that permutation and returns the schedule with the smallest makespan found for that permutation.

Below we describe procedure {\bf PTAS}, the algorithm's top-level routine.
At Phase 0, {\bf PTAS} creates schedule $\sigma^*$ that consists of only long 
non-emerging jobs. Next, for each combined permutation of the 
long emerging jobs, {\bf PTAS} incorporates these emerging jobs into the partial schedule $\sigma^*$, declares new kernels when they arise, and updates the current configuration, accordingly, as it was described in Section 3.

\begin{breakablealgorithm}
{\small
\floatname{algorithm}{{\small PTAS}}
\renewcommand{\thealgorithm}{}
\caption{ }\label{alg:capi}
\begin{algorithmic}[1]
\State $J := $ the set of long jobs
\State $\sigma^* :=$ {\bf Algorithm}\_0$(J,0)$

\State $(\pi,\psi) :=$ the steady combined permutation $(\pi^*,\psi^*)$ of type (1) long jobs
\State $\sigma_{best} := \sigma$ \Comment{the best complete schedule found so far} 
\While{$(\pi,\psi) \not= nil$} \Comment{while exists an unchecked combined permutation of type (1) long jobs}
\State $\sigma_{act} :=$ {\bf Incorporate\_Permutation}$(\sigma^*,(\pi,\psi))$
\State \textbf{foreach} kernel $K$ in $\sigma^{act}$
\If{$K$ is an extended kernel}\Comment{decompose each segment of type (4) jobs}
\State \textbf{for each} segment $K(4)$ of type (4) jobs \textbf{do} \textbf{call Add\_Primary\_Kernel}$(K(4),\sigma_{act})$
\ElsIf{$K$ consists of only type (4) jobs}
\State \textbf{call Add\_Primary\_Kernel}$(K,\sigma_{act})$
\EndIf
\State \textbf{end foreach}
\State $\sigma_{best} :=$ the best schedule among  $\sigma_{best}$ and $\bf BASIC(\sigma_{act})$
\State $(\pi,\psi) :=$ the next combined permutation of type (1) long jobs,  $\mathbf{}{nil}$ if all permutations were considered
\EndWhile

\State \textbf{output} $\sigma_{best}$
\end{algorithmic}
}
\end{breakablealgorithm}

\subsection{Phase 1: Incorporating short jobs}

Phase 1, that incorporates short jobs in partial schedule 
$\sigma(\pi,\psi,\kappa)$ of long jobs created at phase 0, 
is an essential part of the PTAS. 
Combined permutations of long type (1) jobs are enumerated 
in any order. Phase 1, invoked for a combined permutation
$(\pi,\psi)$, incorporates the short jobs into the partial schedule 
$\sigma(\pi,\psi,\kappa)$ of long jobs produced in Phase 0. This insertion phase is the main difference between the PTAS and IEA: whereas IEA works directly with full permutations of type (1) jobs, the PTAS first fixes a combined permutation of the long type (1) jobs and then systematically 
inserts short jobs so as to preserve the $(1+1/k)$-guarantee.

We represent the insertion process for permutation $(\pi,\psi)$
by a simple solution tree $T(\pi,\psi)$. In our Depth-First Search 
we first generate a complete leftmost branch of the tree that
represents the first complete extension of partial schedule 
$\sigma(\pi,\psi,\kappa)$ with all short jobs. Iteratively, 
backtracking to the closest open node can be accomplished to 
create additional complete extensions of schedule $\sigma(\pi,\psi,\kappa)$. 
In each created complete schedule, some optimality conditions 
are verified, based on which either the procedure halts or the
search continues. Each node of corresponds to a partial schedule, 
and each edge corresponds to the decision to insert a particular 
job at a particular scheduling time. We
associate this particular job to that particular edge. 

We denote by $\sigma_h$ the partial schedule at iteration $h$, and by
$t_h$ the scheduling time at which the job added at iteration $h$ 
begins execution; $\sigma_h$ contains long jobs from $\sigma^*$ 
and those short jobs 
that start by time $t_h$. The PTAS enumerates combined permutations 
of the long type (1) jobs in arbitrary order; for each permutation 
$(\pi,\psi)$ it builds $T(\pi,\psi)$, returning the best complete 
schedule found.

\subsubsection{ Procedure {\bf BASIC}} 

In the formal description of the main procedure {\bf BASIC} of phase 1 below, 
at iteration $h > 0$, we denote by $r_h$ the release time of an earliest 
released yet-unscheduled job (of any size), and by $r^s_h$ the release 
time of an earliest released yet-unscheduled {\em short} job. Among all 
short jobs released at 
time $r^s_h$, the {\em short incoming} job $j_h$ is one with the largest 
delivery time; ties are broken arbitrarily in favor of any longest job.

Suppose $r^s_h=r_{j_h}$ falls inside the execution interval of a 
long job $i \in\sigma_{h-1}$ and let $\hat p_i$ denote the remaining processing time of $i$ from time $r^s_h$. If $\hat p_i$ is 
is {\em large} (i.e., of the same order as a long job) and 
$q_i < q_{j_h}$, then we say that job $i$ {\em covers} $j_h$. 

Two alternatives arise:

(i) {\bf Keep long job $i$ in place.} If the execution interval  
job $i$ is left unchanged, it will incur a delay of length $\hat p_i$ 
for (a more urgent short) job $j_h$ which may be significant. 

(ii) {\bf Omit long job $i$ temporarily.} If job $i$ is removed from 
its current execution interval so that job $j_h$ can start immediately, 
we create a gap; in this case we mark $i$ as {\em the pending long job} 
$p$ for the current branch. (The number of pending jobs is at most one 
per branch, and $\sigma_h$ always contains all 
long jobs except at most one pending long job $p$.) 

The subroutine {\bf BASIC} is invoked repeatedly: for a fixed combined permutation of the long emerging jobs, it builds a complete schedule that respects that permutation and returns a schedule with the smallest makespan found for that permutation. When inserting the next yet unscheduled job into the current 
schedule $\sigma_h$ at iteration $h$, if that job overlaps with a job from the 
previous partial schedule $\sigma_{h-1}$, that overlapped job and all subsequent long jobs from $\sigma_{h-1}$ are right-shifted accordingly. 
In the formal description below, we assume such right-shifts are performed at every iteration when the current schedule is extended by a new job. We write
$\sigma_h:=\sigma_{h-1}\setminus\{i\}\cup \{j\}$ to indicate that long 
job $i$ has been omitted in $\sigma_h$ (becoming the pending long job) 
while the incoming short job $j$ is included in that schedule; $p=nil$  
indicates that there is no pending job at the current iteration.

\begin{algorithm}[H]
{\small
\floatname{algorithm}{{\small Initial assignments to global variables}}
\renewcommand{\thealgorithm}{}
\caption{}\label{alg:cap}
\begin{algorithmic}
\State $h := 0$
\State $\sigma_h := \sigma(\pi^*,\psi^*,\kappa)$
\State $p := nil$


\end{algorithmic}
}
\end{algorithm}

\begin{breakablealgorithm}
{\small
\floatname{algorithm}{{\small BASIC}}
\renewcommand{\thealgorithm}{}
\caption{{\small$(\sigma_{h-1})$}}\label{alg:cap} 
\begin{algorithmic}[1]
\If{$\sigma_h$ is a complete schedule}
\State {\bf Backtrack}$(\sigma_h)$
\Else
\State $h:=h+1$
\If{$r_h$ falls in a gap in schedule $\sigma_{h-1}$}
\State \textbf{if} $p=nil$ or $q_{j_h} > q_p$ \textbf{then} $t_h := r_h^s;$ $\sigma_h:=\sigma_{h-1} \cup j_h$
\State \textbf{if} $p\not=nil$ and $q_{j_h} \leq q_p$ \textbf{then} $t_h := r_h;$ $\sigma_h:=\sigma_{h-1} \cup p$
\Else
\If{$r_h$ falls within the execution of a short job}
\State $t_h' := $ completion time of the latest scheduled short job in schedule $\sigma_{h-1}$
\State $j_h := $ short job with the largest delivery time among all short jobs released by time $t_h'$
\If{ the machine is idle immediately after time $t_h'$ }
\State \textbf{if} $p =nil$ or $q_{j_h} > q_p$ \textbf{then} $t_h := t_h';$ $\sigma_h:=\sigma_{h-1} \cup j_h$
\State \textbf{if} $p\not=nil$ and $q_{j_h} \leq q_p$ \textbf{then} $t_h := t_h';$ $\sigma_h:=\sigma_{h-1} \cup p$
\Else \Comment{the machine is not idle immediately after time $t_h'$}
\State $i := $ the long job scheduled at time $t_h'$ in schedule $\sigma_{h-1}$
\State \textbf{if} $p =nil$ and $q_{j_h} > q_i$ \textbf{then} $t_h := t_h';$ $\sigma_h:=\sigma_{h-1} \cup j_h$
\State \textbf{if} $p=nil$ and $q_{j_h} \leq q_i$ \textbf{then} $t_h := t_h';$ $\sigma_h:=\sigma_{h-1}$ \Comment{retain long job $i$}
\State \textbf{if} $p\not=nil$ and $q_{j_h} > q_p$ \textbf{then} $t_h := t_h';$ $\sigma_h:=\sigma_{h-1} \cup j_h$
\State \textbf{if} $p\not=nil$ and $q_{j_h} \leq q_p$ \textbf{then} $t_h := t_h';$ $\sigma_h:=\sigma_{h-1} \cup p$
\EndIf
\Else \Comment{$r_h$ falls within the execution of a long job $i\in\sigma_{h-1}$}
\If{$p\not=nil$}
\State \textbf{if} $q_{j_h} > q_p$ \textbf{then} $t_h := r_h;$ $\sigma_h:=\sigma_{h-1} \cup j_h$
\State \textbf{if} $q_{j_h} \leq q_p$ \textbf{then} $t_h := r_h;$ $\sigma_h:=\sigma_{h-1} \cup p$
\Else \Comment{retain long job $i$}
\State $t_h:= $ the starting time of job $i$ in $\sigma_{h-1}$
\State $\sigma_h := \sigma_{h-1}$
\EndIf
\EndIf 
\EndIf
\State {\bf call} {\bf BASIC}($\sigma_h$)
\EndIf
\end{algorithmic}
}
\end{breakablealgorithm}

\medskip

\comment 

\noindent (0) \{initialization of PTAS\} 
$h:=0$; $\sigma_0:=\sigma(\pi^*,\psi^*,\kappa)$; $p:=nil$; 
call Procedure BASIC$(h)$ 

\smallskip

Procedure BASIC$(h)$:

If at iteration $h$ a complete schedule is generated  

\hskip.2cm then call procedure BACKTRACK$(h)$ else $h:=h+1$ \{and proceed as follows\} 

\noindent (1) If time moment $r_h$ falls in a gap in schedule $\sigma_{h-1}$ then \\
(1.1) If $p = nil$ or $q_{j_h} > q_{p}$ then  
\{schedule short job $j_h$  at time $t_h=r_h=r^s_h$\}\\ 
$t_h:=r^s_h$; $\sigma_h:=\sigma_{h-1} \cup j_h$\\ 
(1.2) If $p \ne nil$ and $q_{j_h} \le q_{p}$ then
\{schedule long job $p$ at time $t_h=r_h$\}\\ 
$t_h:=r_h$; $\sigma_h:=\sigma_{h-1} \cup p$. 

\noindent (2) Else \{time moment $r_h$ falls within the execution interval of 
(2.1) a short job or (2.2) a long job from schedule $\sigma_{h-1}$\}

\noindent (2.1) 
\{Time moment $r_h$ falls within the execution interval of a short job\}\\
Let $t'_h$ be the completion time of the latest scheduled short job in schedule $\sigma_{h-1}$\\ 
\{among all the short jobs released by time $t'_h$, $j_h$ is one with the largest delivery time\}\\ 
(2.1.1) If the machine is idle immediately after time $t'_h$ 
\{there is no right-shifted (long) job scheduled after time $t'_h$ in $\sigma_{h-1}$\} then  
\{schedule either job $j_h$ or job $p$ at time $t'_h$\}\\
(2.1.1.1) If $p=nil$ or $q_{j_h} > q_{p}$ then 
\{schedule short job $j_h$  at time $t_h=t'_h$\}\\
 $t_h:=t'_h$; $\sigma_h:=\sigma_{h-1} \cup j_h$\\          
(2.1.1.2) If $p\ne nil$ and $q_{j_h} \le q_{p}$ then  
\{schedule long job $p$ at time $t_h=t'_h$\}\\ 
 $t_h:=t'_h$; $\sigma_h:=\sigma_{h-1} \cup p$.\\                     
(2.1.2) If the machine is not idle immediately after time $t'_h$,          
let $i$ be the long job scheduled at time $t'_h$ in schedule $\sigma_{h-1}$\\ 
(2.1.2.1) If $p = nil$ and $q_{j_h} > q_{i}$ then
\{schedule short job $j_h$  at time $t_h=t'_h$\}\\
 $t_h:=t'_h$; $\sigma_h:=\sigma_{h-1} \cup j_h$\\          
(2.1.2.2) If $p = nil$ and $q_{j_h} \le q_{i}$ then
\{retain long job $i$ scheduled at time $t_h=t'_h$\}\\ 
 $t_h:=t'_h$; $\sigma_h:=\sigma_{h-1} \cup i$.\\
(2.1.2.3) If $p \ne nil$ and $q_{j_h} > q_{p}$ then
\{schedule short job $j_h$  at time $t_h=t'_h$\}\\
 $t_h:=t'_h$; $\sigma_h:=\sigma_{h-1} \cup j_h$\\  
(2.1.2.4) If $p \ne nil$ and $q_{j_h} \le q_{p}$ then
\{schedule long job $p$ at time $t_h=t'_h$\}\\ 
 $t_h:=t'_h$; $\sigma_h:=\sigma_{h-1} \cup p$.

\noindent (2.2) 
\{Time moment $r_h$ falls within the execution interval of a long job $i\in\sigma_{h-1}$\}\\
(2.2.1) If $p \ne nil$ then \{schedule either job $j_h$ or job $p$ at time $r_h$\}\\
(2.2.1.1) If $q_{j_h} > q_{p}$ then 
\{schedule short job $j_h$  at time $t_h=r_h$\}\\
 $t_h:=r_h$; $\sigma_h:=\sigma_{h-1} \cup j_h$\\          
(2.2.1.2) If $q_{j_h} \le q_{p}$ then 
\{schedule long job $p$ at time $t_h=r_h$\}\\ 
 $t_h:=r_h$; $\sigma_h:=\sigma_{h-1} \cup p$\\  
(2.2.2) If $p = nil$ then       
\{retain long job $i$ scheduled as in schedule $\sigma_{h-1}$\}\\ 
 $t_h:=$the starting time of job $i$ in schedule $\sigma_{h-1}$

call Procedure BASIC$(h)$.  

\endcomment

\subsubsection{ Procedure  {\bf Backtrack}} 

Procedure {\bf Backtrack} explores alternative branches in the solution 
tree $T(\pi,\psi)$  and terminates or branches according to the 
following exhaustive cases.

{\bf Termination check (0).} If the most recently created complete 
schedule $\sigma_h$ contains a kernel whose delaying emerging job is 
short, or contains a long delaying emerging job $e$ whose remaining 
portion $\hat p_e$ (measured from its release inside the kernel) 
is short, then output $\sigma_h$ and stop.

Otherwise exactly one of the following holds:

{\bf (i)} Every kernel in $\sigma_h$ has a long delaying emerging job 
$i$ whose remaining portion $\hat p_i$ is long.

{\bf (ii)} There exists a kernel $K\in \sigma_{h}$
with no delaying emerging job. Two subcases arise:

{\bf (ii.1)} $K$ is an atomic kernel. Then $\sigma_h$ is optimal and 
PTAS halts (Lemma \ref{lb}).

{\bf (ii.2)} $K$ is an extended kernel. Then the construction of 
$T(\pi,\psi)$ for the current permutation $(\pi,\psi)$ is 
complete (Lemma \ref{IAb}).

{\bf Backtracking step (case (i)).} Let $K(\sigma_{h})$ by the first 
kernel in $\sigma_{h}$ (the earliest in schedule order) and let $i$
be its long delaying emerging job. Let $f$ be the predecessor node of 
$h$ in the tree (the node from which $\sigma_{h}$ was created). 
Backtrack to node $f$ and create a second child $g$ of $f$ as follows:
in $\sigma_{g}$ activate job $i$ for kernel $K(\sigma_{h})$
(i.e., schedule $i$ so it no longer delays the incoming short job $j_f$).
Insert the short incoming job $j_f$ at its release time $r_{j_f}=r_f$;
Mark $i$ as the pending long job $p$ for this branch. Invoke 
{\bf BASIC}$(g)$ on node $g$ to continue exploration of the current
branch from $\sigma_{g}$. The pseudo-code of the procedure is given below.


\begin{breakablealgorithm}
{\small
\floatname{algorithm}{{\small {\bf Backtrack}}}
\renewcommand{\thealgorithm}{}
\caption{{\small $(\sigma_{h'})$ }}\label{alg:cap}
\begin{algorithmic}[1]
\If{the complete schedule $\sigma_{h'}$ contains a kernel with either a short delaying emerging job or with a long delaying emerging job $e$ such that $\hat{p_e}$ is short}
\State {\bf halt} \{PTAS\} and {\bf return} $\sigma_{h'}$ \Comment{it is an $(1+1/k)$-approximation schedule}
\EndIf
\If{every kernel in schedule $\sigma_{h'}$ possesses a long delaying emerging job}
\Comment{create the second son $g$ of node $f$}
\State backtrack to the node $f$ of the current branch such that $i$ is a long delaying emerging job for the kernel $K(\sigma_{h'})$ \{$\hat{p}_i$ is long\} 
\State $t_g:=r_{f}$ \Comment{set the scheduling time for iteration $g$}  
\State $\sigma_g := \sigma_{f} \setminus {i} \cup {j_f}$ \Comment{schedule short job $j_f$ at time $r_f$}
\State $p:= i$ \Comment{declare long job $i$ as pending}
\State {\bf call} {\bf BASIC}($\sigma_g$)
\Else \Comment{schedule $\sigma_{h'}$ contains a kernel with no delaying emerging job}
\State \textbf{if} there is an atomic kernel with no delaying emerging job 
in schedule $\sigma_{h'}$
\textbf{then} \textbf{return} schedule $\sigma_{h'}$ and {\bf halt} \{ PTAS\} \Comment{schedule $\sigma_{h'}$ is optimal}
\State \textbf{else} \Comment{any kernel in schedule $\sigma_{h'}$ is an extended one} 
\textbf{return} a schedule from tree $T(\pi)$ with the smallest makespan 
\EndIf
\end{algorithmic}
}
\end{breakablealgorithm}

\comment

Procedure {\bf Backtrack}$(h')$:

(0) if the complete schedule $\sigma_{h'}$ contains a kernel with either a short 
delaying emerging job or  with  a long delaying emerging job $i$ such that 
$\hat p_i$ is short, then finish the construction of tree $T(\pi,\psi)$ and stop
\{$\sigma_{h'}$ is an $(1+1/\kappa)$-approximation schedule\}

(i) if every kernel in  schedule $\sigma_{h'}$ possesses the long delaying emerging job 
$i$ such that $\hat p_i$ is long, then \{continue the construction of tree $T(\pi,\psi)$\}\\
backtrack to the node $h-1$ of the current branch such that $i$ is a long delaying 
emerging job for the kernel $K(\sigma_{h'})$ and $\hat p_i$ is long 
\{create the second alternative son $g$ of node $h-1$ in which 
short job $j_h$ is scheduled  at time $t_h=r_h$\}\ \ 
$t_g:=r_h$; $\sigma_g:=\sigma_{h-1} \setminus\{i\}\cup \{j_h\}$; $p:=i$;  
call Procedure BASIC$(g)$         

(ii) else \{schedule $\sigma_{h'}$ contains kernel $K$ with no delaying emerging job\}

(ii,1) if $K$ is an atomic kernel, return schedule 
$\sigma_{h'}$  and halt \\ \{schedule $\sigma_{h'}$ is optimal\}

(ii,2) else \{$K$ is a secondary kernel with a type (1.1) job\}, 

\hskip 1cm return a schedule from tree $T(\pi,\psi)$ with the smallest makespan.

\endcomment

\subsection{Soundness and time complexity}

{\bf PTAS} repeatedly calls procedure {\bf BASIC} for each next partial schedule $\sigma(\pi,\psi,\kappa)$
 until a halting condition in procedure {\bf Backtrack} is satisfied.
  
\begin{theorem}\label{PTAS}
	Procedure {\bf BASIC} invoked for permutation $(\pi,\psi)$ delivers a
	solution which is at most $(1+1/k)$ times a solution of the
	minimum makespan respecting $(\pi,\psi)$. {\bf BASIC} runs in time 
	bounded by $$\kappa(n-\kappa)O(n\log n).$$ Hence, PTAS delivers an
	$(1+1/k)$-approximation solution and runs in time  bounded by    	
  $$\kappa(n-\kappa)O(n\log n) \kappa_{1.1}! \, 
  \prod_{K\in\mathcal{K}}\kappa_{1.2}(K)!.$$
	\end{theorem}
	
	\comment
Proof. If schedule $\sigma_{h}$ created by Procedure {\bf BASIC} contains a kernel with 
either a short delaying emerging job or with a long delaying emerging job $e$ 
such that $\hat p_e$ is short, it is an $(1+1/k)$-approximation solution
by Property \ref{appr-s}. Otherwise, one of the above mentioned cases 
(i), (ii,1) or (ii,2) occurs. In case (ii,1) ($\sigma_{h}$  contains an atomic 
kernel with no delaying emerging job), schedule $\sigma_{h}$ is optimal, see Lemma \ref{lb}. In case (ii,2) (schedule $\sigma_{h}$ contains an extended kernel), 
no update of the current configuration is required due to an allowable amount of possible right-shift. Then a schedule with the minimum makespan respecting combined permutation $(\pi,\psi)$ must already be in tree $T(\pi,\psi)$, see Lemma \ref{IAb}.  

Consider now  case (i), where every kernel in schedule $\sigma_{h}$ 
possesses a long delaying emerging job. In a schedule with a better 
makespan, the full completion time of the overflow job of the first 
kernel $K(\sigma_{h})$ of schedule $\sigma_{h}$ must be reduced. 
For this, the corresponding long delaying emerging job is to be
activated (note that the activation of other type (1) job would 
violate the order in permutation $(\pi,\psi)$). Let the long incoming 
job $i_{h_1}$ of iteration $h_1$ be the delaying emerging job 
for kernel $K(\sigma_{h})$ in schedule $\sigma_{h}$ (see line 5 in the
description of Procedure {\bf Backtrack}). Let $\sigma_{h'}$
be the complete successor-schedule of schedule $\sigma_{h_1}$ created
after the corresponding call of Procedure {\bf BASIC} (see line 9 in the
description of Procedure {\bf Backtrack}). In schedule $\sigma_{h'}$, 
the first job of kernel $K(\sigma_{h})$ cannot be pushed by a long
job and hence schedule $\sigma_{i_{h_1}}$ already  gives the desired 
approximation if $K(\sigma_{h'})=K(\sigma_{h})$.

If $K(\sigma_{h'})\ne K(\sigma_{h})$, then the two kernels may have a job 
in common or not. In the former case, kernel $K(\sigma_{h'})$ must be
an extended kernel of kernel $K(\sigma_{h})$ containing job $i_{h_1}$. 
Note that kernel $K(\sigma_{h'})$ possesses no long delaying emerging job. 
Then as above no additional complete schedule respecting permutation $(\pi,\psi)$ in 
which job $i_{h_1}$ is in the state of activation for kernel $K(\sigma_{h})$
needs to be created  and Procedure {\bf Backtrack} can halt for permutation 
$(\pi,\psi)$ (line 12 in the formal description, see also Lemma \ref{IAb}). 
In the other case where kernels 
$K(\sigma_{h'})$ and $K(\sigma_{h})$ have no job in common, if there is a kernel 
in schedule $\sigma_{h_1}$ with a short delaying emerging job then again, this
schedule gives the desired approximation. Otherwise, the delaying emerging 
job of kernel $K(\sigma_{h'})$ is the incoming job $i_{h_2}$ of some iteration 
$h_2$. Then an alternative LDT-schedule in which  job $i_{h_2}$ is 
in the state of activation for kernel $K(\sigma_{h'})$ is created and the whole
reasoning can now be repeatedly applied. 

Next, we show that in less than $\kappa_{1.1} n$ steps, one of the above 
halting condition in Procedure {\bf Backtrack} will be satisfied; in other words, 
the total number of times a long delaying emerging job can be 
activated is bounded by the above magnitude. A long delaying emerging job $i_h$, 
once activated for kernel $K(\sigma_h)$, is to remain in the state of 
activation for that kernel at any following iteration in any branch of 
tree $T(\pi,\psi)$. Indeed, if in any complete schedule respecting 
permutation $(\pi,\psi)$ job $i_h$ is scheduled before kernel $K(\sigma_h)$, 
no type (1) job preceding job $i_h$ in permutation $(\pi,\psi)$ can be 
rescheduled after that kernel. But a complete schedule in which all
these jobs are scheduled before kernel $K(\sigma_h)$ cannot clearly have 
a makespan, less than that of schedule $\sigma_h$. 
So, a long delaying emerging job can be activated at most once for the 
same kernel in tree $T(\pi,\psi)$, whereas there are less than 
$n-\kappa$ kernels. Then
the total number of the complete schedules that can be created in 
solution tree $T(\pi,\psi)$ is bounded by $\kappa(n-\kappa)$ since there 
may occur no more than $\kappa$ long delaying emerging jobs. 

Procedure BASIC uses a variation of LDT-heuristics with cost $O(n\log n)$ 
for the generation of each complete schedule. Hence, the overall 
cost of the Procedure {\bf BASIC} invoked for permutation $(\pi,\psi)$ is 
$\kappa(n-\kappa)O(n\log n)$. 

Procedure {\bf PTAS} enumerates all 
$\kappa_{1.1}! \ \sigma_{K\in \cal K}\kappa_{1.2}(K)!$ possible combined 
permutations of type (1) long jobs updating current configuration whenever 
an extended kernel with a type (4) job after the incorporation of the next
combined permutation occurs (lines 9-11). The cost of the overall scheme  
follows.\eop   

\subsection{Soundness and time complexity}

\begin{theorem}\label{PTAS}
For any fixed combined permutation \((\pi,\psi)\), a single invocation of
{\bf BASIC} returns a schedule respecting \((\pi,\psi)\) whose makespan
is at most \(1+1/k\) times the optimum for that permutation. The running
time of {\bf BASIC} for one permutation is

\[
O\big(\kappa(n-\kappa)\,n\log n\big).
\]

Consequently, the running time of {\bf PTAS} is bounded by

\[
O\big(\kappa(n-\kappa)\,n\log n\big)\cdot
\kappa_{1.1}!\;\prod_{K\in\mathcal{K}}\kappa_{1.2}(K)!,
\]

where \(\kappa_{1.1}\) and \(\kappa_{1.2}(K)\) are as defined above and
\(\kappa=\kappa_{1.1}+\sum_{K}\kappa_{1.2}(K)\).
\end{theorem}

\endcomment

\begin{proof}
If the schedule \(\sigma_h\) produced by {\bf BASIC} contains a kernel
whose delaying emerging job is short, or contains a long delaying
emerging job \(e\) whose remaining portion \(\hat p_e\) (measured from
the release that caused the conflict) is short, then Property
\ref{appr-s} implies that \(\sigma_h\) is a \((1+1/k)\)-approximation.
Otherwise, by the Backtrack analysis exactly one of the cases (i),
(ii.1), (ii.2) occurs.

In case (ii.1) there exists an atomic kernel in \(\sigma_h\) 
with no delaying emerging job; 
by Lemma \ref{lb} \(\sigma_h\) is optimal. In case (ii.2)
(\(\sigma_h\) contains an extended kernel with no delaying 
emerging job) no yet unverified schedule may have a makespan
less than that of a schedule already in \(T(\pi,\psi)\), see 
Lemma \ref{IAb}.

Consider case (i), where every
kernel in \(\sigma_h\) has a long delaying emerging job. To improve the
full completion time of the overflow job of the first kernel
\(K(\sigma_h)\) one must activate the corresponding long delaying
emerging job; activation of any other type (1) job would violate the
order imposed by \((\pi,\psi)\). Let \(i_{h_1}\) be the long incoming
job at iteration \(h_1\) that delays \(K(\sigma_h)\) (see line 5 in 
the {\ bf Backtrack}
description). After creating the alternative branch where \(i_{h_1}\) is
activated and invoking {\bf BASIC} on that branch (see line 9), a 
complete schedule  \(\sigma_{h'}\) is created. In  \(\sigma_{h'}\)
one of the following occurs:

\begin{itemize}
  \item  \(K(\sigma_h)\) remains a kernel in schedule \(\sigma_{h'}\). 
  Then kernel \(K(\sigma_h)\) is not pushed by
    a long job because of the activation of \(i_{h_1}\) and \(\sigma_{h'}\) provides  the desired approximation.
  \item  \(K(\sigma_{h'})\neq K(\sigma_h)\) but the two kernels share
    a job. Then \(K(\sigma_{h'})\) is an extended kernel of
    \(K(\sigma_h)\) that contains \(i_{h_1}\). In this case
    \(K(\sigma_{h'})\) has no long delaying emerging job, so no further
    schedule with \(i_{h_1}\) activated for \(K(\sigma_h)\) can improve
    the makespan, see again Lemma \ref{IAb}; so
    {\bf Backtrack} can halt for permutation \((\pi,\psi)\) (line 12 in the formal description).
  \item  \(K(\sigma_{h'})\) and \(K(\sigma_h)\) are disjoint. Then
    either some kernel in schedule $\sigma_{h'}$ already contains a short delaying emerging job (yielding the approximation), or the delaying job of
    \(K(\sigma_{h'})\) is another long incoming job \(i_{h_2}\), and the
    same reasoning applies recursively.
\end{itemize}

We now bound the number of activations of long delaying emerging jobs.
A long delaying emerging job \(i\) that is activated for a kernel
\(K\) remains activated for that kernel in every descendant of the
branch where it was activated. Indeed, if in any complete schedule
respecting \((\pi,\psi)\) job \(i\) were scheduled before \(K\), then
no type (1) job preceding \(i\) in \((\pi,\psi)\) could be rescheduled
after \(K\) without violating the permutation order. But clearly, 
no complete schedule in which these jobs remain before $K$ 
may a strictly smaller makespan than schedule $\sigma_h$. Therefore 
each long delaying emerging job can be activated at most once for a 
given kernel in the tree \(T(\pi,\psi)\).
Since there are at most \(\kappa\) long delaying emerging jobs and at
most \(n-\kappa\) kernels that may require activation, the total number
of distinct complete schedules that can be generated in \(T(\pi,\psi)\)
is bounded by \(\kappa(n-\kappa)\). 

Each complete schedule is produced by a variant of the LDT-heuristic
whose implementation requires \(O(n\log n)\) time per schedule, 
and hence the per-permutation running time bound in {\bf BASIC} is
\((\kappa(n-\kappa)\,O(n\log n)\).  Since {\bf PTAS} enumerates all
\(\kappa_{1.1}!\prod_{K\in\mathcal{K}}\kappa_{1.2}(K)!\) 
combined permutations and for each permutation {\bf BASIC} produces 
a schedule with the desired approximation, {\bf PTAS} finds an 
$(1+1/k)$-approximation schedule in the
claimed overall runtime.\eop
\end{proof}

\comment
--with lemma --
\begin{lemma}\label{lem:activation-once}
Fix a combined permutation \((\pi,\psi)\). Let \(i\) be a long delaying
emerging job and let \(K\) be a kernel for which \(i\) is activated in
some node of the solution tree \(T(\pi,\psi)\). Then \(i\) remains in
the activated state for kernel \(K\) in every descendant node of that
branch of \(T(\pi,\psi)\). In particular, \(i\) can be activated for
the same kernel \(K\) at most once in the construction of \(T(\pi,\psi)\).
\end{lemma}

\begin{proof}
Assume \(i\) is activated for kernel \(K\) at some node \(u\) of
\(T(\pi,\psi)\). Activation means that, in the partial schedule at \(u\),
job \(i\) is postponed (marked pending) so that the short incoming job
that triggered the activation can start immediately; equivalently, in
every complete schedule produced from that branch \(i\) is scheduled
after the short job(s) that were inserted at the activation point.

Suppose, for contradiction, that there exists a descendant node \(v\)
of \(u\) in which \(i\) is no longer activated for \(K\); then in the
complete schedule at \(v\) job \(i\) appears before (or inside) the
execution interval of \(K\). Because the construction always respects
the combined permutation \((\pi,\psi)\), no type (1) job that precedes
\(i\) in \((\pi,\psi)\) can be rescheduled after \(K\) in any schedule
respecting \((\pi,\psi)\). Therefore the relative order of the
type (1) jobs preceding \(i\) with respect to \(K\) is the same in the
schedule at \(v\) as it was before activation. But the activation at
\(u\) was performed precisely because, under the permutation order, the
only way to reduce the overflow completion time of \(K\) was to defer
\(i\); if \(i\) can later be placed before \(K\) without moving any
preceding type (1) jobs after \(K\), then the schedule before
activation already had the same ordering and hence could not have had a
strictly larger makespan than the schedule at \(v\). This contradicts
the premise that activation at \(u\) was necessary to improve the
completion time of \(K\).

Thus no descendant can undo the activation of \(i\) for \(K\); the
activation persists along the branch. Since each activation of a given
job for a given kernel creates a unique branch in \(T(\pi,\psi)\), the
same pair \((i,K)\) cannot be activated more than once during the
construction of \(T(\pi,\psi)\).
\end{proof}

\begin{cor}\label{cor:activation-bound}
For a fixed combined permutation \((\pi,\psi)\) the total number of
distinct activations of long delaying emerging jobs in \(T(\pi,\psi)\)
is at most \(\kappa\,(n-\kappa)\). Consequently, the number of complete
schedules generated for that permutation is \(O\big(\kappa(n-\kappa)\big)\).
\end{cor}

\begin{proof}
By definition there are at most \(\kappa\) long delaying emerging jobs
and at most \(n-\kappa\) kernels that may require activation. By
Lemma \ref{lem:activation-once}, each ordered pair \((i,K)\) (with \(i\)
a long delaying emerging job and \(K\) a kernel) can produce at most
one activation in \(T(\pi,\psi)\). Hence the total number of
activations is bounded by \(\kappa\,(n-\kappa)\). Each activation
produces at most one new complete schedule (via a call to {\bf BASIC}),
so the bound on the number of complete schedules follows.
\end{proof}

\endcomment 

\comment
Let $h'>h$ be any iteration following iteration $h$ in Procedure BASIC. 
By the construction, if the level of node $h'$ in tree $T(\pi,\psi)$ is higher 
than that of node $h$, then node $h'$ is a successor of node $h$ and kernel 
$K(\sigma_{h'})$ succeeds kernel $K(\sigma_h)$. Then our claim  
follows from an easily seen fact that  in a feasible schedule respecting  
permutation $\pi$ in which job $i_h$ is not in the state of activation for kernel $K(\sigma_h)$,
the starting time of the earliest scheduled job of that kernel cannot be less than that in
schedule $\sigma_h$ (in which job $i_h$ is the delaying emerging job). If now the level of node 
$h'$ in tree $T(\pi,\psi)$ is not higher than that of node $h$, then kernel $K(\sigma_{h'})$ precedes 
kernel $K(\sigma_h)$. Once job $i_{h'}$ is activated for kernel $K(\sigma_{h'})$, a gap
is created before that kernel in schedule $\sigma_{h'}$. Furthermore, since any schedule
generated in tree $T(\pi,\psi)$ respects permutation $\pi$, job $i_{h}$ must be included 
before job $i_{h'}$ in a complete schedule generated from node $h'$. Then, because of the
above gap, the completion time of job $i_h$, and hence the full completion time of the 
last job of kernel $K(\pi)$ in the corresponding branch headed by node $h'$ cannot be 
less than that in schedule $\sigma_h$ and our claim follows. 
\endcomment

\section{Earlier PTASs and alternative time complexities}

We refine the time complexity analysis of our PTAS by leveraging techniques from earlier approximation schemes. The runtime of the PTAS from Theorem~\ref{PTAS} can be improved as the term $(n-\kappa)$ can be replaced by $\kappa$. This can be done
by restricting the number of possible job release times, following the approach of \cite{HS,mast}. These schemes rely on the fact that the set of job release times can be reduced to a subset of constant size, depending only on $k$ and not on $n$, without essentially affecting the approximation ratio.

In the first scheme of \cite{HS}, job processing times are rounded to multiples of $P/(nk)$, where $P$ is the total processing time. Release and delivery times are scaled accordingly, yielding an instance with at most $nk$ distinct processing times. This instance is solved by a dynamic programming algorithm that partitions the scheduling horizon into $O(k)$ subintervals and considers all possible distributions of total processing time across them. Since each interval is at most $P$ long, there are at most $P^{O(k)}$ distributions; after rounding, this reduces to $(nk)^{O(k)}$. Jobs within each interval are then scheduled by the LDT heuristic in non‑increasing order of delivery times.

The second scheme of \cite{HS}, and similarly \cite{mast}, distinguishes between short and long jobs. The number of long jobs is bounded by $O(k)$, independent of $n$. The scheduling horizon is divided into $O(\kappa)$ disjoint subintervals according to the reduced set of release times. All possible placements of long jobs into these subintervals are considered, while the set of short jobs in each subinterval is determined by ``guessing'' their total processing time. Each subinterval is then scheduled optimally using the LDT rule.

In the scheme of \cite{mast}, short and long jobs are defined differently, yielding a constant number of small jobs. The instance is converted so that the number of distinct release, processing, and delivery times is bounded by $O(k)$. Jobs are partitioned into subsets with common release and delivery times, giving $O(k^2)$ subsets. Within each subset, jobs are merged into longer composed jobs, leaving at most one small job unmerged. The approximation ratio remains valid under this transformation. Since the number of subsets, small jobs, and large jobs is constant, complete enumeration avoids exponential dependence on $n$. The time complexity is further adjusted by considering only a constant number of possible starting times for large jobs, as in \cite{HS}, while still guaranteeing the desired approximation ratio. Each feasible configuration is scheduled using the LDT heuristic.

Thus, for polynomial‑time approximation, restricting the number of distinct release times incurs essentially no loss in approximation ratio. Both \cite{HS} and \cite{mast} rely on this observation. Define $\eta = \tfrac{1}{2k}\cdot LB$, and round down all release times to the nearest multiple of $\eta$. Without loss of generality, assume $LB=1$ by scaling all job parameters, as in \cite{mast}. In the modified instance, the number of distinct release times is bounded by $2k$. An $(1+1/(2k))$‑approximation for the modified instance can then be transformed into an $(1+1/k)$‑approximation for the original instance, as shown in Lemma~2 of \cite{HS}.

From these observations, the number of different kernels and delaying emerging jobs is bounded by $2\kappa$. Furthermore, the number of iterations in procedure {\bf BASIC} where a long job may cover a short one is at most $\kappa$. Indeed, a long job released simultaneously with a short job cannot cover that short job. Hence, $\kappa$ distinct release times remain vacant, and since a long job can cover at most one simultaneously released short job, the claim follows. Consequently, the time complexity of procedure {\bf BASIC}, when invoked for a permutation of type (1) jobs, is bounded by $\kappa_{1.1}\kappa O(n \log n)$, and we obtain:

\begin{cor}\label{Phase 2}
For a given integer $k > 0$, the PTAS delivers an $(1+1/k)$‑approximation solution in time 
\[
\kappa_{1.1}\kappa O(n\log n)\cdot \kappa_{1.1}! \ \prod_{K\in \cal K}\kappa_{1.2}(K)!.
\]

\eop
\end{cor}


\subsection*{An alternative probabilistic estimation for the running time}

We now provide an intuitive probabilistic bound for the running time of the PTAS. 
Specifically, we estimate the probability that, for a given permutation $\pi$, 
procedure {\bf Backtrack} halts immediately at step (0); in other words, that the 
schedule $\sigma_{h}$ contains a kernel with either (a) a short delaying emerging job or (b) a long delaying emerging job $i$ such that $\hat p_i$ is short. 

Below we give rough probabilistic estimations for events (a) and (b). For (a), 
suppose kernel $K$ possesses a delaying emerging job $e$. Then $r_e < r_o$ and 
$q_e < q_o$ hold, where $o \in K$ is the overflow job. The probability that such 
an $e$ exists is less than $1/4$. For (b), recall that there are $\nu$ type 
(1) jobs in total, of which 
$\kappa$ are long and $\nu-\kappa$ are short. Assuming job processing times are 
uniformly distributed, the probability that the delaying emerging job of 
kernel $K$ is long equals $\kappa/\nu$, while the probability that it is short equals $(\nu-\kappa)/\nu$. Ignoring the event that $e$ can be a long delaying emerging job 
with short $\hat p_e$, the probability of event (b) can be expressed by 
$\kappa/\nu$. Hence,  the probability that {\bf PTAS} does not halt after enumerating permutation $\pi$ is bounded above by $1/4 + \kappa/\nu$, 
and and therefore the probability that PTAS enumerates $\kappa!$ 
permutations is approximately \[(1/4 + \kappa/\nu)^{\kappa!} \]
(note that since $\kappa \ll \nu$, we have $\kappa/\nu \ll 3/4$).  
This probabilistic bound reinforces the central idea of Variable Parameter Analysis: 
the effective running time is governed not by the input size $n$, but by the 
instance‑dependent parameters $\nu$ and $\kappa$, which remain small relative to $n$ 
and thereby explain the observed efficiency of the PTAS in practice.

\section{Experimental study} 

This section complements the theoretical results by reporting empirical 
evidence on the behavior of the variable parameters in practice. 

We briefly overview the results of our experimental study, which aimed to 
establish the values of the variable parameters for real problem instances. 
In total, we tested over 50{,}000 instances across 26 sizes, ranging from 
$n=10$ to $n=50{,}000$ jobs. We report results for the 3{,}900 hardest 
instances and for 900 additional benchmark instances. 

Following standard rules from the literature for instance generation, 
we observed an interesting dependence of the variable parameter 
$\nu_{1.1}$ on the upper limit of the interval from which job processing 
times were drawn. This motivated us to extend the standard derivation rules 
by experimenting with different intervals for processing times. For each 
size $n$, we verified up to 500 upper limits (about 1{,}800 instances) 
to identify the upper limit that maximizes the robustness factor of 
$\nu_{1.1}$. Once this limit was found, we generated 50 additional 
instances with this upper limit, and also 50 instances for each of the 
limits 50 and 100, yielding 150 additional instances per size $n$.

Even for the upper limits maximizing $\nu_{1.1}$, the parameter converges 
to 0 as $n$ grows (see Table~\ref{table type 1.1} and 
Figure~\ref{graphics max type 1.1}). Figures~\ref{graphics average type 1.1} 
and \ref{graphics min type 1.1} illustrate the dependence of the average 
and minimum robustness factors of $\nu_{1.1}$, respectively, on instance size. 
As for the parameters $\nu_{1.2}(K)$, they turned out to be very small 
(see Table~\ref{table type 1.2}). More details on this computational study 
can be found in \cite{kevin}.

\begin{table}
  \centering
    \includegraphics[width=\linewidth]{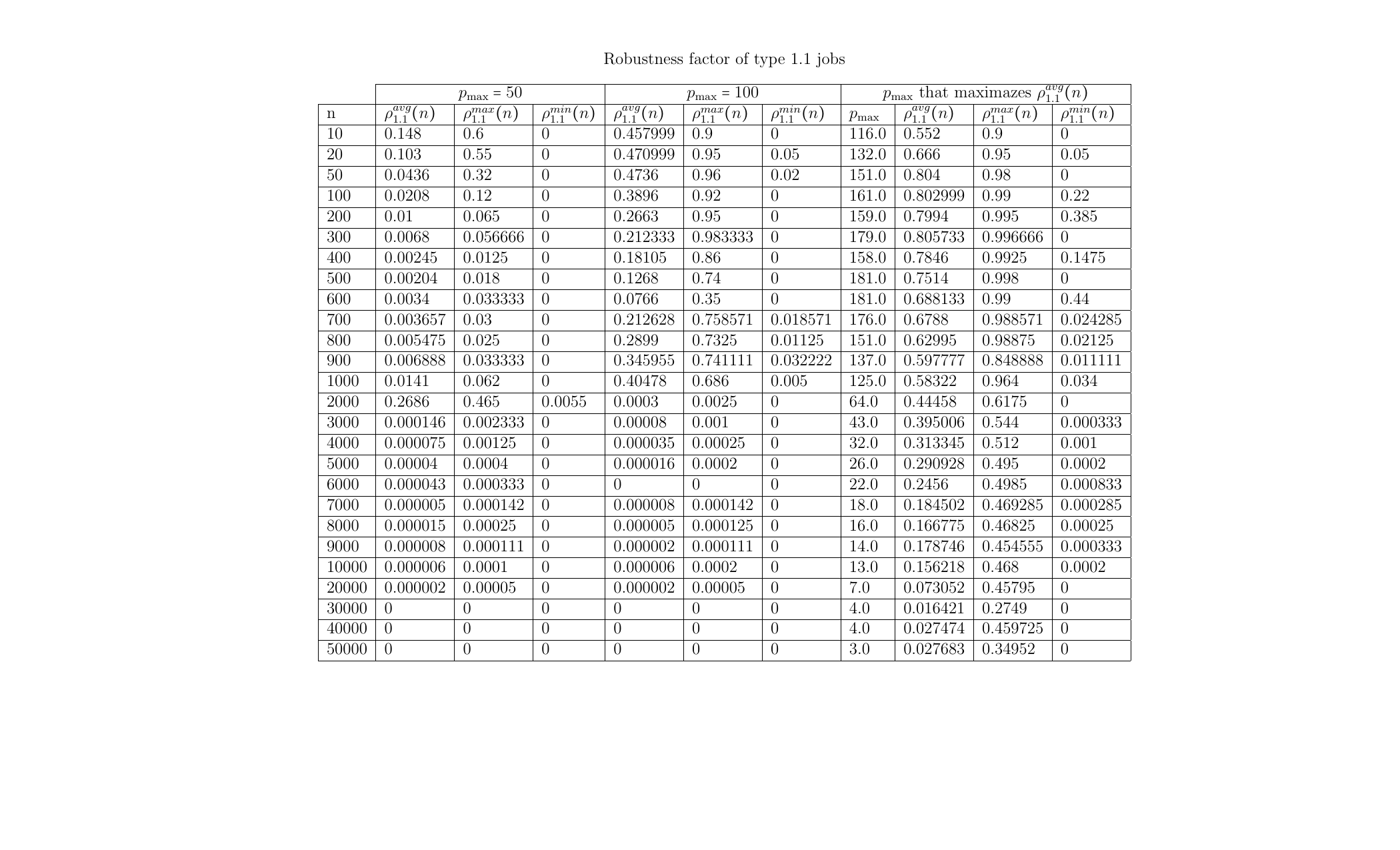} \\
  \caption{The robustness factors for $\nu_{1.1}$}
  \label{table type 1.1}
\end{table}

\begin{figure}
  \centering
    \includegraphics[width=\linewidth]{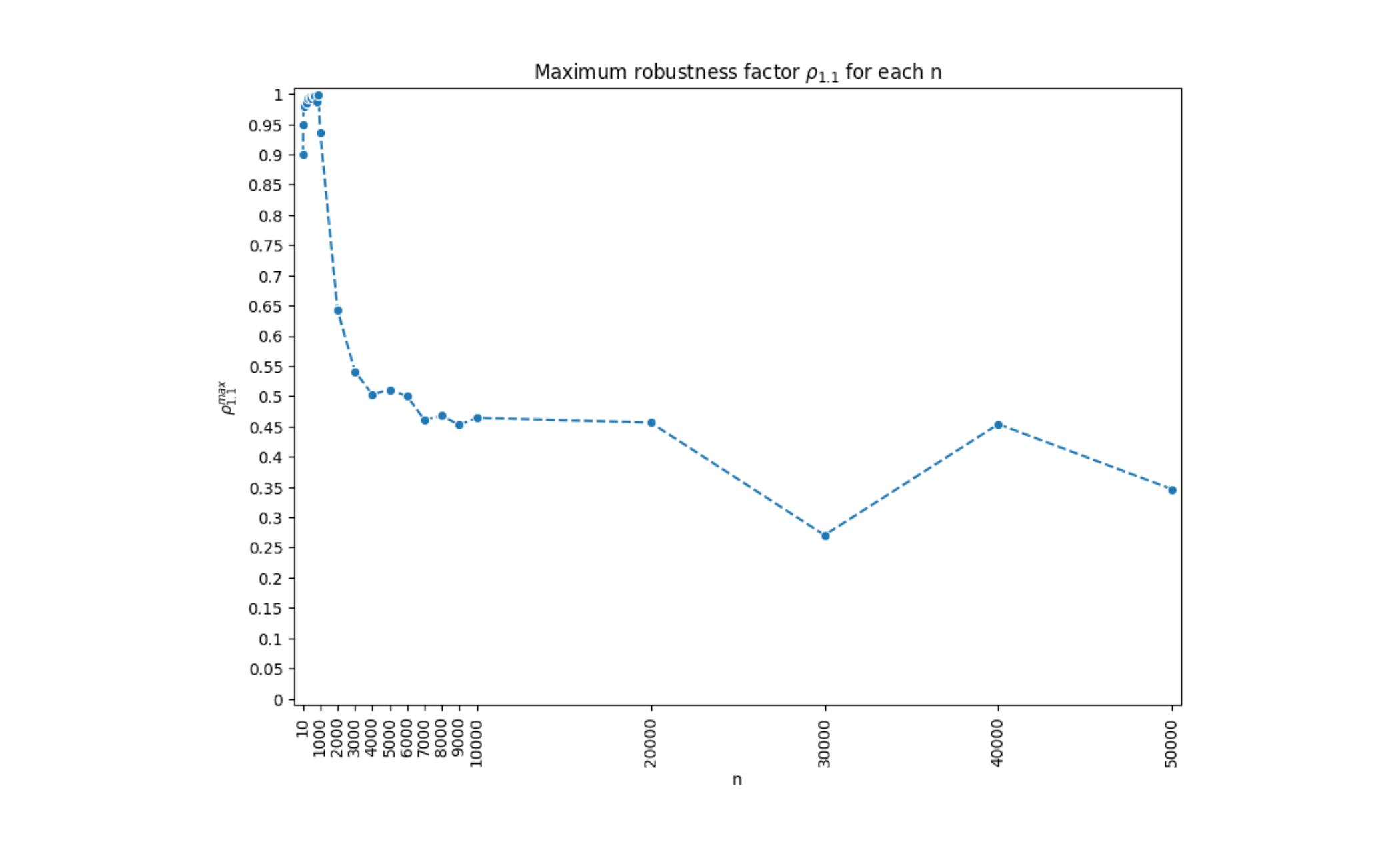} \\
  \caption{Graphics for the maximum number of type 1.1 jobs}
  \label{graphics max type 1.1}
\end{figure}

\begin{figure}
  \centering
    \includegraphics[width=\linewidth]{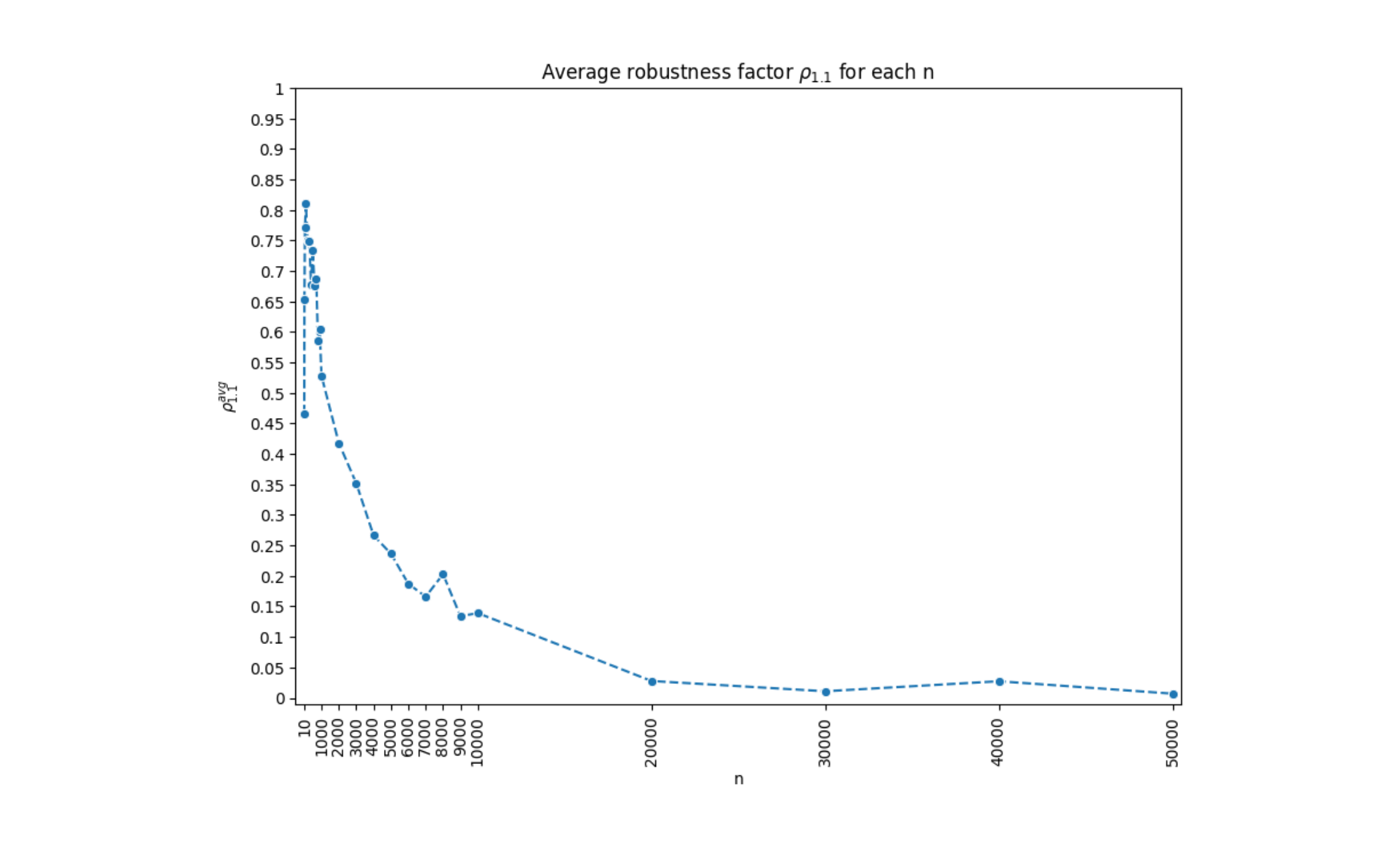} \\
  \caption{Graphics for the average number of type 1.1 jobs}
  \label{graphics average type 1.1}
\end{figure}

\begin{figure}
  \centering
    \includegraphics[width=\linewidth]{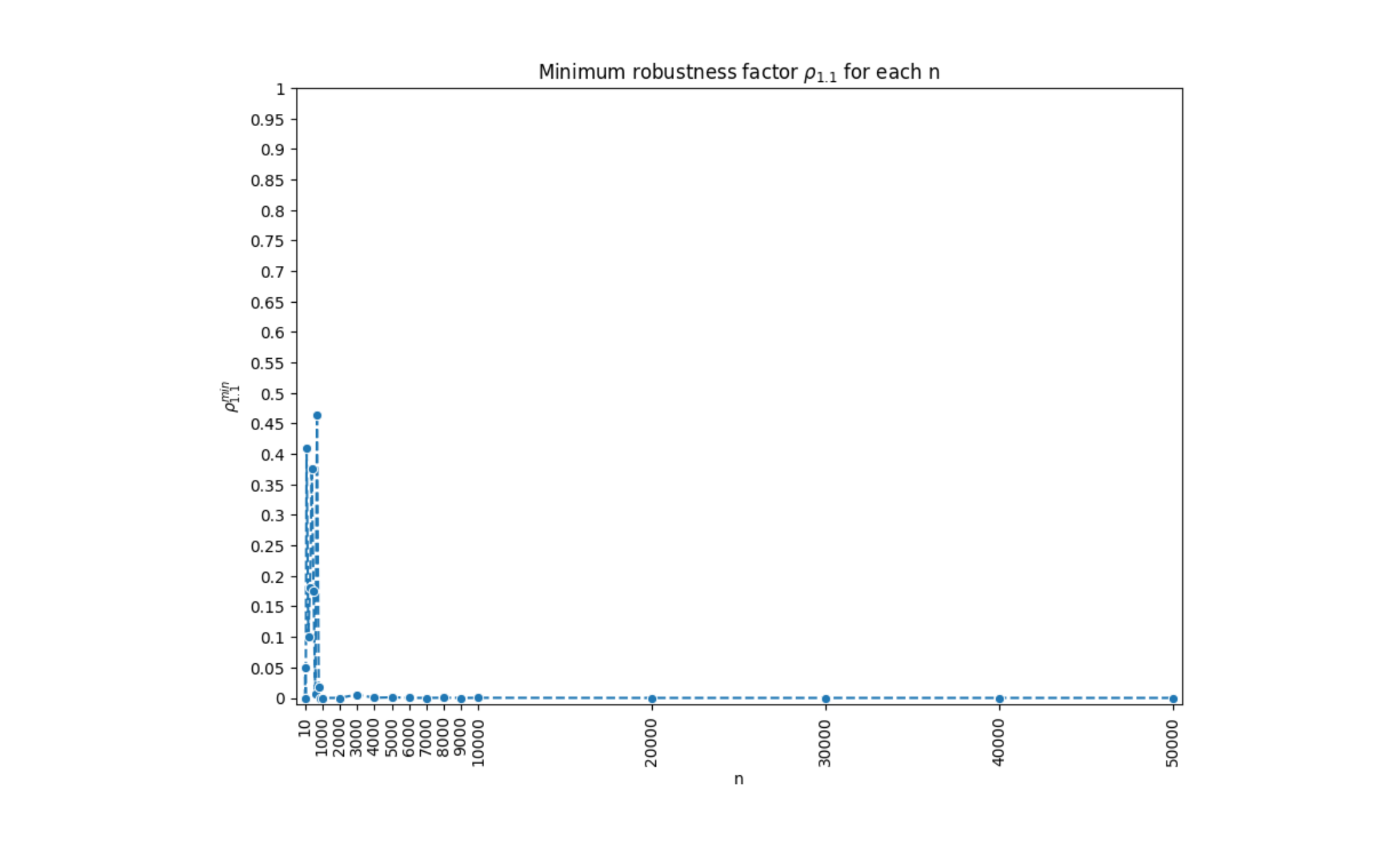} \\
  \caption{Graphics for the minimum number of type 1.1 jobs}
  \label{graphics min type 1.1}
\end{figure}

\begin{table}
  \centering
    \includegraphics[width=\linewidth]{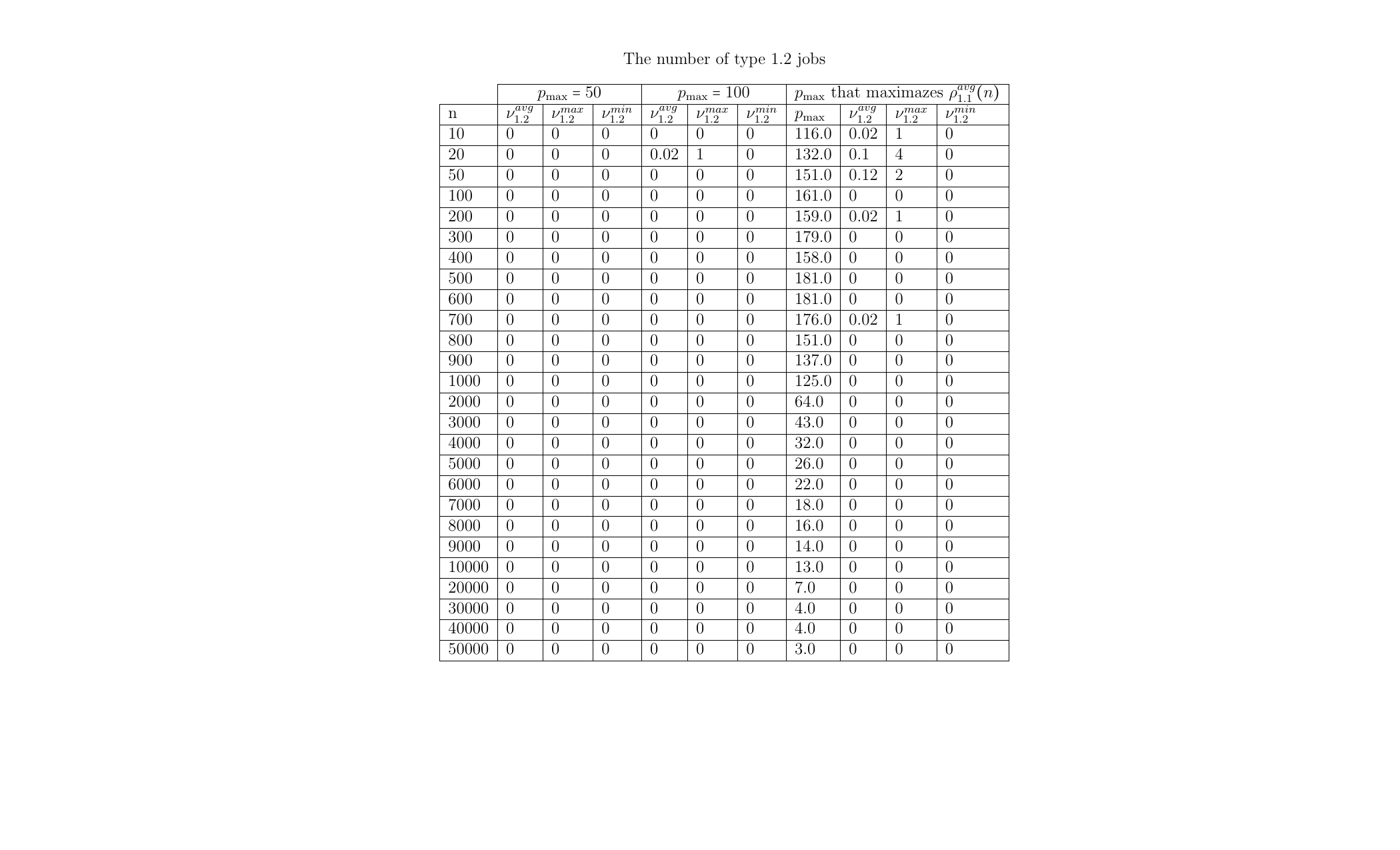} \\
  \caption{The values of $\nu_{1.2}$}
  \label{table type 1.2}
\end{table}

\comment 


\section {Conclusion}

The proposed variable parameter analysis (VPA), conducted for an individual 
instance $I$ of a problem $P$, bears the information on the 
structurally critical elements of that instance in the 
corresponding variable parameter $\nu(I)$. 
Brute-force search for instance $I$ 
is carried out only for the corresponding $\nu(I)$ elements. 
For the studied single-machine scheduling problem,
we proposed an implicit enumeration algorithm (IEA) which 
runtime is exponential only in parameter $\nu(I)$, a polynomial-time
algorithm for the partial scheduling version of the problem, and
a polynomial time approximation scheme (PTAS) which, for any integer 
$k$, creates an $(1+1/k)$-approximation solution to the scheduling 
problem. The runtime of the PTAS is exponential only in variable 
parameter $\kappa  <\nu$, where $\kappa < k$. 
For large amount of randomly generated hard instances of the studied 
scheduling problem, parameter $\nu(I)$ and hence $\kappa(I)$ turned 
out to be arguably smaller 
than the input size  and it showed  a nice asymptotic behavior. 

We emphasize the distinction between variable parameter analysis  
and the traditional fixed parameter analysis (FPA). VPA tailors brute-force enumeration to the instance-specific parameters $\nu(I)$ and $\kappa(I)$,
so the exponential enumeration factor depends on the structurally critical elements of the instance rather than on the worst-case parameters $n$
and $k$. In our experiments $\nu(I)$ and $\kappa(I)$ were typically much smaller than $n$ and $k$, respectively  (Section 6). This, in 
particular, explains the practical efficiency of the IEA and the PTAS 
with VPA; however, when $\nu(I)$ and $\kappa(I)$ approach $n$ and  $k$, 
respectively, VPA and FPA coincide asymptotically and FPA remains 
the natural worst-case framework. We therefore recommend reporting
$\nu(I)$ and $\kappa(I)$ statistics alongside running times in empirical studies and exploring hybrid strategies that switch between VPA and FPA depending on observed instance structure.

One can clearly incorporate lower bounds in the 
solution tree of the IEA and obtain a branch-and-bound algorithm.  
The variable parameter analysis also yielded an 
$(1+1/k)$-approximation PTAS for the scheduling problem, 
which, in terms of the relationship between the run time and the 
guaranteed approximation, essentially benefits 
from the fact that $\kappa<k$. The fast polynomial-time
algorithm that we proposed for partial scheduling setting of
the studied scheduling problem, omits the necessity
of parameterization with the number of scheduled jobs. This result 
might be extendable for parallel machine environment as well. 

The future research can obviously be directed towards the development 
of VP-algorithms for optimization problems. Another interesting
line of research would look for VP-algorithms  
with function $\nu(I)$ that represents the minimum 
size up to which the reduction of instance $I$ is possible. 
This would involve the 
development of new formal tools in fine-grained complexity theory.  

\endcomment 

\newpage

\section{Conclusion}

We introduced a variable parameter analysis (VPA) framework that captures
the structurally critical elements of an instance \(I\) of a problem
\(P\) via an instance‑specific parameter \(\nu(I)\). Brute‑force
enumeration is then confined to the \(\nu(I)\) critical elements rather
than to the entire input. For the single‑machine scheduling problem
studied here we presented three main contributions: an implicit
enumeration algorithm (IEA) whose runtime is exponential only in
\(\nu(I)\); a polynomial‑time algorithm for a natural partial‑scheduling
variant; and a polynomial‑time approximation scheme (PTAS) that, for any
integer \(k\), produces a \((1+1/k)\)-approximate schedule. The PTAS’s
enumeration factor depends only on the variable parameter
\(\kappa(I)<\nu(I)\), and we ensure \(\kappa<k\) in our construction.

Empirically, on a large set of randomly generated hard instances both
\(\nu(I)\) and \(\kappa(I)\) were typically much smaller than the input
size and exhibited favorable asymptotic behavior (Section 6). This
explains the practical efficiency of IEA and the PTAS when VPA is used.

We emphasize the distinction between VPA and traditional fixed
parameter analysis (FPA). VPA tailors enumeration to instance‑specific
parameters \(\nu(I)\) and \(\kappa(I)\), so the exponential factor
reflects structural hardness of the instance rather than worst‑case
values of \(n\) or \(k\). When \(\nu(I)\) and \(\kappa(I)\) are small,
VPA yields substantial practical gains; when they approach \(n\) and
\(k\), respectively, VPA and FPA coincide asymptotically. We therefore
recommend reporting \(\nu(I)\) and \(\kappa(I)\) statistics alongside
running times in empirical studies and exploring hybrid strategies that
switch between VPA and FPA based on observed instance structure.

The solution tree of IEA readily admits lower bounds, yielding a
branch‑and‑bound variant. The PTAS benefits from the fact that
\(\kappa<k\), and the fast polynomial‑time algorithm for the partial
scheduling setting removes the need to parameterize by the number of
scheduled jobs. Extending these results to parallel‑machine models is a
promising direction.

Future work includes developing VP-algorithms for other optimization
problems and investigating whether, for a given problem \(P\), one can
construct a VP‑algorithm whose parameter \(\nu(I_P)\) is the minimum
size to which instance \(I_P\) can be reduced. Pursuing this line will
likely require new tools in fine‑grained complexity theory.


\comment

\section{Conclusion}

We introduced a variable parameter analysis (VPA) framework that captures
instance‑specific structural hardness via a parameter \(\nu(I)\), and
confines brute‑force enumeration to the \(\nu(I)\) critical elements.
For the single‑machine scheduling problem we presented an implicit
enumeration algorithm (IEA) exponential only in \(\nu(I)\), a
polynomial‑time algorithm for a natural partial‑scheduling variant, and
a PTAS that for any integer \(k\) produces a \((1+1/k)\)-approximate
schedule; the PTAS’s enumeration factor depends only on the variable
parameter \(\kappa(I)<\nu(I)\), and our construction ensures
\(\kappa<k\). Empirically \(\nu(I)\) and \(\kappa(I)\) were typically
much smaller than \(n\) and \(k\) (Section~6), which explains the
practical efficiency of IEA and the PTAS under VPA; when \(\nu(I)\) and
\(\kappa(I)\) approach \(n\) and \(k\), respectively, VPA and FPA
coincide asymptotically. We recommend reporting \(\nu\)- and
\(\kappa\)-statistics alongside running times, exploring hybrid VPA/FPA
strategies, and pursuing extensions to parallel machines and formal
tools in fine‑grained complexity.


\endcomment

\comment
Problem instances with $\nu$ being close to $n$ can be constructed.  
In Section 5 we describe a scheduling instance derived from the PARTITION 
problem, where $\nu$, roughly, is the number of items in the 
PARTITION instance (the PARTITION problem is a typical example of 
a non-VP-tractable problem).

Finally, it is worth of mentioning that there can be constructed ``hard'' 
scheduling instances, for which our variable
parameters will approach the total number of objects. 
Such instances can be generated using a 
PARTITION instance. We associate with every element in a PARTITION instance
a partition job with the same weight (duration). We also create a separator job 
$j$ that forces all jobs from  a subset of partition jobs to complete at a given 
time moment in any optimal solution (due to the release 
time of the separator job). In such a scheduling instance, there occurs a single 
kernel containing (type (2)) job $j$ and the remaining jobs are type (1.1) 
jobs.  In an optimal solution, the type (1.1) jobs need to fill in intervals before
and after the separator job in an optimal fashion, so that either to create a
schedule without a delaying emerging job or to create an appropriate forced delay 
for the separator job. This is where a solution to the PARTITION instance can be useful. 
So, consider  a complete schedule $\sigma(\pi)$ created for permutation $\pi$ of the 
type (1.1) partition jobs and the corresponding two fragments in that schedule, the 
first one consisting of the jobs included before  job $j$ and the second one 
consisting of the jobs included after job $j$. If in schedule $\sigma(\pi)$  job 
$j$ starts at its release time and no type (1.1) job realizes the maximum full job 
completion time in that schedule, then it is optimal. Otherwise, in an optimal schedule, 
the first above non-critical fragment is filled in by some partition jobs 
of permutation $\pi$ so that job $j$ is pushed by an appropriate amount of time 
units. If all type (1.1) jobs are released at time 0, the time interval of 
the first fragment in an optimal solution is one of the following time intervals 
$[0,r_j], [0,r_j+1], [0,r_j+2],\dots,[0,r_j+\delta(K)]$, i.e., job $j$ is 
pushed by an integer magnitude $\Delta\in[0,\delta(K)]$. So, a solution to PARTITION 
gives a desired packing of the first fragment and hence also of the second one. The 
application of the standard dynamic programming
algorithm yields the time complexity $O((r+\Delta)\nu)$ for a given $\Delta$. 
Using binary search, we restrict the possible values for $\Delta$ and obtain an overall 
cost $O((r+\Delta)\nu)\log(\delta(K))$ for a procedure that creates  
the corresponding schedules and selects a schedule with the minimum makespan. 
\endcomment

\comment 
\begin{obs}\label{delaying}
The overflow job of a kernel $\bar K$ in LDT-schedule $S$ is either the 
overflow job of an atomic kernel $K^*$ or a type (1.1) or a type (1.2) job. 
\end{obs}
Proof. First note that no type (4) job scheduled 
after kernel $\bar K$ will be right-shifted in schedule $S$ by more time 
units than the last scheduled job of kernel $K$, and hence it cannot become 
the overflow job in the former kernel. We claim that neither a type (3) 
job $i$ associated with kernel $K$ may become  the overflow job in 
kernel $\bar K$. It will suffice to show that job $i$, 
pushed by a type (1.2) job $j$ associated with kernel $K$, cannot be 
converted to an overflow job in schedule $S$. 
Suppose 
first that there is a gap between job $i$ and an atomic kernel $K^*$ in 
schedule   $S$. Then clearly, the claim holds as
the overflow job in schedule $S^*[K]$ is from  kernel $K^*$. Suppose
there is no such gap. Since $i$ is a type (3) job, the last job of 
kernel $K^*$ should have been the overflow job in an earlier generated
complete schedule, whereas there was no gap 
between job $i$ and kernel $K^*$ in that schedule. 
Then if job $i$ is pushed by job $j$, 
the jobs between  job $i$ and kernel $K^*$ will be right-shifted by the
same amount. Then the last job of kernel $K^*$ will remain to 
be the overflow job in that schedule. This completes the proof. \eop
\endcomment

\small


\end{document}